\PassOptionsToPackage{table,dvipsnames}{xcolor}
\documentclass[final,3p,times]{elsarticle}

\usepackage{lineno,hyperref}
\modulolinenumbers[5]

\journal{Journal of \LaTeX\ Templates}

\usepackage{amssymb}
\usepackage{listings}
\usepackage{float}
\usepackage{subfig}
\usepackage{algorithm}
\usepackage[noEnd=true,indLines=true]{algpseudocodex}
\usepackage[fleqn]{amsmath}
\usepackage{graphicx}
\usepackage{showkeys}
\usepackage{amsthm}
\usepackage{psfrag}
\usepackage{pstricks}
\usepackage[abs]{overpic}
\usepackage{epsfig}
\usepackage{dsfont}
\usepackage{comment}
\usepackage{caption}
\usepackage{booktabs}
\definecolor{CeruleanRef}{RGB}{12,127,172}
\hypersetup{colorlinks=true}

\usepackage{tikz}
\usepackage{pgfplots}
\usepackage{pgfmath}
\usetikzlibrary{calc}

\DeclareMathAlphabet{\mycal}{OMS}{cmsy}{m}{n}
\newcommand{\V}[1]{\mathbf{#1}}  %
\newcommand{\M}[1]{\mathbf{#1}}  %
\newcommand{\OP}[1]{\mycal{#1}} %

\newcommand{\RoundRect}[4]{%
\draw[
  rounded corners=5,
  black!60!white,
  fill=black!5!white,
] (#1,#2) rectangle ++(#3,#4);
}

\newcommand{\DofSquare}[5]{%
\draw[black] (#1,#2) rectangle ++(#3,#4);
\draw node[fill,circle,inner sep=0pt,minimum size=2.5pt,#5] at (#1, #2) {};
\draw node[fill,circle,inner sep=0pt,minimum size=2.5pt,#5] at ({#1+#3/2}, #2) {};
\draw node[fill,circle,inner sep=0pt,minimum size=2.5pt,#5] at ({#1+#3}, #2) {};
\draw node[fill,circle,inner sep=0pt,minimum size=2.5pt,#5] at (#1, {#2+#4/2}) {};
\draw node[fill,circle,inner sep=0pt,minimum size=2.5pt,#5] at ({#1+#3/2}, {#2+#4/2}) {};
\draw node[fill,circle,inner sep=0pt,minimum size=2.5pt,#5] at ({#1+#3}, {#2+#4/2}) {};
\draw node[fill,circle,inner sep=0pt,minimum size=2.5pt,#5] at (#1, {#2+#4}) {};
\draw node[fill,circle,inner sep=0pt,minimum size=2.5pt,#5] at ({#1+#3/2}, {#2+#4}) {};
\draw node[fill,circle,inner sep=0pt,minimum size=2.5pt,#5] at ({#1+#3}, {#2+#4}) {};
}

\newcommand{\QuadSquare}[5]{%
\draw[black] (#1,#2) rectangle ++(#3,#4);
\draw node[fill,circle,inner sep=0pt,minimum size=2.5pt, #5] at (#1+#3/3, #2+#4/3) {};
\draw node[fill,circle,inner sep=0pt,minimum size=2.5pt, #5] at (#1+2*#3/3, #2+#4/3) {};
\draw node[fill,circle,inner sep=0pt,minimum size=2.5pt, #5] at (#1+#3/3, #2+2*#4/3) {};
\draw node[fill,circle,inner sep=0pt,minimum size=2.5pt, #5] at (#1+2*#3/3, #2+2*#4/3) {};
}

\newcommand{\DofGrid}[5]{%
\DofSquare{#1}{#2}{#3}{#4}{#5}
\DofSquare{#1+#3}{#2}{#3}{#4}{#5}
\DofSquare{#1}{#2+#4}{#3}{#4}{#5}
\DofSquare{#1+#3}{#2+#4}{#3}{#4}{#5}
}

\def\spacingfactor{0.85}

\newcommand{\DofElem}[5]{%
\DofSquare{#1}{#2}{#3*\spacingfactor}{#4*\spacingfactor}{#5}
\DofSquare{#1+#3/\spacingfactor}{#2}{#3*\spacingfactor}{#4*\spacingfactor}{#5}
\DofSquare{#1}{#2+#4/\spacingfactor}{#3*\spacingfactor}{#4*\spacingfactor}{#5}
\DofSquare{#1+#3/\spacingfactor}{#2+#4/\spacingfactor}{#3*\spacingfactor}{#4*\spacingfactor}{#5}
}

\newcommand{\QuadElem}[5]{%
\QuadSquare{#1}{#2}{#3*\spacingfactor}{#4*\spacingfactor}{#5}
\QuadSquare{#1+#3/\spacingfactor}{#2}{#3*\spacingfactor}{#4*\spacingfactor}{#5}
\QuadSquare{#1}{#2+#4/\spacingfactor}{#3*\spacingfactor}{#4*\spacingfactor}{#5}
\QuadSquare{#1+#3/\spacingfactor}{#2+#4/\spacingfactor}{#3*\spacingfactor}{#4*\spacingfactor}{#5}
}

\newcommand{\RoundRectGrid}[2]{%
\RoundRect{0}{0}{#1}{#2}
\RoundRect{#1 + 0.075}{0}{#1}{#2}
\RoundRect{0}{#2 + 0.075}{#1}{#2}
\RoundRect{#1 + 0.075}{#2 + 0.075}{#1}{#2}
}

\usepackage{lipsum}                     %
\usepackage{xargs}                      %
\usepackage[colorinlistoftodos,prependcaption,textsize=tiny, textwidth=18mm]{todonotes}
\newcommandx{\unsure}[2][1=]{\todo[linecolor=red,backgroundcolor=red!25,bordercolor=red,#1]{#2}}
\newcommandx{\change}[2][1=]{\todo[linecolor=blue,backgroundcolor=blue!25,bordercolor=blue,#1]{#2}}
\newcommandx{\info}[2][1=]{\todo[linecolor=olive,backgroundcolor=olive!25,bordercolor=olive,#1]{#2}}
\newcommandx{\improvement}[2][1=]{\todo[linecolor=violet,backgroundcolor=violet!25,bordercolor=violet,#1]{#2}}

\journal{arXiv}

\begin{document}
\hypersetup{allcolors=CeruleanRef} %

\begin{frontmatter}

 \title{Finite elements for Mat\'ern-type random fields: Uncertainty in computational mechanics and design optimization}

 \affiliation[llnl]{
  organization={Lawrence Livermore National Laboratory},%
  country={United States of America}}
 \affiliation[cern]{organization={CERN},%
  city={Geneva},
  country={Switzerland}}
 \affiliation[tum]{
  organization={School for Computation, Information, and Technology,
    Technical University of Munich},%
  country={Germany}}
 \affiliation[brown]{
  organization={Division of Applied Mathematics,
    Brown University},%
  country={United States of America}}

 \author[cern,tum]{Tobias Duswald}
 \ead{tobias.duswald@tum.de}
 \author[brown]{Brendan Keith}
 \ead{brendan_keith@brown.edu}
 \author[llnl]{Boyan Lazarov}
 \ead{lazarov2@llnl.gov}
 \author[llnl]{Socratis Petrides}
 \ead{petrides1@llnl.gov}
 \author[tum]{Barbara Wohlmuth}
 \ead{wohlmuth@cit.tum.de}

 \begin{abstract}
  This work highlights an approach for incorporating realistic uncertainties into scientific computing workflows based on finite elements, focusing on  prevalent applications in computational mechanics and design optimization.
  We leverage Matérn-type Gaussian random fields (GRFs) generated using the SPDE method to model aleatoric uncertainties, including environmental influences, variating material properties, and geometric ambiguities.
  Our focus lies on delivering practical GRF realizations that accurately capture imperfections and variations and understanding how they impact the predictions of computational models as well as the shape and topology of optimized designs.
  We describe a numerical algorithm based on solving a generalized SPDE to sample GRFs on arbitrary meshed domains.
  The algorithm leverages established techniques and integrates seamlessly with the open-source finite element library MFEM and associated scientific computing workflows, like those found in industrial and national laboratory settings.
  Our solver scales efficiently for large-scale problems and supports various domain types, including surfaces and embedded manifolds.
  We showcase its versatility through biomechanics and topology optimization applications, emphasizing the potential to influence these domains.
  The flexibility and efficiency of SPDE-based GRF generation empowers us to run large-scale optimization problems on 2D and 3D domains, including finding optimized designs on embedded surfaces, and to generate design features and topologies beyond the reach of conventional techniques.
  Moreover, these capabilities allow us to model and quantify geometric uncertainties on reconstructed submanifolds, such as the interpolated surfaces of cerebral aneurysms provided by postprocessing CT scans.
  In addition to offering benefits in these specific domains, the proposed techniques transcend specific applications and generalize to arbitrary forward and backward problems in uncertainty quantification involving finite elements.
 \end{abstract}


\end{frontmatter}
\setcounter{page}{1}

\section{Introduction}

This article demonstrates a scalable and expressive approach for generating \textit{Gaussian random fields} with applications to modeling uncertainties in computational mechanics and design optimization.
The goal is to illustrate the importance and the necessity of incorporating random fields with widely varying properties, such as correlation length, smoothness, and anisotropy, on arbitrary computational domains and embedded surfaces.
The ability to easily control these properties and generate corresponding samples significantly reduces epistemic uncertainties in computational models.
Therefore, we focus on examples with aleatoric uncertainties, describing the systems’ intrinsic randomness,  assigned to the three categories described below.
\textit{Environmental and external excitations} describe external influences such as loads or other constraints.
Random wind forces \cite{mann1994spatial, mann1998wind, Keith2021, keith2021learning} acting on buildings~\cite{Kodakkal2022} and wind turbines~\cite{nybo2021analysis,nybo2020evaluation} serve as compelling examples in this category.
\textit{Material uncertainties} are associated with spatially varying material properties such as Young moduli, diffusion coefficients, or damping coefficients \cite{Torquato2002,Robertson2022}.
Such variations may be caused by material heterogeneity, changes occurring during manufacturing, or environmental degradation.
Moreover, \textit{geometric uncertainties} describe disturbances in the system's geometry, for instance, the shape or the positioning of specific components.
Examples arise in additive manufacturing processes where smooth surfaces may show artifacts and imperfections from manufacturing (cf.~\cite{Khristenko2021}).
Computational models that fail to account for aleatoric uncertainties are prone to inaccurate predictions, poor risk assessment, and suboptimal designs \cite{Kiureghian2009}.

The examples considered here leverage spatially correlated noise models.  Mathematically, such noise is described by \textit{random fields}, a generalization of a stochastic process modeling spatial variability.
The fields are defined on a domain $D \subset \mathbb{R}^{d}$, $d = 1, 2, 3$, and possess local notions of length scale, smoothness, and orientation (anisotropy).
Practical system analyses and optimizations embed these fields into the physical model, e.g., as a random coefficient in a partial differential equation (PDE), and different field realizations appear in the inner loop of stochastic optimization algorithms and uncertainty quantification (UQ) workflows.

Among various types, \textit{Gaussian random fields} (GRFs) are particularly notable as their mean value and covariance functions characterize them completely.
The mean is a single-variable function $m \colon D \to \mathbb{R}$, and the covariance can be expressed as a two-variable function $\mathcal{C} \colon D\times D \to \mathbb{R}$.
As espoused below, we focus on \textit{Mat\'ern-type} random fields that can be defined on an arbitrary Lipschitz domain $D \subset \mathbb{R}^{d}$.

\begin{figure}
 \centering
 \includegraphics[width=0.6\textwidth]{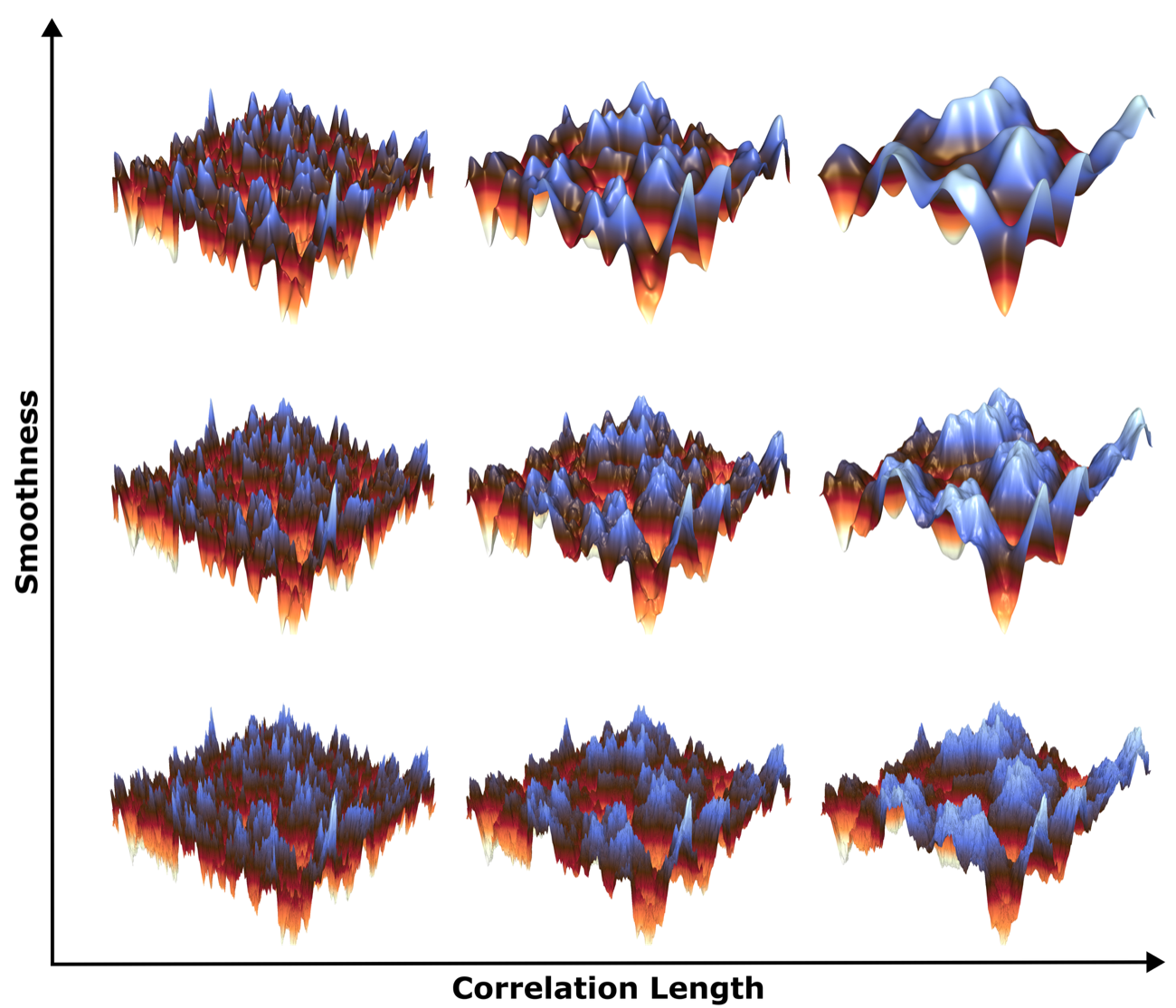}
 \caption{
  Isotropic, homogeneous, Mat\'ern-type Gaussian random fields of different smoothness and correlation length generated with the SPDE method.
  The \textit{smoothness} and \textit{correlation-length} increase along their respective axis.
  The examples are generated with homogeneous Neumann boundary conditions on $D=(0,1)^2$ using identical white noise realizations for all samples.
 }
 \label{fig:rf-matrix-annotated}
\end{figure}

Classically, Mat\'ern random fields are a special class of homogeneous GRFs with covariance defined over the domain $D = \mathbb{R}^d$ as
\begin{equation}
 \label{eq:MaternCovariance}
 \mathcal{C}(\V{x},\V{y})
 =
 \frac{\sigma^2}{2^{\nu-1}\Gamma(\nu)}(\kappa r)^\nu \mathcal{K}_\nu (\kappa r)
 \,,
 \quad
 r = \|\V{x}-\V{y}\|_2
 \,,
 \quad
 \kappa = \frac{\sqrt{2\nu}}{l}
 \,,
 \quad
 \V{x},\V{y} \in \mathbb{R}^d
 \,,
\end{equation}
where $\sigma^2, \nu, l > 0$ are the variance, smoothness parameter, and correlation length, respectively.
Moreover, in this definition, we have the gamma function $\Gamma$ and the modified Bessel function of the second kind $\mathcal{K}_\nu$.
The Mat\'ern covariance~\eqref{eq:MaternCovariance}, known for its flexibility and effectiveness, describes spatial uncertainties arising across diverse fields, including astronomy, health, engineering, imaging, environmental studies, geostatistics, and econometrics~\cite[Section~1.2]{Lindgren2022}.
Its versatility across numerous applications stems from its ability to interpolate between the exponential and squared-exponential covariance function.
Moreover, it facilitates precise control over the field's smoothness and its characteristic length scales via the parameters $\nu$ and $l$, respectively.
Mat\'ern-type random fields generalize these attractive features to a larger space of covariance kernels on domains beyond $D=\mathbb{R}^d$.
Figure~\ref{fig:rf-matrix-annotated} illustrates random fields of Mat\'ern-type covariance with different length scales and smoothness on the domain $D=(0,1)^2$.

We consider problems requiring samples to be represented on the same discrete mesh utilized for modeling and analyzing the physics.
The most popular techniques for generating such samples of GRFs employ \textit{Karhunen--Loève} (KL) type series expansions or filter-based Fourier methods.
Both techniques represent the GRF as a linear combination of spatially varying basis functions with uncorrelated Gaussian random variables as coefficients.
The size and type of basis determines the computational complexity of the random field generator. For correlation lengths that are relatively small compared to the characteristic size of the computational domain, the number of terms in KL-based expansions becomes large, and the associated computational cost is prohibitively expensive. On the other hand, Fourier-based methods are very computationally efficient for homogeneous GRFs but non-trivial to apply to the intricate design domains that are ubiquitous in computational mechanics.

The \textit{Stochastic Partial Differential Equation} (SPDE) methodology proposed in~\cite{Lindgren2011} alleviates these limitations.
The idea is to model a GRF as the solution to a fractional SPDE that smooths spatial Gaussian white noise $\mathcal{W}$, i.e., the inverse of the differential operator plays the role of a filter for convolving $\mathcal{W}$.
In particular, Whittle~\cite{Whittle1954,Whittle1963} showed that the full-space ($D = \mathbb{R}^d$) solution to
\begin{equation}\label{eq:SPDEWhittle}
 (\mathcal{I} - \kappa^{-2} \Delta)^{k} u = \eta \mathcal{W} \, ,
 \quad
 k = \nu/2 + d/4
 \,,
\end{equation}
is a GRF with covariance~\eqref{eq:MaternCovariance}.
Here, $\mathcal{I}$ denotes the identity and $\eta$ is a normalization constant depending on $\sigma$, $\nu$, and $l$; see, e.g.,~\eqref{eq:eta_normalization} below.
Lindgren et al.~\cite{Lindgren2011} understood that one might exploit this link between~\eqref{eq:MaternCovariance} and~\eqref{eq:SPDEWhittle} to efficiently generate random field realizations for downstream applications and coined the term \textit{SPDE method}.
The method does not impose any special limitations on the domain geometry, provides excellent flexibility for controlling the RF regularity, and allows for specifying particular behavior at the boundaries.

The method recently began to gain popularity in computational mechanics.
For instance, Koh et al.~\cite{Koh2023} used the SPDE method in the context of Bayesian modeling and Gaussian process regression.
Additionally and of significant relevance to this work, Guilleminot et al.~\cite{Guilleminot2019} and Ben-Yelun et al.~\cite{BenYelun2024} applied the methodology to optimize the topology of cantilevers and lattice structures, respectively.
Here we demonstrate its efficacy in simulating complex uncertainties prevalent in scientific computing workflows for forward and inverse problems on modern distributed computing platforms.
We present a numerical algorithm solving a generalized form of~\eqref{eq:SPDEWhittle} to sample GRFs of Mat\'ern covariance for computational workflows involving arbitrary meshed domains.
The algorithm combines established techniques introduced in the context of the SPDE method;
most notably, it promotes the general setup of~\cite{Lindgren2011}, employs a rational approximation~\cite{Nakatsukasa2018} to handle the fractional operator~\cite{bolin2019rational,harizanov2018positive}, and uses the efficient white noise sampling suggested in~\cite{Croci2018}.
We provide a performant algorithm implementation, available as part of the open-source finite element library MFEM~\cite{Anderson2021,andrej2024mfem} (as of version 4.6).
Standard domain decomposition techniques ensure perfect parallel scalability for large-scale optimization problems.
The solver can be applied to 1D, 2D, or 3D domains, a surface part of the domain's boundary, or an arbitrary manifold embedded in the 3D space.
The GRF discretization naturally utilizes the mesh of the physical problem and does not require additional discretization libraries and tools.

Highlighting the versatility of this methodology, we present its applications in two key areas we anticipate to benefit substantially: biomechanics and topology optimization.
We show applications in modeling micro-structures with post-processed GRFs, illustrate the effects of uncertainties on complex geometries by considering GRFs on a cerebral aneurysm, and show comprehensive numerical experiments for topology optimization under uncertainty (thermal compliance and bridge design).
In the latter set of examples, we observe new design features and topologies that are out of reach with established random field discretization methods.
The paper proceeds with Section~\ref{sec:GeneratingRandomFields}, which is devoted to methods generating GRFs, and presents our argument for adopting the SPDE method.
Next, Section~\ref{sec:SPDE_discretization} contains an overview of the SPDE method and its numerical implementation.
Sections~\ref{sec:aneurysm} and~\ref{sec:num_ex} are devoted to biomechanics and topology optimization applications, respectively, demonstrating the potential of the method in these fields.
The paper ends in Section~\ref{sec:conclusions} with a summary of our findings.

\section{Numerical methods for generating Gaussian random fields}
\label{sec:GeneratingRandomFields}

Gaussian random fields can be sampled through various techniques, each characterized by distinct theoretical underpinnings and practical applications.
For a zero-mean GRF, a direct approach begins with specifying the covariance matrix $\M{C}$ between points at different locations and factorizing it with a Cholesky decomposition.
Such matrix decomposition methods have been explored by numerous authors~\cite{Davis1987,Dietrich1995,Blanchard2015,Feischl2018}.
The so-called spectral method, utilizes the association between the covariance matrix and the power spectrum to perfrom this factorization in frequency space~\cite{Mejia1974,Shinozuka1971, Shinozuka1972,Gutjahr1997,Ruan1998,Lang2011,Ravalec2000}.
The centerpiece of this strategy is the Fourier transform, which facilitates the transformation to frequency space and the conversion of white noise into spatially-correlated random fields.
Another approach is the \textit{Karhunen--Loève} (KL) expansion~\cite{Loeve1977,Schwab2006,Zhang2004}, which leverages an optimal decomposition into eigenfunctions of the covariance kernel.
Finally, PDE-based sampling techniques~\cite{Lindgren2011,Lindgren2022} have been developed, leveraging the connection between stochastic, fractional PDEs and certain GRFs~\cite{Whittle1954,Whittle1963}.
Additional methods, such as the circulant embedding~\cite{Dietrich1997,Graham2018,Park2015}, also deserve recognition, as extensively reviewed in~\cite{Liu2019}.

This section highlights some of the most prominent methodologies for generating GRFs.
First, the \textit{Expansion Optimal Linear Estimation} (EOLE) method~\cite{Li1993}, a variant of the KL expansion, is reviewed in Section~\ref{sec:KLforRF}.
Second, the spectral method, employing the \textit{fast Fourier transform} (FFT), is summarized in Section~\ref{sec:FFTforGRF}.
Section~\ref{sec:SPDEforGRF} is dedicated to the SPDE method~\cite{Whittle1954,Whittle1963,Lindgren2011}, the method of choice in this work.
The selection of a particular method is contingent upon the specific objectives and constraints of the problem at hand, the desired characteristics of the random field (such as its correlation structure), and the computational resources available.
While some methods are more suited to specific applications, others offer more flexibility regarding the attributes of the generated field.
This section concludes by justifying and promoting the choice of the SPDE method for finite element-based applications.

\subsection{The Karhunen–-Loève expansion and the {EOLE} method} \label{sec:KLforRF}

The EOLE method~\cite{Li1993} is a KL-type series expansion method.
In general, the KL expansion provides Gaussian random field samples by forming a linear combination of normalized eigenfunctions $\phi_i(\V{x})$ of the covariance kernel $\mathcal{C}(\V{x},\V{y})$ with i.i.d.\ coefficients drawn
from a normal distribution. In particular, for a GRF with mean $m(\V{x})$, we have the
expansion
\begin{equation}\label{eq:KL}
 u(\V{x}) = m(\V{x}) +
 \sum_{i=1}^{\infty} \alpha_i \sqrt{\lambda_i} \phi_i(\V{x})
 \,,
\end{equation}
where $\int_D \mathcal{C}(\V{x},\V{y}) \phi_i (\V{y}) \operatorname{d}\! \V{y} = \lambda_i \phi_i(\V{x})$, $\|\phi_i\|_{L^2(D)} = 1$, and $\alpha_i \sim \mathcal{N}(0,1)$ for each $i = 1, 2, \ldots$~\cite{ghanem2003stochastic}.

The EOLE method is a popular version of this expansion where the covariance kernel is sampled in a covariance matrix at a representative set of points $\{\V{x}^{(i)} \in D \mid i = 1,\dots, M \}$ in the domain $D$, and the eigenfunctions are approximated by their values at these points.
For the selected set of points, the elements of the covariance matrix are computed as $C_{ij} = \mathcal{C}(\V{x}^{(i)},\V{x}^{(j)})$.
The eigenvalues and the eigenvectors of the resulting matrix $C \in \mathbb{R}^{M\times M}$ are denoted as $\lambda_i$ and $\V{e}^{(i)}$, respectively.
The eigenvalues are ordered in non-increasing order, i.e., $\lambda_i \geq \lambda_j$ for all $j > i$.
Following~\cite{Li1993}, the EOLE approximation of a GRF with mean $m(\V{x})$ and covariance $\mathcal{C}(\V{x},\V{y})$ is given as
\begin{equation}\label{eq:EOLE}
 u(\V{x}) \approx
 m(\V{x}) +
 \sum_{i=1}^{K} \alpha_i
 \sum_{j=1}^M \mathcal{C}(\V{x},\V{x}^{(j)})e_j^{(i)} / \sqrt{\lambda_i}
 \,,
 \quad
 K \leq M
 \,,
\end{equation}
where $\alpha_i \sim \mathcal{N}(0,1)$ are again independent Gaussian random variables, and $e_j^{(i)}$ denotes the $j$-th component of the $i$-th eigenvector.

In applications, the sum over the index $i$ in~\eqref{eq:EOLE} is truncated after $K \ll M$ terms, i.e., the field is approximated using the $K$ largest eigenvalues.
For smooth random fields with correlation lengths comparable to the characteristic size of the domain $D$, the ordered sequence of eigenvalues decays fast~\cite{Maitre2010}, and the above expression provides a good approximation of the random field with only a relatively small number of terms $K$.
Conveniently, this truncation reduces the computational complexity of evaluating $u(\V{x})$.
However, for rough oscillatory GRFs or computational domains much larger than the correlation lengths of the fields, an accurate representation requires a considerable number of terms, resulting in significantly increased computational costs.

\subsection{Spectral methods}
\label{sec:FFTforGRF}

A \textit{homogeneous} random field in $\mathbb{R}^d$ ($d>1$) is the generalization of a stationary stochastic process in 1D.
The statistical properties of such a field are invariant under parallel translations and, therefore, independent of the actual positions in space.
In this case, the covariance simplifies to
\begin{equation}
 \label{eq:HomogeneousCovariance}
 \mathcal{C}(\V{x}, \V{y}) =
 \widehat{\mathcal{C}}(\V{x} - \V{y})
 \,,
 \quad
 \V{x},\V{y} \in \mathbb{R}^d
 \,
\end{equation}
showing that the covariance between two points $\V{x},\V{y}$ may only depend on the length and direction of their connection.
Further demanding invariance of the field's statistical properties under all isometric transformations yields the important special case of an \emph{isotropic}, homogeneous random field, i.e., the covariance function exclusively depends on the distance $\mathcal{C}(\V{x}, \V{y}) = \widetilde{\mathcal{C}}(\vert \V{x} - \V{y} \vert)$.

Homogeneous random fields can be constructed by convolving spatial Gaussian white noise $\mathcal{W}$ with a function
$T \in L^1(\mathbb{R}^d)$~\cite{Ravalec2000,Mejia1974,Shinozuka1971,Shinozuka1972,Journel1974,Oliver1995};
namely,
\begin{equation}
 \label{eq:Convolution}
 u(\V{x}) = (T * \mathcal{W})(\V{x})
 \,.
\end{equation}
Here, the function $T$ is often referred to as a \textit{filter} and the convolution is often interpreted as $T$ smoothing (filtering) the white noise input.
The Fourier transform $\mathcal{F}$ is a computationally efficient way to evaluate the convolution in~\eqref{eq:Convolution} because it becomes a pointwise multiplication in frequency space.
Thus,~\eqref{eq:Convolution} is computed as
\begin{equation}\label{eq:FFT1}
 u
 = \mathcal{F}^{-1}( \mathcal{F} (T * \mathcal{W}) )
 = \mathcal{F}^{-1}(
 \mathcal{F}( T )\,
 \mathcal{F}( \mathcal{W} ) )
 \,.
\end{equation}
The filter $T$ determines the covariance structure of the random field $u$.
According to the Wiener--Khinchin theorem~\cite{Wiener1930,Khintchine1934}, the power spectrum $\mathcal{P}(\V{k})$, which characterizes the distribution of variance and power across frequencies, is the Fourier transform of the covariance function $\widehat{\mathcal{C}}(\V{r})$, where $\V{r}=\V{x}-\V{y}$.
Setting $\mathcal{F}\left( T \right) = \sqrt{\mathcal{P}}$ results in a random field with a desired mean $m$ and homogeneous covariance $\widehat{\mathcal{C}}\left(\V{r}\right)$,
\begin{equation}\label{eq:FFT2}
 u
 = m + \mathcal{F}^{-1}(
 \sqrt{\mathcal{P}} \cdot
 \mathcal{F}( \mathcal{W} ) )
 \,.
\end{equation}
In practice, the FFT is utilized to evaluate~\eqref{eq:FFT2}, which requires accommodating artificial boundary conditions (e.g., periodic boundary conditions) on a truncated computational domain~\cite{Khristenko2019}.

\subsection{The SPDE method}
\label{sec:SPDEforGRF}

The above approaches, i.e., the Karhunen–Loève expansion and the spectral method, require an explicit definition of the covariance function $\mathcal{C}$.
On the other hand, the SPDE method~\cite{Lindgren2011} defines $\mathcal{C}$ implicitly via the SPDE in~\eqref{eq:SPDEWhittle}.
For an infinite computational domain, the Green's function of the differential operator determines the covariance of the random field, thus replacing the filter function in the spectral method.
As the method is based on the Mat\'ern-type differential operator~\eqref{eq:SPDEWhittle}, we say that the proposed method is limited to random fields of \emph{Mat\'ern-type covariance}.
Generating GRFs with other types of covariance functions via a PDE-based methodology involves the derivation of alternative PDEs, which we do not consider here but refer the reader to~\cite[Section~1.3]{lindgren2023diffusionbased} for further details.
However, notice that~\eqref{eq:SPDEWhittle} can be solved on arbitrary domains and manifolds by replacing the Laplace operator with the Laplace--Beltrami operator.
This helps generalize random fields of \textit{Mat\'ern-type} covariance to arbitrary domains.
More details follow in Section~\ref{sec:spde_on_manifolds}.
Additionally, the method allows for easy prescription of different boundary conditions and handling complicated domains, especially when the discretization process is performed using the Finite Element Method.

While the statistics literature typically uses the isotropic parametrization of the SPDE in~\eqref{eq:SPDEWhittle}, this work and the associated C++ implementation parametrize the SPDE following~\cite{Khristenko2020}.
Recalling the smoothness
parameter $\nu > 0$ used to specify the random field's regularity and introducing the
matrix $\M{\Theta}$ defining its correlation lengths in different directions, the SPDE reads
\begin{equation}\label{eq:SPDE_fullspace}
 \left(
 \mathcal{I}
 -\frac{1}{2\nu} \nabla \cdot
 \left( \M{\Theta} \nabla \right)
 \right)^{\frac{2\nu+d}{4}}
 u = \eta \mathcal{W}
 \,,\quad
 \M{\Theta} = \M{R}^T
 \begin{pmatrix}
  l_1^2 &        &       \\
        & \ddots &       \\
        &        & l_d^2 \\
 \end{pmatrix} \M{R}
 \, ,
\end{equation}
with $\eta$ being a normalization constant.
Here, the rotation matrix and the identity operator are denoted by $\M{R} \in \mathrm{SO}(d)$ and $\mathcal{I}$, respectively.
Examples of such Mat\'ern-type random fields with $\M{\Theta} = l^2\, \M{I}$ and homogeneous Neumann boundary conditions on the domain $D = (0,1)^2$ are depicted in Figure~\ref{fig:rf-matrix-annotated}, giving intuition for interpreting the regularity parameter $\nu$ and correlation length $l$.

As demonstrated below, this parameterization also allows for a natural definition of anisotropy in engineering applications.
Taking $D = \mathbb{R}^d$, the generalized Mat\'ern covariance resulting from~\eqref{eq:SPDE_fullspace} is given as
\begin{equation}\label{eq:GerneralizedMaternCovariance}
 C(\V{x},\V{y}) = \sigma^2M_\nu \left(\sqrt{2\nu}\, d(\V{x},\V{y}) \right)\,
 \,,\quad
 M_\nu(z) = \frac{2^{1-\nu}}{\Gamma(\nu)} z ^{\nu} K_\nu \left( z \right) \,,
\end{equation}
with the anisotropic distance defined as $d(\V{x},\V{y}) = \sqrt{(\V{x}-\V{y})\M{\Theta}^{-1} (\V{x}-\V{y})}$.
Moreover, the normalization constant $\eta$ leads to unit variance ($\mathcal{C}(\V{x},\V{x})=1$) and is defined
\begin{equation}\label{eq:eta_normalization}
 \eta = \left(
 \frac{
   (2 \pi)^{\frac{d}{2}}
   \sqrt{\det (\M{\Theta})} \Gamma(\nu + \frac{d}{2})
  }{
   \nu^{\frac{d}{2}} \Gamma(\nu)
  }
 \right)^\frac{1}{2} \, ,
\end{equation}
with $\Gamma$ denoting the gamma function.
The corresponding GRF is homogeneous since~\eqref{eq:GerneralizedMaternCovariance} only depends on the anisotropic distance.
However, taking $D$ to be any Lipschitz domain in $\mathbb{R}^d$, \eqref{eq:SPDE_fullspace} easily generalizes to a more general class of inhomogeneous Mat\'ern-type GRFs upon replacing the matrix $\M{\Theta}$ with the diffusion tensor $\M{\Theta}(\V{x})$.

\subsection{Selection criteria and arguments for the SPDE method}
\label{sec:WhySPDE}

Each of the random field generation methods summarized above has strengths and weaknesses that must be evaluated for the specific problem scenario.
In this work, we are concerned with finite element workflows for computational mechanics and PDE-constrained optimization under uncertainty.
The target problems are often defined on complex domains and can impose a wide range of requirements on the random field model.
Typical workflows involve sampling multiple forward simulations of the physical model, with every sample corresponding to a different realization of the GRF and corresponding quantity of interest (QoI).
The sampled QoIs may be used to estimate a posterior probability distribution (cf.\ Section~\ref{sec:aneurysm}) or an associated risk measure (cf.\ Section~\ref{sec:num_ex}).
For an optimization workflow (cf.\ Section~\ref{sec:num_ex}), the gradients of the objective function and constraint functions can be estimated by solving an adjoint problem in additional to every forward problem.

We now match requirements of the random field generation method to the general workflows described above.
Our underlying assumption is that the GRF model~\eqref{eq:SPDE_fullspace}, with appropriate boundary conditions, is sufficiently general to capture the underlying uncertainty.
First, the method must be able to generate random fields of Mat\'ern-type covariance and handle all choices for its parameters $\nu$ and $\M{\Theta} = \M{\Theta}(\V{x})$.
We emphasize the possible spatial dependence of $\M{\Theta}$ because of the practical relevance of inhomogeneous GRFs~\cite{Keith2021}.
Second, the method must be able to handle complicated domains $D \subset \mathbb{R}^3$ and allow for defining general boundary conditions.
Third, since the underlying physical processes will be simulated with finite elements, the random field discretizations should conveniently map to such data structures.
Fourth, we demand (relative) scalability of the method.
In particular, denoting the times to solve the physical problem and sample the random field as $T_p$ and $T_r$, respectively, we demand that $T_r \leq c T_p$ with $c = \mathcal{O}(1)$.
Here, we pause to note that demanding a great difference between $T_r$ and $T_p$ (i.e., $c \ll 1$) is usually not necessary because there is typically no more than one random field generation step per physics solve.
Thus, the dominant cost will always be determined by $T_p$.
Lastly, we require the method to be memory efficient relative to the physics.
In this case, letting $M_{r}$ and $M_p$ denote the memory footprint of the random field and the physical problem, respectively, we demand $M_r \leq \tilde{c} M_p$ with $\tilde{c} = \mathcal{O}(1)$.
In practice, limited computing resources may sometimes impose tighter upper bounds for $c$ and $\tilde{c}$ than $\mathcal{O}(1)$.

Karhunen–Loève-type expansions such as EOLE are algorithmically simple and effective for generating random fields and have been used successfully in engineering workflows, e.g., in topology optimization~\cite{Schevenels2011,Lazarov2012,MartinezFrutos2018}.
However, as mentioned in Section~\ref{sec:KLforRF}, methods in this class scale unfavorably as the correlation length diminishes.
In~\eqref{eq:EOLE}, the number of points $M$ sampling the space must be increased, yielding a large covariance matrix $\M{C} \in \mathbb{R}^{M \times M}$ whose decomposition becomes expensive.
Additionally, its eigenvalues decay slowly, making evaluating the expansion increasingly challenging because the memory footprint grows linearly with the number of eigenvectors $K$ and the overall computational cost scales roughly like $\mathcal{O}(K M)$.
Decomposing the computational domain, evaluating the decomposition on each subdomain, and requiring continuity of the samples between the partitions reduces the size of the covariance matrix and helps the method to scale~\cite{DeCarvalhoPaludo2019}.
However, such an approach can introduce errors on the interface between the subdomains and requires the utilization of regular geometries or advanced data structures not supported by general FEM libraries.
Therefore, we rule out EOLE and similar approaches.

Spectral methods are excellent alternatives.
FFT's $\mathcal{O}(n \log n)$-scaling enables fast generation of random fields with limited resources.
Modern FFT libraries such as \textit{heFFTe}~\cite{Ayala2020} have proven that the algorithm scales on top-tier computing systems and complicated domains can be addressed with windowing techniques \cite{Khristenko2019}.
However, FFT-based methods become difficult, or infeasible, when approaching problems on two-dimensional manifolds embedded in $\mathbb{R}^3$.
Moreover, the FFT is restricted to regular grids, and enforcing boundary conditions on anything but trivial domains requires expertise.
Projecting the regular grid solution onto the finite element discretization of the physical problem is non-trivial, particularly in the context of distributed computing for which the data distribution of a graph-partitioned finite element problem does not coincide with the one used by the FFT libraries and extensive MPI-communication may be necessary for the mapping process.
Finally, generating inhomogeneous fields requires space-dependent filter functions~$T$.
While such filters may be defined, the convolution theorem no longer holds, making the filtering of the white noise considerably more complex and resource intensive.

Here, we advocate for using the SPDE method.
While we mainly restrict ourselves to examples from biomechanics and topology optimization under uncertainty, our arguments for employing the SPDE method extend to general application areas whose problems are encoded in PDEs and may be solved with the finite element method, provided that the Matérn-type covariance class accurately describes the uncertainties.
FFT-based methods are arguably faster; however, even significantly shorter sampling times barely influence a PDE-constrained optimization workflow's overall runtime, and the SPDE method's convenience outweighs the FFT's computational advantages.
The SPDE method allows arbitrary choices of regularity, orientation, and correlation lengths and samples random fields on arbitrary meshed domains.
As the SPDE solution method here is based on finite elements, the random field can be computed on the same (distributed) mesh as the physical problem, and matching the random fields with the physical problem becomes a trivial task, even in setups involving thousands of CPU cores.
We further note that the numerical solution of~\eqref{eq:SPDE_fullspace} is efficient and scalable (see Section~\ref{sec:SPDE_discretization}) and, thus, our relative scalability demands, $T_r \leq c T_p$ and $M_r \leq \tilde{c} M_p$, are satisfied.
The SPDE method enables us to easily define boundary conditions for the GRF on complicated domains, and the method extends to a general class of inhomogeneous fields with $\M{\Theta} = \M{\Theta}(\V{x})$.
Hence, the SPDE method satisfies all requirements and carries the potential to become the method of choice in finite-element-driven application areas.

\section{Implementing the SPDE method}
\label{sec:SPDE_discretization}

The purpose of this section is to present an implementation-based overview of the SPDE approach to generating M\'atern-type GRFs.
For ease of presentation, we first choose to focus on the simplified PDE model~\eqref{eq:SPDEWhittle} in the case where $k$ is a positive integer.
As shown in Section~\ref{sub:FEM_discretization}, the approach reduces to a Galerkin finite element method.
Section~\ref{sub:sampling_the_white_noise_vector} and~\ref{sub:FractionExponents} discuss an efficient sampling of the spatial white noise vector and the handling of the fractional operator for real-valued $k > d/4$.
From this point, the generalization to~\eqref{eq:SPDE_fullspace} is straightforward.
Lastly, Section~\ref{sec:spde_on_manifolds} explains how the SPDE generalizes to Mat\'ern-type random fields on manifolds.
We refer the reader to~\ref{apx:implementation} for a small note on our MFEM implementation and its scalability.

\subsection{FEM discretization}
\label{sub:FEM_discretization}

Our description mainly follows~\cite[Sections 3.1]{Croci2018}.
Solving~\eqref{eq:SPDEWhittle} with integer exponent $k\geq 1$ on a domain $D \subset \mathbb{R}^d$ reduces to a sequence of second-order PDEs:
\begin{equation}
 \label{eq:SecondOrderPDESequence}
 \left\{
 \begin{alignedat}{5}
  u_1 - \kappa^{-2}\Delta u_1 & = \eta \mathcal{W} ~~ &  & \text{in }
  D \,,                                                               \\ u_{i+1} - \kappa^{-2}\Delta u_{i+1} &= u_{i} &&\text{in } D \,, ~~&&\text{for } i = 1,\ldots,k-1,\end{alignedat} \right.
\end{equation}
with $u := u_k$ and appropriate boundary conditions (e.g., Dirichlet or Neumann) on each $u_i$, $i = 1,\ldots,k$; cf.~\cite{Lindgren2011}.

The equations in~\eqref{eq:SecondOrderPDESequence} for $u_i$, $i = 2,\ldots,k$, lead to weak formulations that can be treated using standard finite element techniques; the interested reader is referred to~\cite{hughes2012finite} for further details.
We focus on the case $k = \eta = 1$ and, for simplicity, assume homogeneous Dirichlet boundary conditions.
In this case,~\eqref{eq:SecondOrderPDESequence} reduces to
\begin{equation}
 \label{eq:SecondOrderSPDE}
 -\kappa^{-2} \Delta u + u = \mathcal{W}~~\text{in }
 D \,, \quad u = 0 ~~\text{on } \partial D \,.
\end{equation}

To solve~\eqref{eq:SecondOrderSPDE} with finite elements, we proceed as follows.
Let $V_h = \mathop{\mathrm{span}} ( \phi_1, \phi_2, \ldots, \phi_m ) \subset H^1_0(D)$ be a finite element subspace spanned by the basis functions $\phi_i$, $i = 1, \ldots, m$.
The Galerkin approximation $u_h \approx u$ is then written
\begin{equation}
 \label{eq:DiscreteWeakFormSPDE}
 \text{Find } u_h \in V_h \text{ such that~~}
 \kappa^{-2}(\nabla u_h, \nabla v) + (u_h,v) = \langle \mathcal{W}, v \rangle
 \,,
 ~~\text{for all } v \in V_h
 \,,
\end{equation}
where the notation $(\cdot, \cdot)$ denotes the $L^2(D)$-inner product and $\langle \mathcal{W}, \phi \rangle = \mathcal{W}(\phi)$ denotes the action of the generalized stochastic field $\mathcal{W}$ onto the function $\phi \in L^2(D)$.

Our task is to rewrite~\eqref{eq:DiscreteWeakFormSPDE} as a linear equation
\begin{equation}
 \label{eq:LinearEquationSPDE}
 \M{A} \V{x} = \V{b}\,,
\end{equation}
with a normally-distributed random right-hand side vector $\V{b}$.
We begin by referring to~\cite[Example 1.2 and Lemma 1.10]{hida2013white}, where it is shown that for the spatial white noise $\mathcal{W}$ and any collection of functions $\{\phi_i\}$ in $L^2(D)$, the actions
\begin{subequations}
 \label{eqs:WhiteNoiseActionOnTestFunction}
 \begin{equation}
  \label{eq:WhiteNoiseActionOnTestFunction}
  b_i = \langle \mathcal{W}, \phi_i \rangle
 \end{equation}
 are joint Gaussian random variables with zero mean and covariance given by
 \begin{equation}
  \label{eq:WhiteNoiseActionOnTestFunction_covariance}
  \mathbb{E}[b_ib_j] = (\phi_i,\phi_j)
  \,.
 \end{equation}
\end{subequations}
In turn, we find that expressing $u_h = \sum_{i=1}^m x_i \phi_i(\V{x})$ leads to the identities
\begin{equation}
 A_{ij} = \kappa^{-2} (\nabla\phi_i,\nabla\phi_j) + (\phi_i,\phi_j) \quad \text{and} \quad b_i = \langle \mathcal{W}, \phi_i \rangle \,.
\end{equation}
Moreover, by~\eqref{eqs:WhiteNoiseActionOnTestFunction}, we find that $\V{b}$ is a zero-mean Gaussian vector,
\begin{equation}
 \label{eq:WhiteNoiseVector}
 \V{b} \sim \mathcal{N}(0, \M{M})
 \,,
\end{equation}
with covariance matrix equal to the standard finite element mass matrix $M_{ij} = (\phi_i,\phi_j)$.

Efficient procedures to assemble the finite element stiffness matrix $\M{A}$ are well-documented in the literature.
It remains to state a technique to efficiently generate samples of the spatial white noise vector $\V{b}$.
This is done in Subsection~\ref{sub:sampling_the_white_noise_vector} below.
Once $\V{b}$ is sampled, the linear system~\eqref{eq:LinearEquationSPDE} can be solved efficiently with standard FE system techniques, e.g., via various preconditioned linear equation solvers.
The average cost can be further reduced if multiple samples of $u_h$ are required simultaneously.
Indeed, the matrix equation
\begin{equation}
 \M{A} \M{X} = \M{B}
 \,,
 \quad
 \M{X} = \begin{bmatrix}
  \vert   &        & \vert      \\
  \V{x}_1 & \cdots & \V{x}_\ell \\
  \vert   &        & \vert
 \end{bmatrix}
 \,,
 \quad
 \M{B} = \begin{bmatrix}
  \vert   &        & \vert      \\
  \V{b}_1 & \cdots & \V{b}_\ell \\
  \vert   &        & \vert
 \end{bmatrix}
 \,,
\end{equation}
is easily formed by first drawing $\ell > 0$ samples of $\V{b} \sim \mathcal{N}(0, \M{M})$ and then solved with the same preconditioner as if $\ell = 1$.
Likewise, no new finite element solver is required if $k > 1$ since the same elliptic operator appears in each of the equations in~\eqref{eq:SecondOrderPDESequence}.

\subsection{Sampling the spatial white noise vector}
\label{sub:sampling_the_white_noise_vector}

We understand from~\eqref{eq:WhiteNoiseVector} that spatial white noise realizations $\V{b} \in \mathbb{R}^{m}$ can be generated by sampling a Gaussian vector whose covariance $\M{C}$ is given by the mass matrix $\M{M} \in \mathbb{R}^{m\times m}$.
Sampling such a vector can be thought of as executing a linear transformation that converts an $n$-dimensional Gaussian random vector $\V{z} = (z_1,\ldots,z_n)^\top \sim \mathcal{N}(0,\M{I})$ with zero mean $\mathbb{E}[\V{z}] = \V{0}$ and identity covariance $\mathbb{E}[z_iz_j] = \delta_{ij}$, to a new random vector
\begin{equation}
 \V{b} = \M{H} \V{z}
 \,,
\end{equation}
where $\M{H} \in \mathbb{R}^{m\times n}$ decomposes the mass matrix $\M{M} = \M{H} \M{H}^\top$.
Indeed, observe that for any such matrix $\M{H}$, it holds that $\mathbb{E}[\V{b}] = \M{H} \mathbb{E}[\V{z}] = \V{0}$ and, furthermore,
\begin{equation}
 \mathbb{E}[\V{b}\V{b}^\top]
 =
 \mathbb{E}[\M{H} \V{z}( \M{H} \V{z})^\top]
 =
 \M{H} \mathbb{E}[\V{z}\V{z}^\top] \M{H}^\top
 =
 \M{H} \M{I} \M{H}^\top
 =
 \M{M}
 \,.
\end{equation}

Generating samples of $\V{z}$ is easy; pseudorandom number generators are available in all scientific software languages and may be used to independently sample each component $z_i \sim \mathcal{N}(0,1)$, $i = 1, \ldots, n$.
The main task is to construct $\M{H}$ efficiently.
A tempting choice $\M{H} = \M{L} \in \mathbb{R}^{m\times m}$ follows from the Cholesky factorization $\M{L} \M{L}^\top = \M{M}$.
However, this choice may be prohibitively expensive if $m$ is large since it involves an $\mathcal{O}(m^3)$ factoring cost (from factoring $\M{L} \M{L}^\top = \M{M}$) and $\mathcal{O}(m^2)$ sampling cost (from matrix-vector multiplication $\V{b} = \M{L}\V{z}$).
Instead, we promote the construction suggested in~\cite[Section 4.1]{Croci2018} whose factoring and sampling costs both scale like $\mathcal{O}(m)$.

The idea is to leverage the assembly procedure that is used to construct global matrices in any typical finite element code (cf.~Figure~\ref{fig:Assembly}) and perform the Cholesky decomposition in parallel at the element level.
For the mass matrix, we have the factorization
\begin{equation}
 \M{M} = \M{P}^\top \M{G}^\top \M{B}^\top \M{D} \M{B} \M{G} \M{P}
 \,.
\end{equation}
In this expression, $\M{P} \in \mathbb{R}^{m_{\mathrm{local}} \times m}$ projects the (true) global degrees of freedom (DOFs) onto a space of local DOFs with redundancies at the intersections of local subdomains (usually restricted to separate processors) and $\M{G} \in \mathbb{R}^{m_{\mathrm{elem}} \times m_{\mathrm{local}}}$ further distributes these local DOFs to individual elements.
Meanwhile, the matrix $\M{B} \in \mathbb{R}^{m_{\mathrm{quad}} \times m_{\mathrm{elem}}}$ and the diagonal matrix $\M{D} \in \mathbb{R}^{m_{\mathrm{quad}} \times m_{\mathrm{quad}}}$ involve function evaluation (at quadrature points) and perform multiplication with quadrature weights, respectively.

\begin{figure}[t]
 \centering
 \begin{tikzpicture}[scale=1.75]

  \RoundRect{0}{0}{0.975}{0.975}
  \RoundRect{1.05}{0}{0.725}{0.975}
  \RoundRect{0}{1.05}{0.975}{0.725}
  \RoundRect{1.05}{1.05}{0.725}{0.725}

  \DofGrid{0.1}{0.1}{0.4}{0.4}{red!75!black}
  \DofGrid{0.9}{0.1}{0.4}{0.4}{red!75!black}
  \DofGrid{0.1}{0.9}{0.4}{0.4}{red!75!black}
  \DofGrid{0.9}{0.9}{0.4}{0.4}{red!75!black}

  \draw[->, line width=0.5pt] (1.85, 0.9) -- node[midway,above] {$\M{P}$} ++(.3,0.0);
  \draw[<-, line width=0.5pt] (1.85, 0.8) -- node[midway,below] {$\M{P}^T$} ++(.3,0.0);
  \node[align=center] at (0.9, 2) {\small Global true DOFs};

  \begin{scope}[shift={(2.2,0)}]
   \RoundRectGrid{0.85}{0.85}

   \DofGrid{0.1}{0.1}{0.325}{0.325}{black}
   \DofGrid{1.025}{0.1}{0.325}{0.325}{black}
   \DofGrid{0.1}{1.025}{0.325}{0.325}{black}
   \DofGrid{1.025}{1.025}{0.325}{0.325}{black}

   \draw[->, line width=0.5pt] (1.85, 0.9) -- node[midway,above] {$\M{G}$} ++(.3,0.0);
   \draw[<-, line width=0.5pt] (1.85, 0.8) -- node[midway,below] {$\M{G}^T$} ++(.3,0.0);
   \node[align=center] at (0.9, 2) {\small Local subdomain DOFs};
  \end{scope}

  \begin{scope}[shift={(4.4,0)}]
   \RoundRectGrid{0.85}{0.85}

   \DofElem{0.1}{0.1}{0.325}{0.325}{blue!65!black}
   \DofElem{1.025}{0.1}{0.325}{0.325}{blue!65!black}
   \DofElem{0.1}{1.025}{0.325}{0.325}{blue!65!black}
   \DofElem{1.025}{1.025}{0.325}{0.325}{blue!65!black}

   \draw[->, line width=0.5pt] (1.85, 0.9) -- node[midway,above] {$\M{B}$} ++(.3,0.0);
   \draw[<-, line width=0.5pt] (1.85, 0.8) -- node[midway,below] {$\M{B}^T$} ++(.3,0.0);
   \node[align=center] at (0.9, 2) {\small Element DOFs};
  \end{scope}

  \begin{scope}[shift={(6.6,0)}]
   \RoundRectGrid{0.85}{0.85}

   \QuadElem{0.1}{0.1}{0.325}{0.325}{green!50!black}
   \QuadElem{1.025}{0.1}{0.325}{0.325}{green!50!black}
   \QuadElem{0.1}{1.025}{0.325}{0.325}{green!50!black}
   \QuadElem{1.025}{1.025}{0.325}{0.325}{green!50!black}

   \draw [->] (1.85, 1.1) to[out=0,in=0,looseness=1.5] node[midway,right] {$\M{D}$} (1.85, 0.7) ;
   \node[align=center] at (0.9, 2) {\small Quadrature point values};
  \end{scope}

 \end{tikzpicture}
 \caption{The finite element assembly procedure in a typical finite element code reduces to matrix multiplication.
  For the mass matrix, we multiply through the factorization $\M{M}  = \M{P}^\top \M{G}^\top \M{B}^\top \M{D} \M{B} \M{G} \M{P}$ depicted above.
  Figure modified from~\cite{Anderson2021,andrej2024mfem}.
 }
 \label{fig:Assembly}
\end{figure}

The block-diagonal matrix $\widetilde{\M{M}} = \mathop{\mathrm{diag}}(\M{M}_T) = \M{B}^\top \M{D} \M{B}$ is comprised of the local mass matrices indexed by each individual element $T$ in the mesh $\mathcal{T}$.
Note that each $\M{M}_T$ is symmetric positive definite and, therefore, we can use a local Cholesky decomposition to factor each local mass matrix $\M{L}_T \M{L}^\top_T = \M{M}_T$ independently.
Moreover, defining $\M{H} = \M{P}^\top \M{G}^\top \mathop{\mathrm{diag}}(\M{L}_T) \in \mathbb{R}^{m\times m_{\mathrm{elem}}}$, we find that
\begin{equation}
 \M{M} = \M{P}^\top \M{G}^\top \widetilde{\M{M}} \M{G} \M{P}
 = \M{P}^\top \M{G}^\top \mathop{\mathrm{diag}}(\M{L}_T \M{L}_T^\top) \M{G} \M{P}
 = \M{P}^\top \M{G}^\top \mathop{\mathrm{diag}}(\M{L}_T) (\M{P}^\top \M{G}^\top \mathop{\mathrm{diag}}(\M{L}_T))^\top
 = \M{H} \M{H}^\top
 \,.
\end{equation}
In summary, it is only necessary to factor the mass matrix locally, because of this, the cost of sampling the white noise vector $\V{b}$ scales linearly in the number of elements and DOFs.
We encourage the interested reader to inspect our parallel and scalable implementation of the above sampling strategy in the latest release of MFEM~\cite{Duswald2023SPDEMiniapp,andrej2024mfem}.

\subsection{Fractional exponents}
\label{sub:FractionExponents}

The SPDE~\eqref{eq:SPDEWhittle} includes a non-integer exponent, specifically $k = \frac{2\nu+d}{4}$.
Solving boundary value problems with such fractional exponents, as they are known, presents unique challenges and demands substantial computational resources due to the non-local nature of the associated operator.
However, the past decade has witnessed significant progress in developing efficient techniques to tackle these problems~\cite{bonito2018numerical,lischke2020fractional}.
A notable approach is based on the Caffarelli--Silvestre extension, in which the fractional PDE solution is identified as a restriction of the solution to an integer-order boundary value problem set in a higher-dimensional semi-infinite cylinder~\cite{caffarelli2007extension}.
This identification allows the higher-dimensional integer-order PDEs to be discretized using the Galerkin finite element method~\cite{meidner2018hp, banjai2019tensor, banjai2020exponential}.
Another technique involves using an integral representation of the inverse fractional operator, combined with a specific quadrature rule, to break down the problem into a series of integer-order PDEs~\cite{balakrishnan1960fractional,bonito2015numerical,bonito2019sinc}.
This technique has recently been applied to the SPDE~\cite{Bonito2024}.
An alternative strategy suggests modeling the fractional PDE as a transient pseudo-parabolic problem, where numerical integration over a 'time' parameter results in a sequence of integer-order PDEs~\cite{vabishchevich2015numerically,lazarov2017numerical}.
In our implementation~\cite{Duswald2023SPDEMiniapp}, we adopt a rational approximation method~\cite{harizanov2018positive,bolin2019rational,Khristenko2023}, which involves approximating the operator's spectrum with a rational function.
The multipole expansion of this function then leads to a set of independent integer-order PDEs that can be solved simultaneously.
This method is summarized below.

The rational approximation method can be applied to a wide variety of elliptic differential operators $\OP{A}$, with spectrum~$\sigma(\OP{A})\subset[\lambda_\mathrm{min}, \lambda_\mathrm{max}]$, $0<\lambda_\mathrm{min}< \lambda_\mathrm{max}$.
However, in~\eqref{eq:SPDE_fullspace}, we are specifically interested in the fractional PDEs of the form
\begin{equation}
 \OP{A}^k u = \eta \mathcal{W}
 \,,
 \quad
 k > d/4
 \,,
\end{equation}
where $\OP{A} := \mathcal{I} - \kappa^{-2} \Delta$ with its associated boundary conditions.
If the rational function~$\lambda \mapsto \sum_{n=1}^{N}\frac{c_n}{\lambda + d_n}$ approximates the function~$\lambda \mapsto \lambda^{-\alpha}$ on the interval~$[\lambda_\mathrm{min}, \lambda_\mathrm{max}]$, i.e.,
\begin{equation}
 \label{eq:RA}
 \sum_{n=1}^{N}\frac{c_n}{\lambda + d_n}
 \approx
 \lambda^{-\alpha}
 \quad
 \text{for all }
 \lambda \in [\lambda_\mathrm{min}, \lambda_\mathrm{max}]
 \,,
\end{equation}
then the solution~$u$ can be approximated as the weighted average of the solutions of $N$ problems written in terms of $A$; namely,
\begin{equation}\label{eq:RA_approximant}
 u \approx
 \sum_{n=1}^{N} c_n u_n
 \,,
 \quad
 \text{where~~}
 \Big(d_n \mathcal{I} + \OP{A} \Big)u_n = \eta \mathcal{W}
 \,.
\end{equation}
Because $A$ is an integer-order differential operator, each of the $N$ problems above can be solved using the finite element method as described previously in this section.
It remains to explain how to generate the weights~$c_n$ and the poles~$-d_n$ of the rational function~\eqref{eq:RA}.
These can be obtained with one of the various rational approximation algorithms; see, e.g.,~\cite{harizanov2018positive,bolin2019rational,Nakatsukasa2018}.
In our MFEM implementation, we used the adaptive Antoulas--Anderson (AAA) algorithm proposed in~\cite{Nakatsukasa2018}.
From this point on, the generalization of the numerical scheme from~\eqref{eq:SPDEWhittle} to~\eqref{eq:SPDE_fullspace} is straightforward and merely requires replacing the scalar number $\kappa^{-2}$ with the matrix $\M{\Theta}/2\nu$ in the weak form~\eqref{eq:DiscreteWeakFormSPDE} and the corresponding matrix assembly.

\begin{figure}[h]
 \centering
 \begin{minipage}[b]{0.07\textwidth}\centering
  \scriptsize{$\mathcal{W}(\omega_1)$}
  \vspace{1.6cm}
 \end{minipage}
 \begin{minipage}[b]{0.22\textwidth}\centering
  \includegraphics[width=0.95\textwidth]{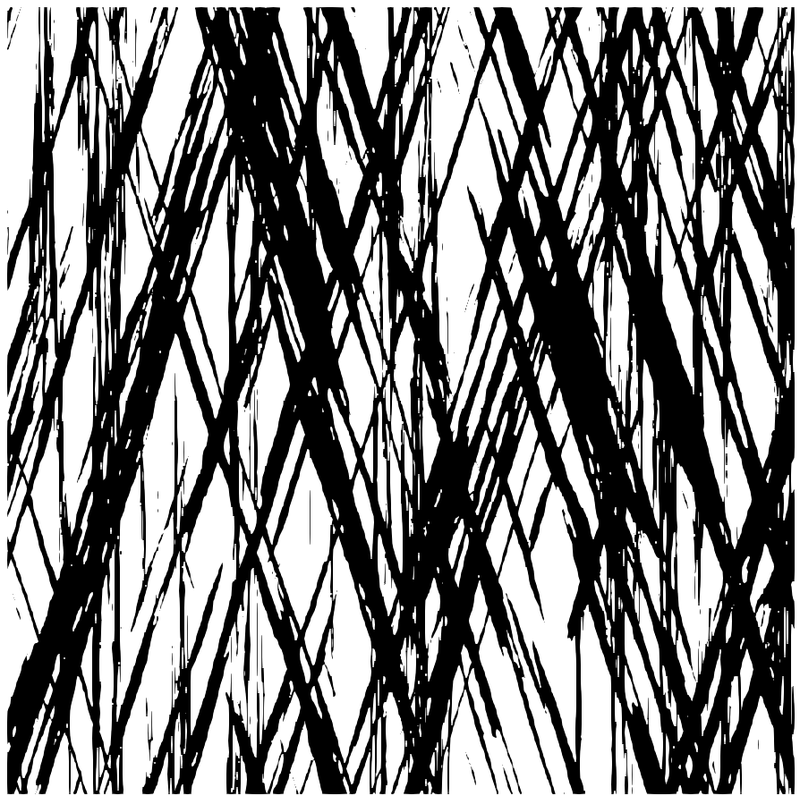}
 \end{minipage}
 \begin{minipage}[b]{0.22\textwidth}\centering
  \includegraphics[width=0.95\textwidth]{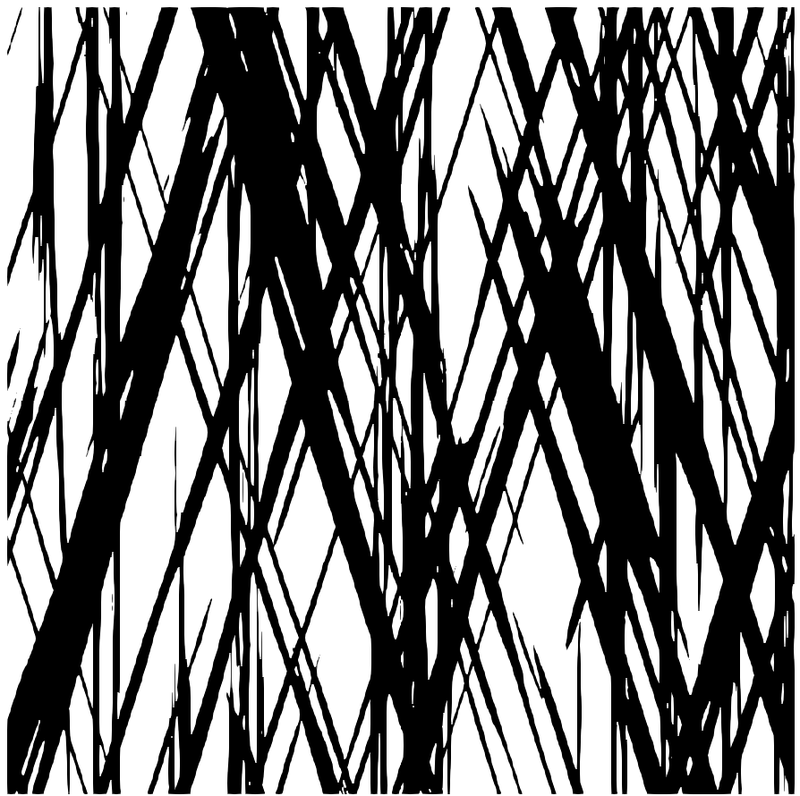}
 \end{minipage}
 \begin{minipage}[b]{0.22\textwidth}\centering
  \includegraphics[width=0.95\textwidth]{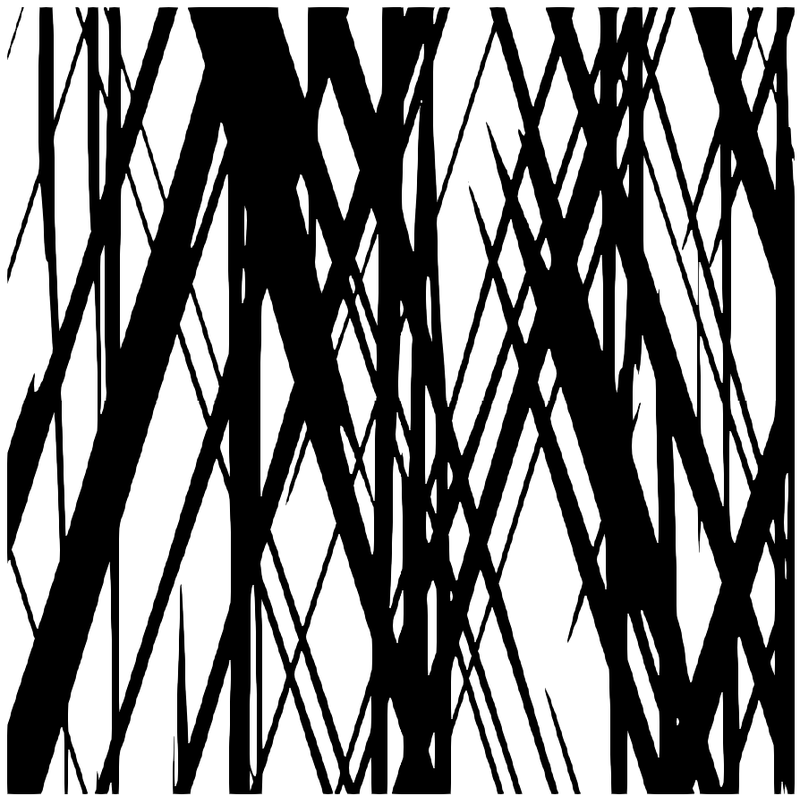}
 \end{minipage}
 \begin{minipage}[b]{0.22\textwidth}\centering
  \includegraphics[width=0.95\textwidth]{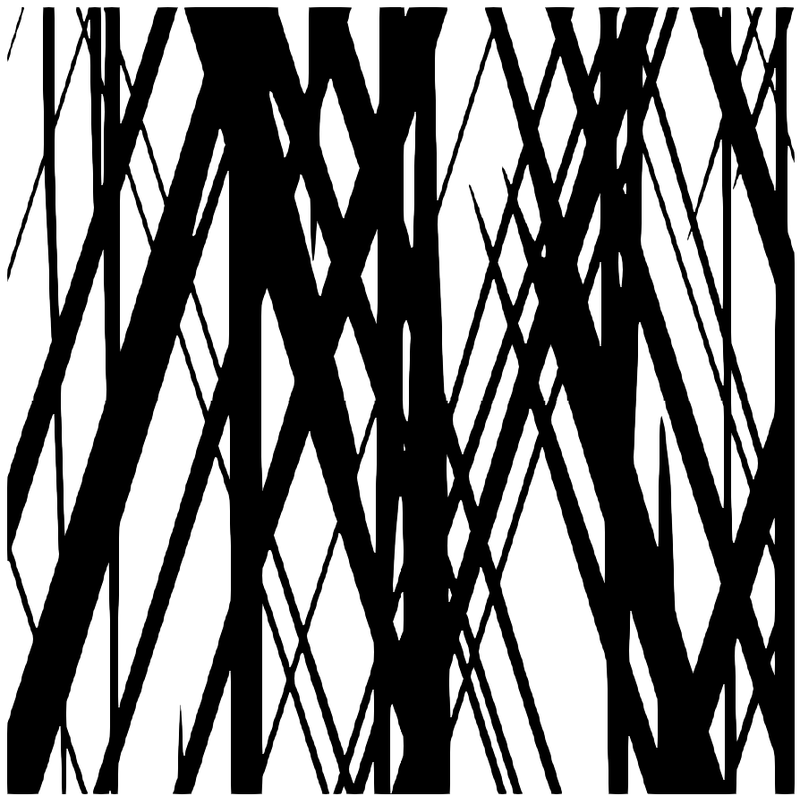}
 \end{minipage} \\
 \begin{minipage}[b]{0.07\textwidth}\centering
  \scriptsize{$\mathcal{W}(\omega_2)$}
  \vspace{2.3cm}
 \end{minipage}
 \begin{minipage}[b]{0.22\textwidth}\centering
  \includegraphics[width=0.95\textwidth]{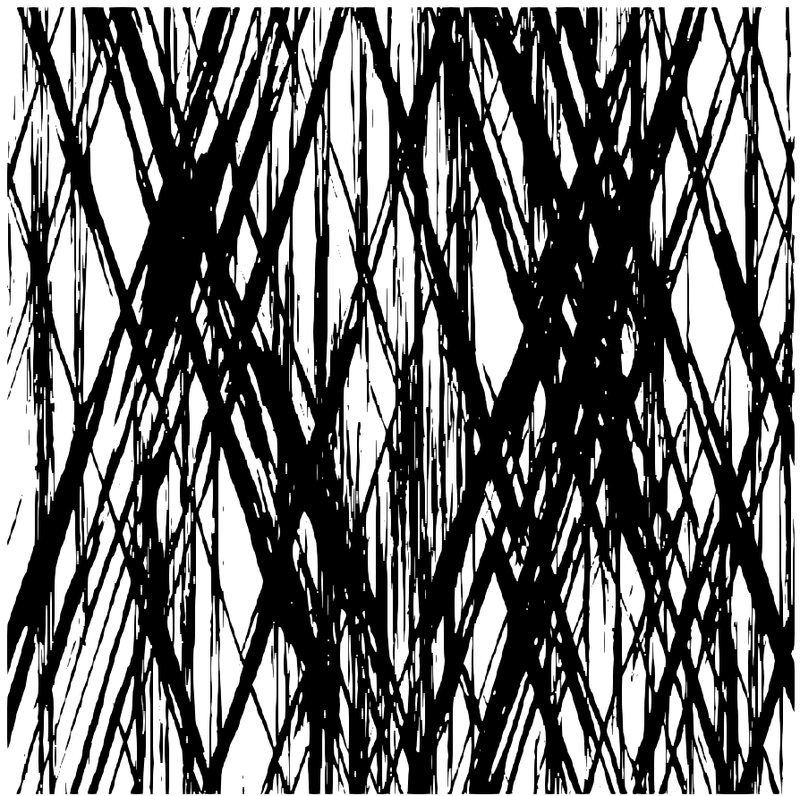}
  \caption*{(a) $\nu = 0.5$}
 \end{minipage}
 \begin{minipage}[b]{0.22\textwidth}\centering
  \includegraphics[width=0.95\textwidth]{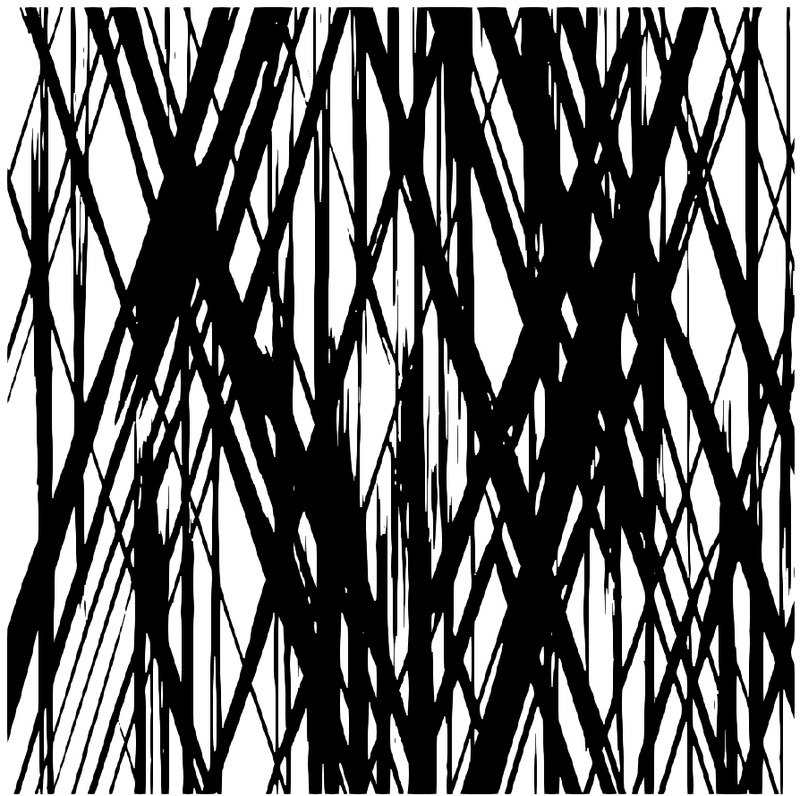}
  \caption*{(b) $\nu = 1.0$}
 \end{minipage}
 \begin{minipage}[b]{0.22\textwidth}\centering
  \includegraphics[width=0.95\textwidth]{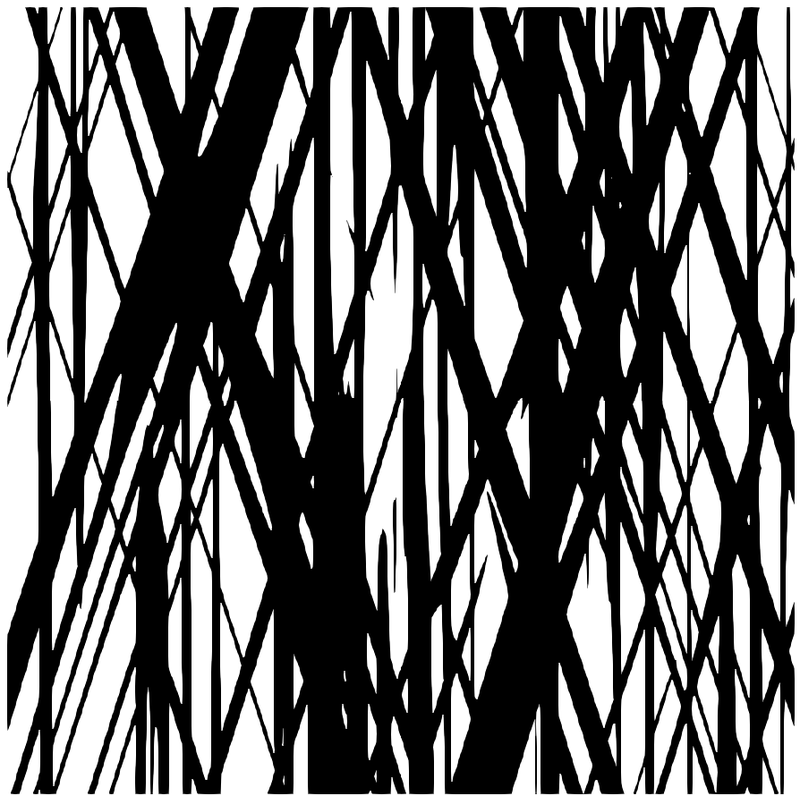}
  \caption*{(c) $\nu = 2.0$}
 \end{minipage}
 \begin{minipage}[b]{0.22\textwidth}\centering
  \includegraphics[width=0.95\textwidth]{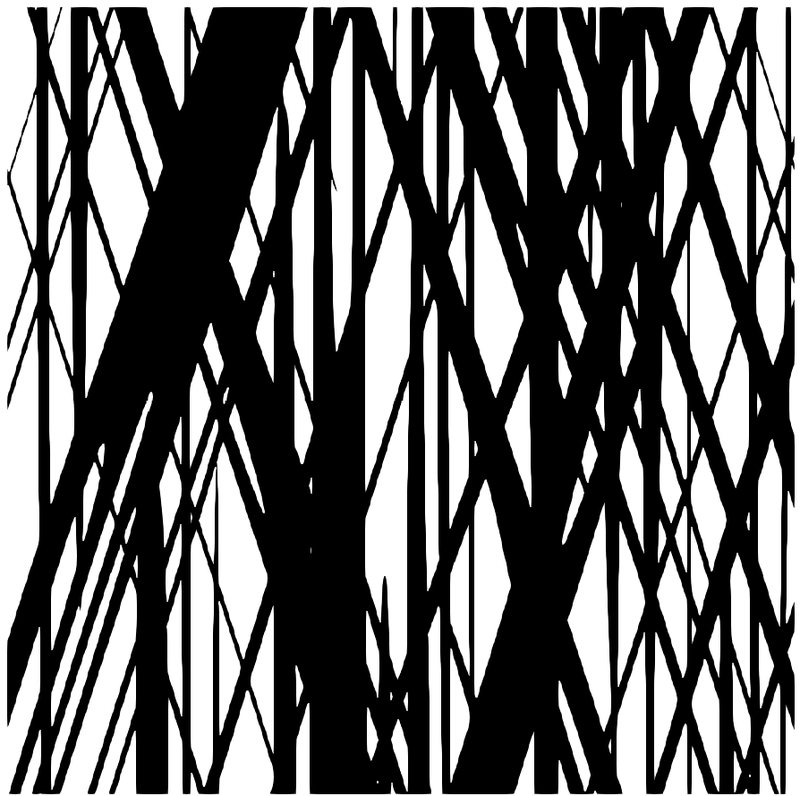}
  \caption*{(d) $\nu = 4.0$}
 \end{minipage}
 \caption{
  Combination of three, thresholded random fields to mimic material uncertainties for different smoothness (columns)
  and different white noise samples (rows).
  Thresholds are set to 0 and 1.
 }
 \label{fig:thresholded_fields}
\end{figure}

The fractional exponent gives fine control over the smoothness of the GRF, which may not be achieved with integer-order exponents.
Indeed, we remind the reader of Figure~\ref{fig:rf-matrix-annotated}, which depicts a variety of isotropic random fields that can be generated from the SPDE approach.
In Figure~\ref{fig:thresholded_fields}, we illustrate the impact of the regularity parameter $\nu$ on GRFs generated via the generalized SPDE~\eqref{eq:SPDE_fullspace} with strong anisotropy.
We use a pointwise non-linear transformation of the generated samples to demonstrate the effect.
Such transformations are common in modeling material properties~\cite{Guilleminot2020}, loads, and inputs with marginal distributions~\cite{Grigoriu2002} different from Gaussian.
In the random field samples presented in Figure~\ref{fig:thresholded_fields}, we threshold the generated GRF.
All values of the GRF sample between zero and one are projected to one and all others to zero.
We generated three GRF samples with enforced strong anisotropy and rotated them with different angles.
The final image is obtained by pointwise summation and thresholding the result at one.
The presented model can model the random distribution of fibers in composite materials.
Close inspection of the images reveals that smoother fields, i.e., larger $\nu$, produce thicker fibers with well-defined spaces between them.
The fibers' size relative to the domain's characteristic length can be scaled by modifying the correlation length $l$.
On the other hand, less regular Gaussian fields, i.e., smaller values of the parameter $\nu$, result in well-separated clumps of thinner fibers.
The clumping phenomena can be modeled only by increasing the roughness of the field.
Thus, varying the parameter $\nu$ combined with a non-linear transformation provides another control parameter for developing models of physically realistic distributions.
Further examples for microstructure models based on the post-processing of Matérn-type random fields can be found in~\cite{Khristenko2021}.

\subsection{The SPDE method on manifolds}
\label{sec:spde_on_manifolds}

The SPDE approach is especially desirable to generate random fields on complex geometries and embedded surfaces.
Indeed, the representation~\eqref{eq:SecondOrderPDESequence} provides a natural generalization to any sufficiently smooth manifold if the symbol $\Delta$ is interpreted as the corresponding Laplace--Beltrami operator.
In turn, instead of~\eqref{eq:MaternCovariance}, the general formula for the covariance involves the eigenvalues $\lambda_j$ and eigenfunctions $\Psi_j$ of the Laplace--Beltrami operator,
\begin{equation}\label{eq:CovarianceOfSPDE_eigen}
 C(\V{x},\V{y}) = \sum_{j=1}^\infty \frac{c_k}{(\kappa^2 + \lambda_j)^\alpha} \Psi_j(\V{x}) \Psi_j(\V{y}) \, .
\end{equation}
Typically, a closed from for~\eqref{eq:CovarianceOfSPDE_eigen} cannot be derived because the eigenfunctions are unknown.
Notable exceptions are intervals, rectangles, parallelepipeds, disks, sectors, spheres, spherical shells, ellipses, equilateral triangles, and circular and elliptical annuluses~\cite{Grebenkov2013}.
Truncated and spherical domains are particularly interesting in the context of the SPDE method and our examples.
Khristenko and co-workers~\cite{Khristenko2019} thoroughly analyzed the effect of the boundaries on the SPDE covariance on $D = (0,1)$.
They derived closed from solutions for~\eqref{eq:CovarianceOfSPDE_eigen} with homogeneous Dirichlet and Neumann boundary conditions resulting in a sum over sine and cosine terms, respectively.
They then found that the difference between the Mat\'ern covariance~\eqref{eq:MaternCovariance} and the different expressions for~\eqref{eq:CovarianceOfSPDE_eigen} rapidly decays as the distance to the boundary increases and that the covariance is practically indistinguishable for points further than one to two correlation lengths away from the boundary.
The authors additionally considered periodic boundaries, where they observe wrap-around effects due to the topology.
Such effects are also present in the sphere.
Here, the eigenfunctions are the \textit{spherical harmonics} but~\eqref{eq:CovarianceOfSPDE_eigen} can be further simplified to a sum over the \textit{Legendre} polynomials~\cite{Lang2015,lindgren2023diffusionbased}.
The Mat\'ern-type covariance on the sphere is very similar to the Mat\'ern covariance~\eqref{eq:MaternCovariance} if $l/R \ll 1$, where $R$ denotes the radius of the sphere.
Intuitively this can be understood from realizing that decreasing $l$ on the sphere restricts the action of the filter defined by the Laplace-Beltrami operator to small local patches that are increasingly similar to $\mathbb{R}^2$.
Finding the eigenfunctions and, therefore, the covariance of more intricate domains or manifolds requires numerical methods~\cite{Levy2006}.
For further details regarding the SPDE on manifolds, we refer to Lindgren and co-worker's original work~\cite{Lindgren2011} and the associated ten-year review~\cite{Lindgren2022}.
Additional insights regarding the Mat\'ern covariance on Riemannian manifolds can be found in the work of Borovitskiy and co-workers~\cite{Borovitskiy2020}.
Lastly, we wish to highlight the recent work of Bonito et al.~\cite{Bonito2024}, providing a complete error analysis for GRFs on closed surfaces defined via the SPDE.

\begin{figure}[ht]
 \centering
 \begin{minipage}[b]{0.33\textwidth}\centering
  \includegraphics[height=4cm]{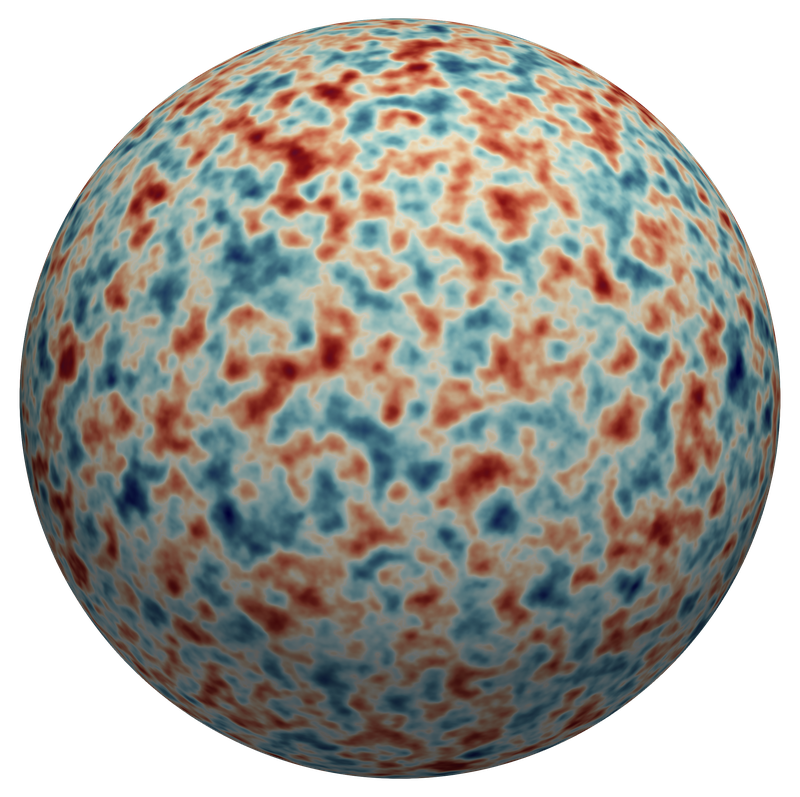}
 \end{minipage}
 \begin{minipage}[b]{0.33\textwidth}\centering
  \includegraphics[height=4cm]{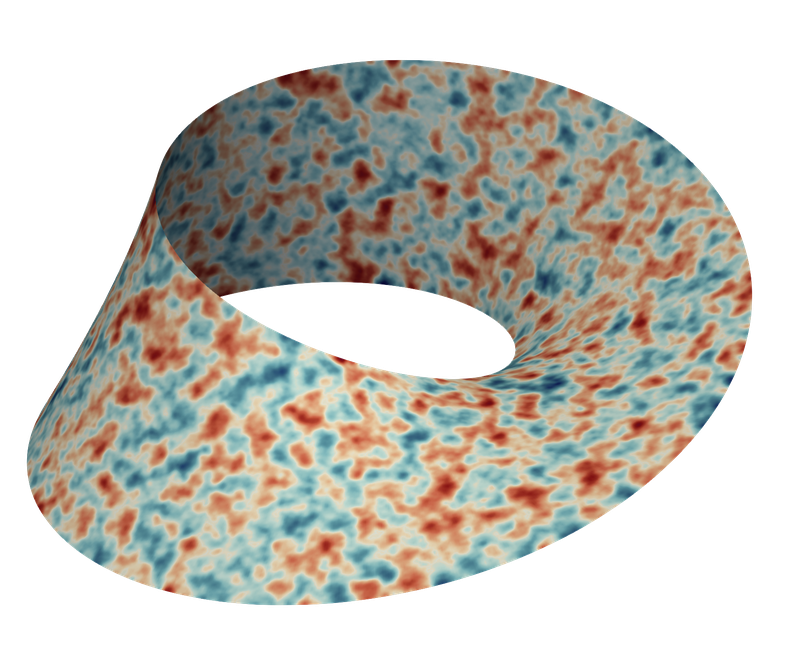}
 \end{minipage}
 \begin{minipage}[b]{0.33\textwidth}\centering
  \includegraphics[height=4cm]{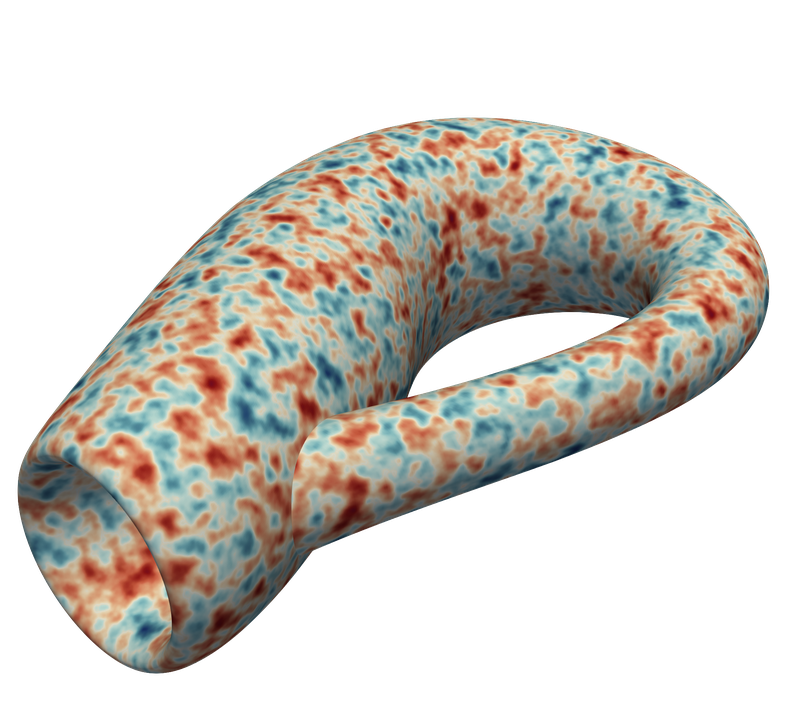}
 \end{minipage}
 \caption{Mat\'ern random fields on sphere, mobius strip, and Klein bottle geometries.}
 \label{fig:SurfaceGRFs}
\end{figure}

Figure~\ref{fig:SurfaceGRFs} displays a series of images depicting random field realizations on sphere, mobius strip, and Klein bottle geometries.
These independent random fields were generated with our MFEM implementation of the SPDE method~\cite{Duswald2023SPDEMiniapp}.
Additionally, we note that MFEM has long-standing native support for FEM discretizations of elliptic surface PDEs~\cite{Anderson2021}.
This feature was easy to leverage, and provided immediate support for the SPDE method on embedded surfaces without any code duplication.

\section{Simulating random fields on complex domains}
\label{sec:aneurysm}

To demonstrate the relevance of the SPDE method for geometric uncertainties, we borrow an example from biomedical engineering; namely, cerebral aneurysms.
Cerebral aneurysms are weakened areas in the brain's blood vessel walls that bulge and sometimes rupture, potentially leading to fatality.
Various factors influence cerebral aneurysm formation, including hemodynamics, genetics, and weakened vessel walls; see~\cite{Schievink1997,Brisman2006,Sforza2009} and references therein.
Once an aneurysm has formed, endovascular coiling has become a popular treatment option.
In this procedure, medical professionals insert a thin wire into the aneurysm, forming a tiny coil to fill the space and prevent rupture~\cite{Hui2014,Pierot2013,Zhao2018}.

Mathematical models for cerebral aneurysms and their treatment describe a complex interplay between fluid dynamics, biomechanics, and aneurysm geometry~\cite{Sforza2009,Frank2024}.
The computational domain, represented by a numerical mesh, is the foundation for these simulations.
However, geometric uncertainties arise from the limitations of CT scans and subsequent post-processing techniques from which the meshes are extracted.
These uncertainties propagate through the subsequent flow field and coiling simulations, potentially influencing predictions and conclusions that may be drawn from these analyses.

In the following, we assess uncertainties arising because of geometric ambiguities of the computational domain, i.e., the boundary of the synthetic aneurysm.
Section~\ref{subsec:SyntheticAneurysm} demonstrates the generation of synthetic aneurysms via the SPDE method followed by two forward problems for UQ presented in Section~\ref{subsec:AneurysmDefomration}.
We investigate how endovascular coiling deforms and stretches the aneurysm surface and complete the examples with blood flow simulations through perturbed geometries via the Stokes equation.
The overarching goal of these forward problems is to illustrate the relationship between uncertainties in the aneurysm geometry and QoIs that influence the risk of rupture, such as the wall shear stress and pressure.
This information can provide valuable medical insights and inform the development of more robust and reliable treatment strategies.

\subsection{Synthetic Aneurysms}
\label{subsec:SyntheticAneurysm}

\begin{figure}
 \centering
 \begin{minipage}[b]{5.56cm}\centering
  \includegraphics[height=6cm]{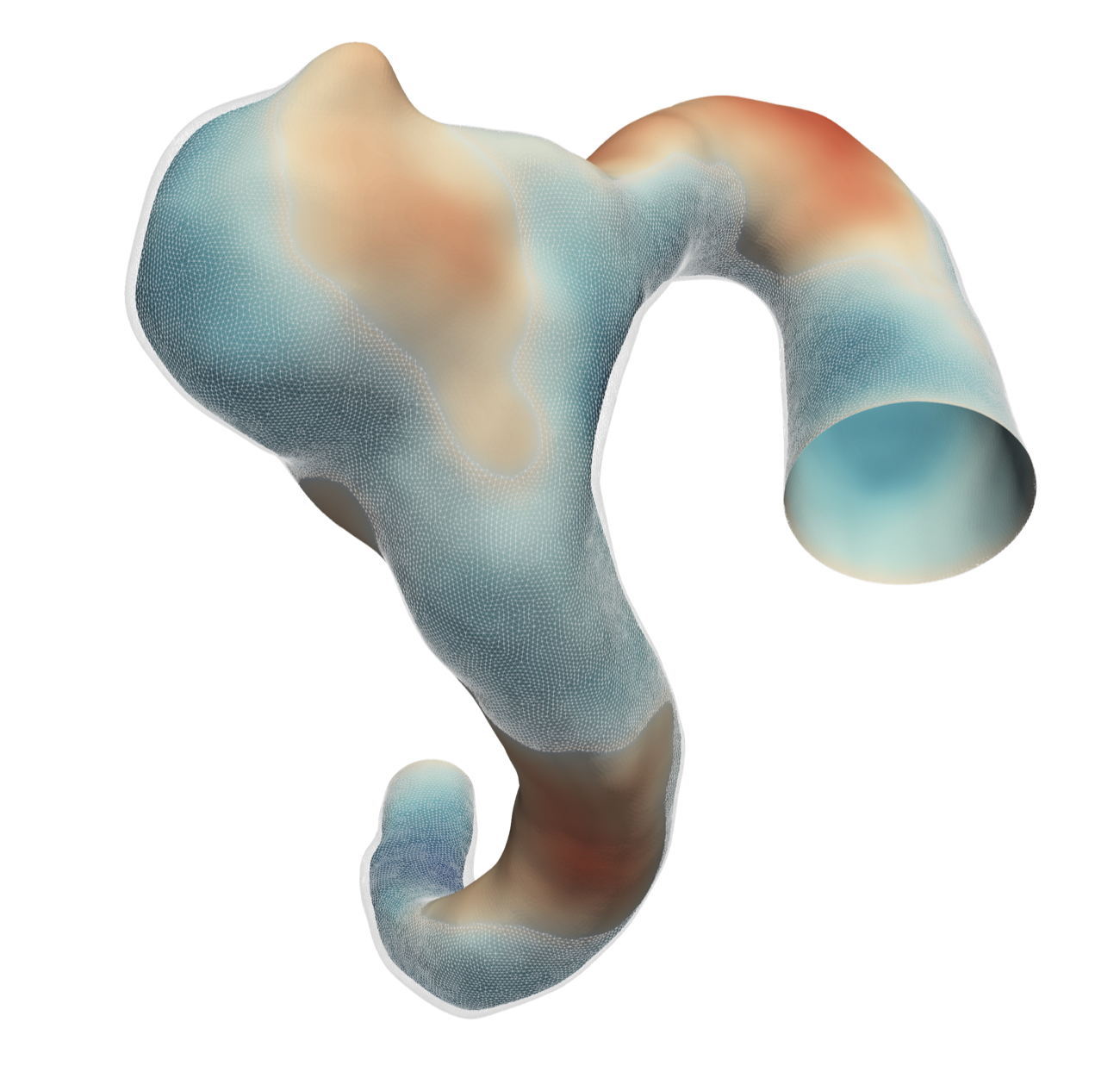}
 \end{minipage}
 \begin{minipage}[b]{2cm}\centering
  \includegraphics[height=3.5cm]{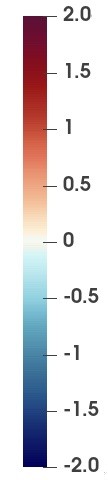}
  \vspace*{1.25cm}
 \end{minipage}
 \caption{
  The synthetic
  aneurysms are obtained from a reference geometry (grey mesh)
  \cite{yang2020intra} via the SPDE
  method and shifting the respective vertices in normal direction proportionally to the
  random field values (red-blue color bar) multiplied by a scaling parameter $\alpha_S > 0$.
  Regions with positive random field values are lifted while negative values are lowered.
  Mat\'ern-type GRF parameters: $l =5.0$, $\nu=2.0$, and $\alpha_S = 0.2$.
  We choose a larger scaling parameter $\alpha_S$ than in Table~\ref{tab:aneurysm-uq-param} so that the geometric variations are easier to observe.
 }
 \label{fig:synthetic-aneurysm}
\end{figure}

We generate synthetic aneurysms to quantify the influence of uncertainties in an aneurysm geometry.
We proceed in three steps.
First, we employ the SPDE method to sample an isotropic (i.e., $\M{\Theta} = l^2 \M{I}$) Mat\'ern-type random field $u$ on the surface of a reference aneurysm reconstructed from a CT scan~\cite{yang2020intra}.
Next, we estimate the outward-facing unit normal vector $\V{n}_k$ at each vertex $\V{v}_k$ of the triangulated aneurysm surface.
To achieve this, we average the normals of the bordering elements on the triangulated surface.
Lastly, we shift the vertices of the triangulated aneurysm surface proportionally to the values of the generated random fields; i.e., we translate the vertices $\V{v}_k \mapsto \V{v}_k + \alpha_S u(\V{v}_k)\, \V{n}_k$, where $\alpha_S \in \mathbb{R}^+$.
This process introduces physical variations in the aneurysm geometry capable of modeling the uncertainty introduced by errors in the CT scans and associated post-processing techniques.

Figure~\ref{fig:synthetic-aneurysm} shows an example of a synthetic aneurysm.
It displays the grey mesh of the reference geometry~\cite{yang2020intra} and the synthetic aneurysm colored in red and blue to indicate the values of the random field sample.
The computational mesh comprises approximately $71 \cdot 10^3$ nodes that define $142 \cdot 10^3$ triangular elements.
The shape of the random synthetic aneurysms depends in a natural way on the model parameters $l$, $\nu$, and $\alpha_S$.
Here, and throughout this section, we choose relatively long correlation lengths $l$ and high smoothness parameters $\nu$ to avoid creating unrealistic, rough, and highly-oscillatory aneurysm surfaces.
The SPDE method allows us to generate as many synthetic aneurysm geometries as we wish in parallel.
Computing a single random field realization as depicted in Figure~\ref{fig:synthetic-aneurysm} ($\nu = 2.0$, $l=5.0$) takes slightly less than one second on a standard notebook using 4 MPI ranks.

\subsection{Aneurysm deformation through coiling}
\label{subsec:AneurysmDefomration}

We may now focus on uncertainty quantification with the synthetic aneurysm geometries.
We consider an example motivated by endovascular coiling \cite{Holzberger2024}.
Here, we are interested in estimating how a medical coil inside the aneurysm stretches the surface.
The mathematical model, introduced by Nobile and Vergara in \cite{Nobile2008}, describes the displacement $\eta$
of the aneurysm surface in the presence of a load $f$ by a diffusion equation,
\begin{equation}\label{eq:membrane}
 \left( \nabla \cdot T \nabla  + \beta (\V{x}) \right) \eta (\V{x})
 = f (\V{x}) \ ,
\end{equation}
where the coefficient $T$ describes the stiffness of the membrane and $\beta$ is
constructed from weighted curvature terms.
Denoting the mean and Gaussian curvature as $\rho_M$ and $\rho_G$, the latter
coefficient reads
\begin{equation}\label{eq:membrane_beta}
 \beta (\V{x})= 2 T h \left( 2 \rho_M(\V{x})^2 - \rho_G(\V{x}) \right) +
 2 g \left( 2 \rho_M(\V{x})^2 - (1 - \nu_m)\rho_G(\V{x}) \right)
 \,,
\end{equation}
with the parameters $h$ and $\nu_M$ denoting the membrane thickness and Poisson coefficient, respectively.
The original authors derived a second-order time-dependent membrane equation from the Koiter shell model~\cite{Koiter1970,Koiter1996}.
Our model further simplifies their pre-stressed model by assuming a stationary solution and an isotropic, spatially invariant coefficient $T$.

A detailed representation of the load $f$ in~\eqref{eq:membrane} would require simulating the coiling~\cite{Holzberger2024} and coupling this to the membrane equation.
Instead, we focus on the effect of geometric uncertainties on complicated manifolds and limit ourselves to an explicit model for the internal force $f$ arising from the coiling.
We sample a set of $M$ points $\V{x}_{1,\dots, M}$
near the aneurysm center and model the force at a point $\V{x}$ on the aneurysm
surface as
\begin{equation}\label{eq:membrane_force}
 f(\V{x}) = \max_{k \in \{ 1, \dots, M \}} \left(
 b_k \exp
 \left(- a_k^{-1} (\V{x} - \V{x}_k)^T \M{\Theta}_k (\V{x} - \V{x}_k)  \right)
 \right)
 \ \ \text{with} \ \
 \M{\Theta}_k = \M{R}_k^T
 \begin{pmatrix}
  t_1 & 0   & 0   \\
  0   & t_2 & 0   \\
  0   & 0   & t_3 \\
 \end{pmatrix} \M{R}_k \ ,
\end{equation}
where $a_k$, $b_k$, and $t_i > 0$ parametrize the force.
The matrices $\M{R}_k \in \mathrm{SO}(3)$ are random rotation matrices obtained by sampling Euler angles.

\begin{figure}[ht]
 \centering
 \begin{minipage}[b]{0.32\textwidth}\centering
  \includegraphics[width=\textwidth]{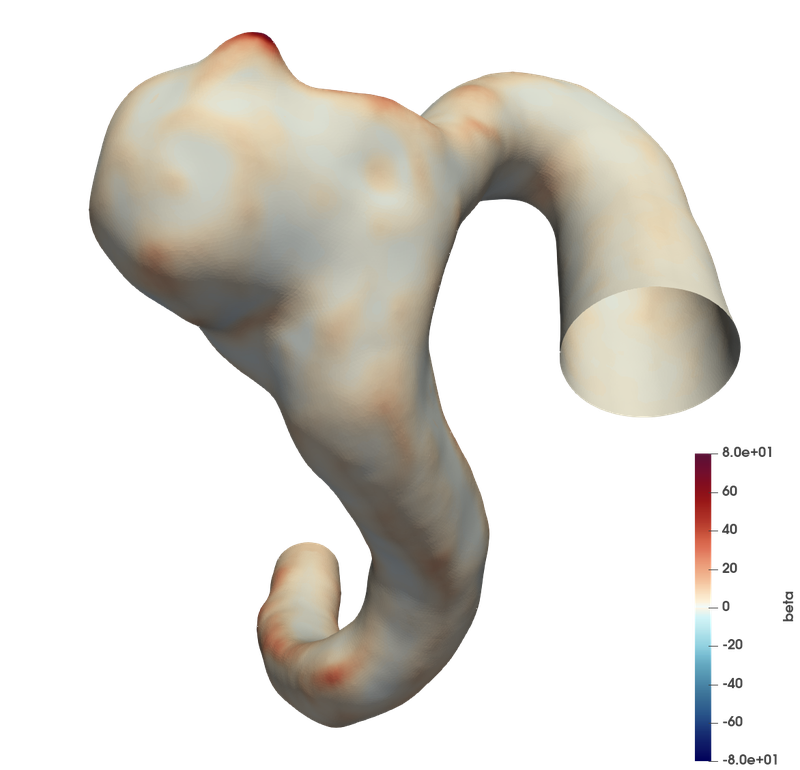}
  \caption*{(a) Mass $\beta(\V{x})$}
 \end{minipage}
 \begin{minipage}[b]{0.32\textwidth}\centering
  \includegraphics[width=\textwidth]{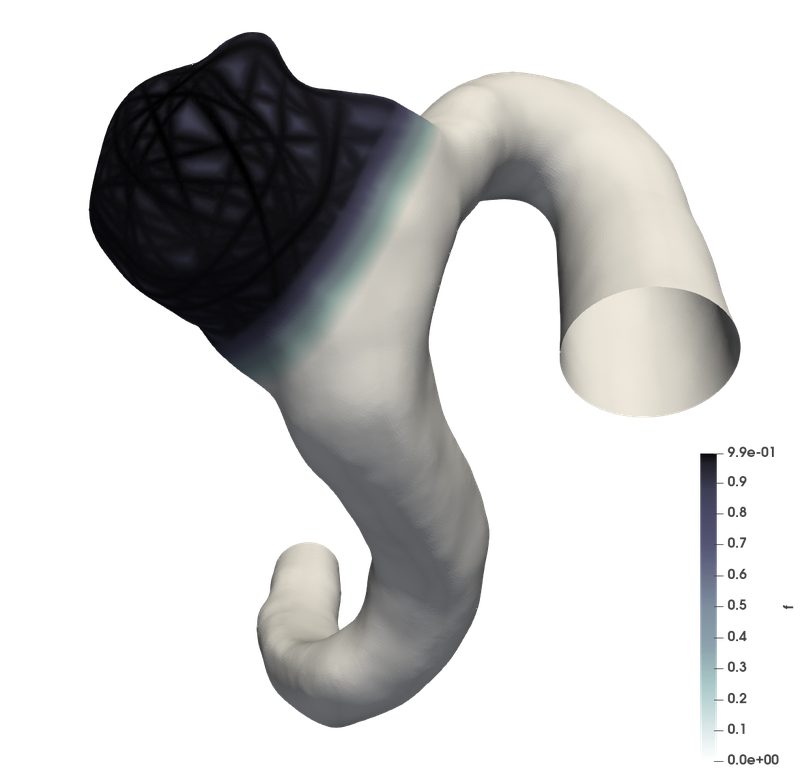}
  \caption*{(b) Load $f(\V{x})$}
 \end{minipage}
 \begin{minipage}[b]{0.32\textwidth}\centering
  \includegraphics[width=\textwidth]{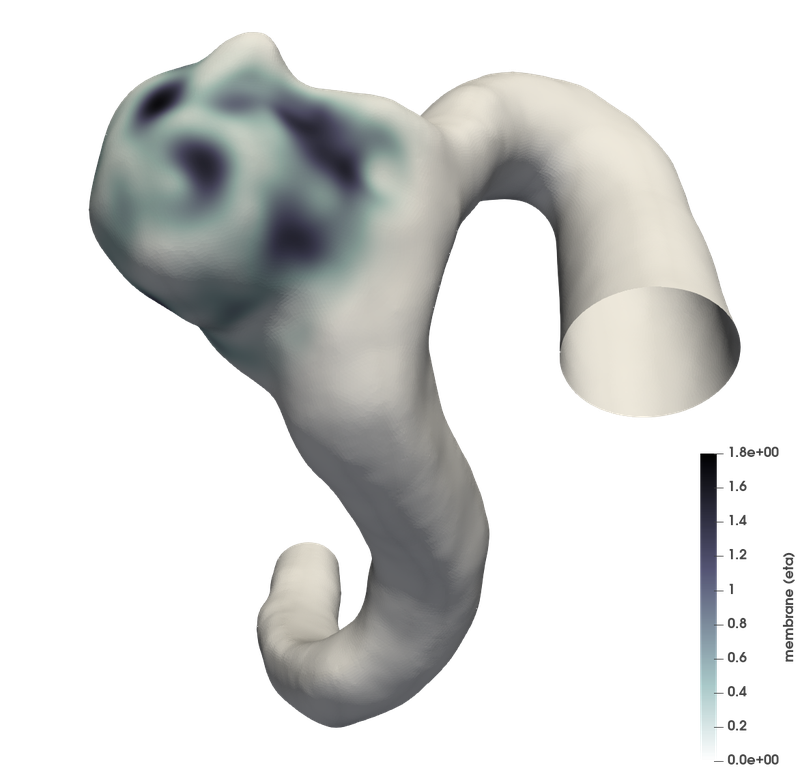}
  \caption*{(c) Membrane displacement $\eta(\V{x})$}
 \end{minipage}
 \caption{
  The solution of \eqref{eq:membrane} through~\eqref{eq:membrane_force} on the reference aneurysm.
  Images (a) and (b) show the functions defined in
  \eqref{eq:membrane_beta} and~\eqref{eq:membrane_force}, respectively.
  (c) shows the solution $\eta$ of~\eqref{eq:membrane} obtained with the
  finite element method.
  The model parameters are given in Table~\ref{tab:aneurysm-uq-param}.
 }
 \label{fig:membrane-solution}
\end{figure}

Conveniently, the mathematical problem defined via \eqref{eq:membrane} through~\eqref{eq:membrane_force} is similar to the SPDE problem.
We discretize it using the finite element method and solve the arising linear system with the conjugate gradient method preconditioned with an algebraic multi-grid.
For more details, we refer to Section~\ref{sec:SPDE_discretization}.
Computationally, the time for solving it is dominated by the approximate computation of the curvature terms in~\eqref{eq:membrane_beta}~\cite{CohenSteiner2003}, such that $c=T_r/T_p \ll 1$ holds.
Figure~\ref{fig:membrane-solution} shows the solution of the forward problem for the parameter choices given in Table~\ref{tab:aneurysm-uq-param} and used throughout this section.

\begin{table}[h]
 \centering
 \begin{tabular}{@{}l|lll|lllllllll@{}}
  \toprule
  Problem   & \multicolumn{3}{c|}{SPDE} & \multicolumn{9}{c}{Membrane}                                                                                               \\ \midrule
  Parameter & $\nu$                     & $l_i$                        & $\alpha_{S}$ & $T$ & $h$  & $\nu_M$ & $g$  & $a_k$ & $b_k$ & $t_1$ & $t_2=t_3$ & $\alpha_M$ \\
  Value     & 2.0                       & 5.0                          & 0.1          & 0.1 & 0.05 & 0.5     & 10.0 & 4.0   & 1.0   & 20.0  & 0.05      & 0.05       \\ \bottomrule
 \end{tabular}
 \caption{Parameters for the UQ aneurysm experiments.}
 \label{tab:aneurysm-uq-param}
\end{table}

Based on the solution $\eta$ of~\eqref{eq:membrane} depicted in Figure~\ref{fig:membrane-solution} (c), we define our quantity of interest (QoI) as the surface distortion resulting from shifting the vertices $\V{v}_k \mapsto \V{v}_k + \alpha_M \eta(\V{v}_k)\, \V{n}_k$, where $\alpha_M \in \mathbb{R}^+$.
More specifically, we compute the \textit{relative surface change} (RSC),
\begin{equation}
 q_j = \frac{A^{new}_j - A^{old}_j}{A^{old}_j}
 \,,
\end{equation}
an element-wise QoI, where $A_j$ denotes the $j$-th face surface area and $j$ ranges over the number of faces in the mesh.
Regions experiencing high stretches are prone to ruptures and, thus, are of significant interest for treatment planning~\cite{Levy2001}.
Consequently, we are interested in the largest (or $k$ largest) observed RSC values per sample.

Having defined the uncertainties, the mathematical forward model, and the QoIs, we now lay out the UQ workflow.
We generate a set of 5000 synthetic aneurysms following the procedure described in Section~\ref{subsec:SyntheticAneurysm} and using the parameters given in Table~\ref{tab:aneurysm-uq-param}.
The force field defined via \eqref{eq:membrane_force} is generated once before generating the synthetic aneurysms and left unchanged for all samples.

\begin{figure}[h]
 \centering
 \begin{minipage}[b]{0.33\textwidth}\centering
  \includegraphics[width=\textwidth]{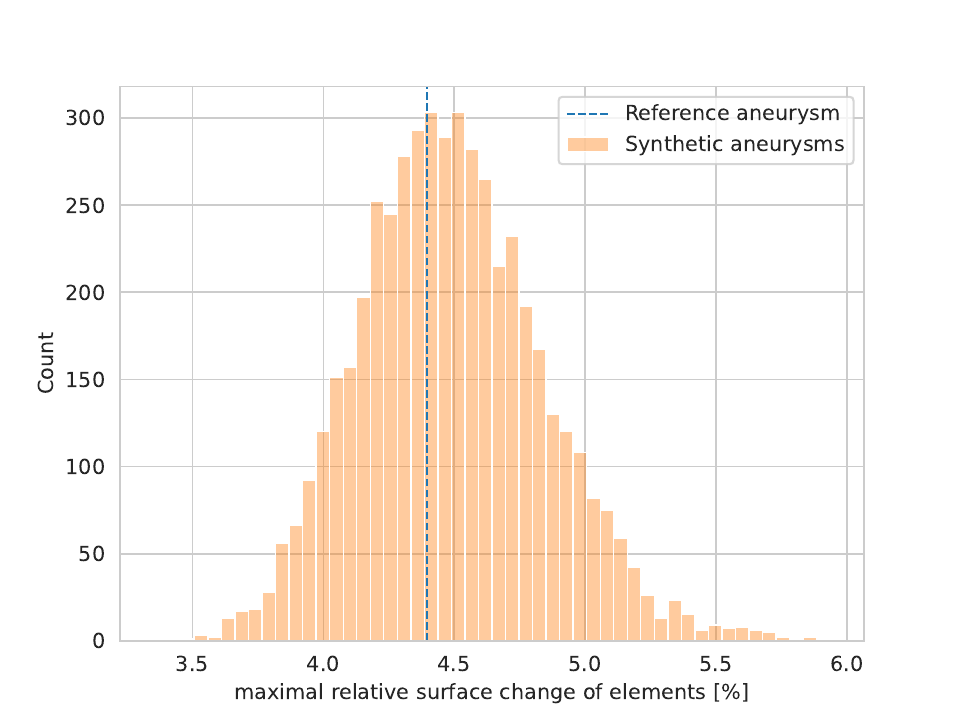}
  \caption*{(a) Maximal element distortion}
 \end{minipage}
 \begin{minipage}[b]{0.33\textwidth}\centering
  \includegraphics[width=\textwidth]{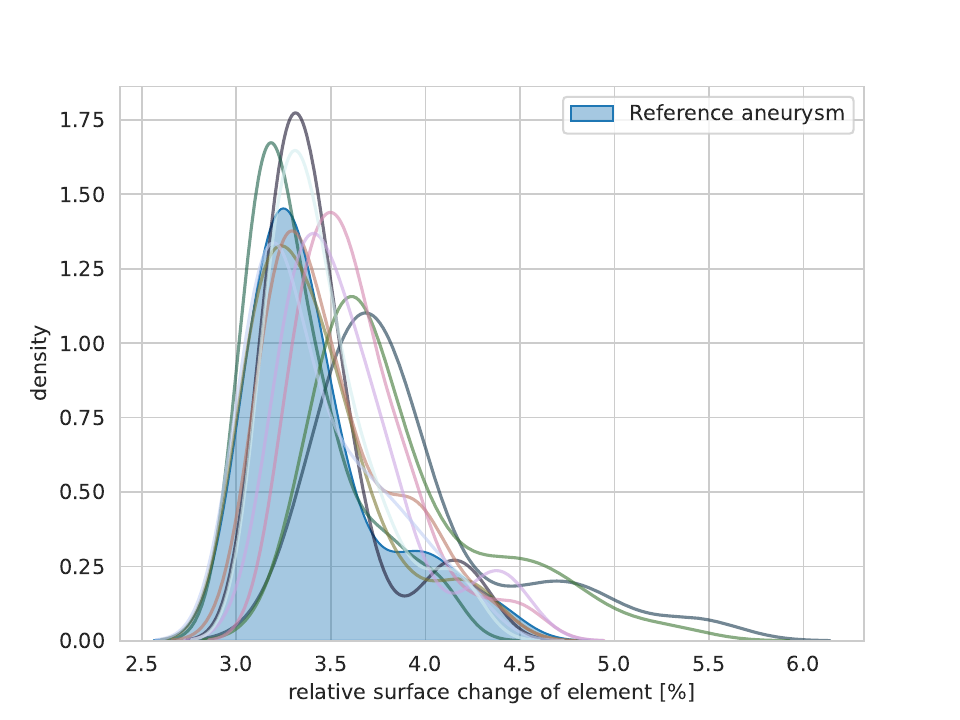}
  \caption*{(b) KDEs of top 50 distortions}
 \end{minipage}
 \begin{minipage}[b]{0.33\textwidth}\centering
  \includegraphics[width=\textwidth]{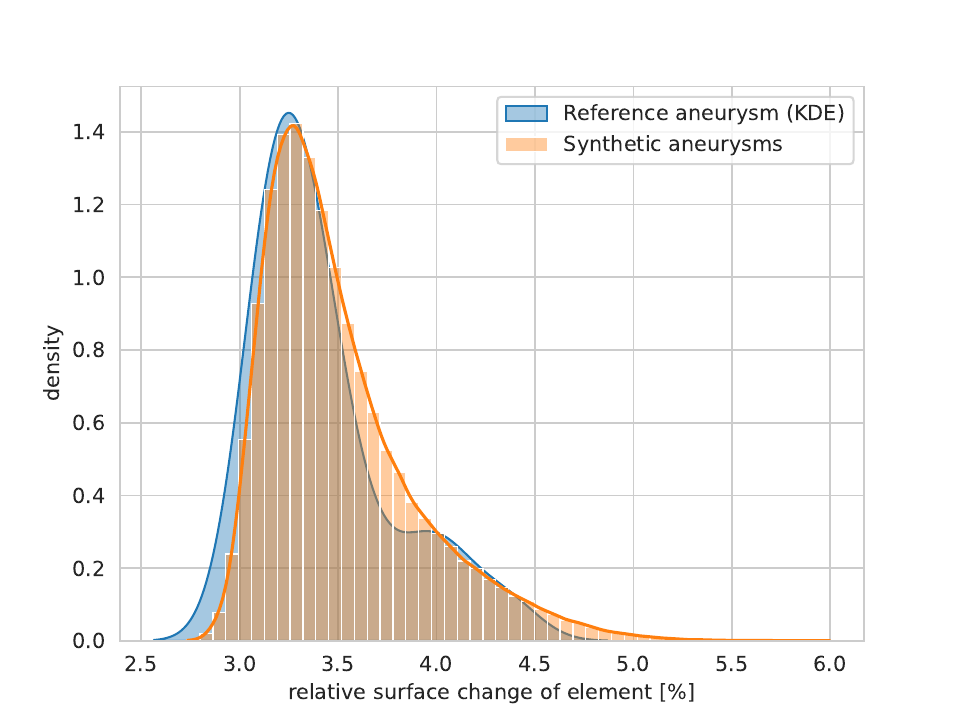}
  \caption*{(c) Distribution of top 50 distortions}
 \end{minipage}
 \caption{
  Effects of uncertainties in the aneurysm geometry on the
  stretching of elements caused by a coil force.
  The \textit{reference aneurysm} is obtained via a CT scan and depicted in Figure~\ref{fig:synthetic-aneurysm}.
  (a) Histogram of the maximal
  relative surface change of an element observed across 5000 synthetic
  samples. (b) KDE estimates of the largest 50 relative surface changes for
  the reference aneurysm (blue) and ten synthetic aneurysms. (c) Histogram of
  the largest 50 relative surface changes accumulated over 5000 samples.
 }
 \label{fig:aneurysm-deformation}
\end{figure}

Figure~\ref{fig:aneurysm-deformation} shows the results of this representative study.
The histogram in Figure~\ref{fig:aneurysm-deformation}~(a) displays the distribution of the maximal RSC value in each sample.
While the reference aneurysm obtained via a CT scan shows a maximal RSC of roughly $4.4$\% (blue), the synthetic aneurysms (orange) show a wide spread of the QoI, with some aneurysms showing RSCs of over $5.5$\%.
This suggests that only simulating on the reference aneurysm may underestimate the risk of rupture during treatment.
Figure~\ref{fig:aneurysm-deformation}~(b,c) provides additional insights.
Both figures show the distribution of the 50 largest RSC values for various samples.
Figure~\ref{fig:aneurysm-deformation}~(b) shows the kernel density estimate~(KDE) of the 50 largest RSC values for the reference aneurysm~(blue, filled) and ten randomly chosen synthetic aneurysms.
Figure~\ref{fig:aneurysm-deformation}~(c) presents a histogram of all of the top 50 RSC values across all synthetic RSC distributions.
The heavy tails in both (b) and (c) provide additional support to the conclusion that only simulating on the reference aneurysm may underestimate the probability of a rupture.

\begin{figure}[H]
 \centering
 \begin{minipage}{0.9\textwidth}
  \begin{minipage}[b]{0.33\textwidth}\centering
   \includegraphics[width=\textwidth]{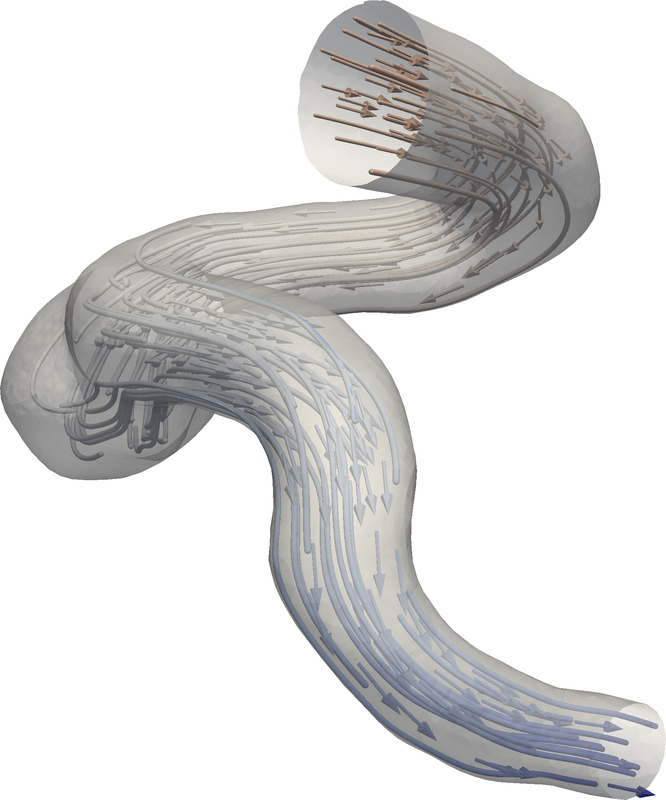}
  \end{minipage}
  \begin{minipage}[b]{0.33\textwidth}\centering
   \includegraphics[width=\textwidth]{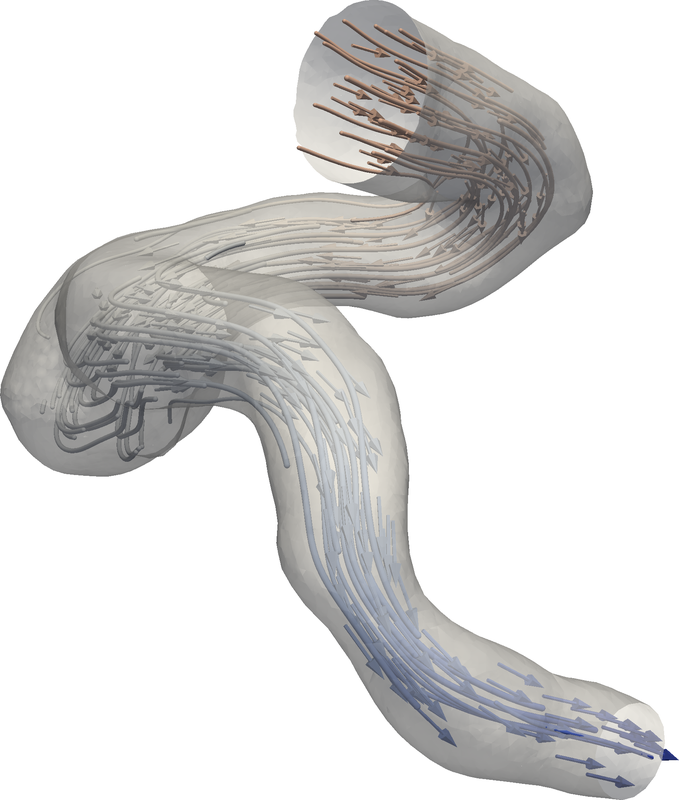}
  \end{minipage}
  \begin{minipage}[b]{0.33\textwidth}\centering
   \includegraphics[width=\textwidth]{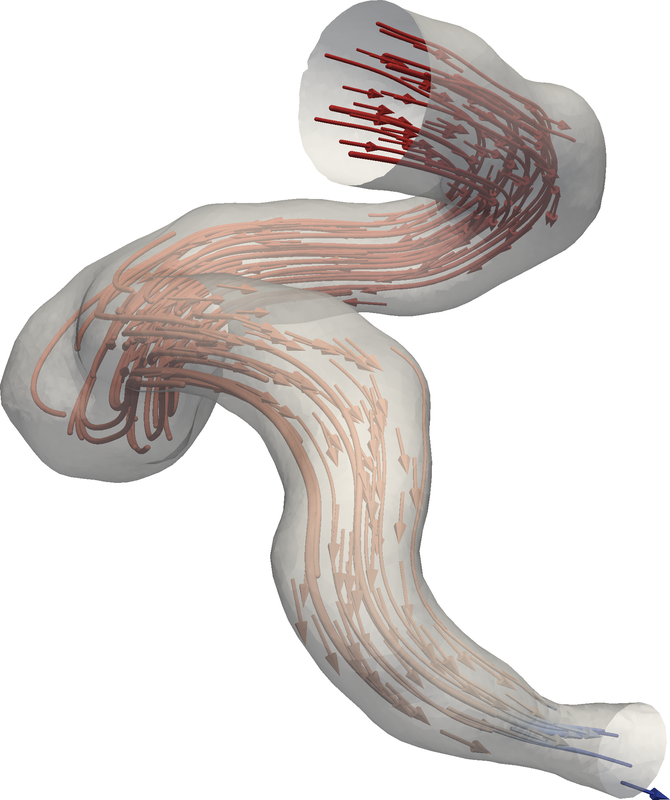}
  \end{minipage}
 \end{minipage}
 \begin{minipage}{0.06\textwidth}
  \includegraphics[width=\textwidth]{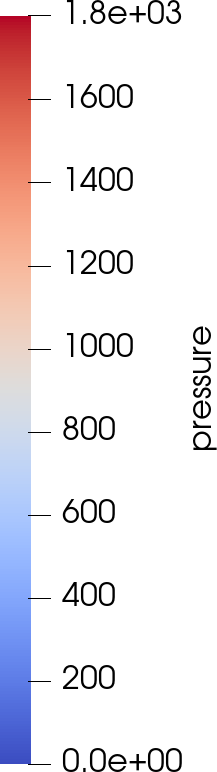}
 \end{minipage}
 \caption{Effects of geometric uncertainties for three realizations on a fluid flow throughout the aneurysm.
  The streamlines are colored by the pressure in the fluid using the color scale shown in the figure.
 }
 \label{fig:aneurysm-flow}
\end{figure}

The same GRF model can also be used for analyzing blood flow variations through different realizations of the aneurysm geometry.
From basic fluid dynamics considerations, we expect the flow, particularly the pressure drop, to be very sensitive to geometric uncertainties.
This can be understood from the Hagen--Poiseuille law, which states that the pressure drop is inversely proportional to the fourth power of the vessel's radius~\cite{Sutera1993}.
However, because this law is derived for a cylindrical pipe instead of the complicated vessel geometry, we choose to numerically probe this hypothesis with a steady state Stokes flow model.
We use a parabolic velocity profile prescribed at the inlet, free flow boundary conditions at the outlet, and no-slip boundaries at the vessel walls.
The solutions for three different random vessels are shown in Figure~\ref{fig:aneurysm-flow}, and significant differences in the flow velocity and pressure distributions can be observed.
Indeed, the first two vessels show similar pressure losses and velocity distributions.
However, the third realization has a narrow diameter close to the outlet, which results in much more significant pressure loss and higher velocities towards the outlet.
This change to the velocity profile will result in larger shear stresses on the aneurysm, increasing its risk of rupturing.
The considered Stokes model is only utilized here for illustration purposes, and a more realistic blood flow model with a non-trivial Reynolds number and pulsating boundary conditions is outside of the scope of the current article, yet necessary for further medical analyses~\cite{Sforza2009,Frank2024,horvat2024lattice}.

\section{Topology optimization under uncertainty}
\label{sec:num_ex}

The ability to sample Gaussian random fields with a wide range of correlation lengths opens the possibility of investigating the impact of realistic operational, manufacturing, and material uncertainties on topology-optimized designs. Thus, the main goal of this section is to demonstrate these capabilities, discuss the computational complexity, and share details on the method's performance. In the following, Section~\ref{sec:topology_optimization} provides a concise introduction to the standard techniques and notations of topology optimization.
The experienced practitioner may directly proceed to the selected examples, demonstrating the effect of such uncertainties.
Section~\ref{sub:heat_sink} considers a heat sink design problem subject to operational and material uncertainties.
We treat the problem in 2D and on a surface embedded in 3D, and investigate how different correlation lengths, anisotropy, and symmetry affect the design.
Section~\ref{sub:bridge} presents the optimized design of bridges under stochastic excitation in 2D and 3D.
If not explicitly stated to be otherwise, all GRFs generated via the SPDE~\eqref{eq:SPDE_fullspace} obey homogeneous Neumann boundaries.
The vast collection of examples illustrates the flexibility and parallel scalability of the SPDE method, as well as the importance of incorporating uncertainties into the design problem.

\subsection{Topology optimization}
\label{sec:topology_optimization}

Topology optimization is an iterative design process aiming to find a material distribution by optimizing a specified performance measure and fulfilling a set of constraints~\cite{Bendsoe2003}.
Mathematically, the associated optimization problems can be written in the following form:
\begin{subequations}
 \label{eqopt01}
 \begin{align}
  \label{eqopt01_objective}
  \text{minimize}   & \quad j\left(\rho,\V{u}\right) \text{~over~} \left(\rho, \V{u}\right)\in Z\times V               \\
  \label{eqopt01_stateeq}
  \text{subject to} & \quad  a\left(\V{u}, \V{v}; \rho \right)= \ell\left(\V{v}\right)\,\,  \text{for all~} \V{v}\in V
  \,.
 \end{align}
\end{subequations}
Here, $j \colon Z\times V\rightarrow \mathbb{R}$ is an integral functional subject to a set of constraints.
The first constraint is a weak formulation of a partial differential equation (PDE).
If the PDE is linear, then $a \colon V\times V\rightarrow \mathbb{R}$ is a parametric bilinear form,  $\ell \colon V\to \mathbb{R}$ is a linear form, $V$ is the admissible solution space of the weak formulation, and $Z$ is the so-called design space defined as
\begin{equation}
 Z=\left\{\rho\in {L}^1\left(D\right)\,:\,0\leq \rho \leq 1\,\,{\rm{a.e.
   }}, \int_D \rho \operatorname{d}\!\V{x} \leq \gamma \left|D\right| \right\}\,,
\end{equation}
for some fixed volume fraction parameter $\gamma \in (0,1)$.
Equation~\eqref{eqopt01_stateeq} is called the state equation, whose coefficients depend on a material distribution $\rho \in Z$, which in the topology optimization literature is referred to as the \textit{density field}.

Assuming that~\eqref{eqopt01_stateeq} is uniquely solvable for every $\rho\in Z$, the state variable $\V{u} = \V{u}(\rho)$ in~\eqref{eqopt01_objective} is often eliminated and the design problem is reduced to
\begin{equation}
 \text{minimize} \quad J(\rho) := j\left(\rho,\V{u}\left(\rho\right)\right) \text{~~over~} \rho\in Z\,.
\end{equation}
The above form provides the basis of the so-called nested design space approach in topology optimization \cite{Bendsoe2003}.
It is often advantageous compared to the original formulation~\eqref{eqopt01} because the state equations are always satisfied during the optimization process, often allowing the users to stop the iterative optimization process early.

For most practical problems, the density $\rho \in Z$ should take on discrete values, zero or one, at every point in the design domain $D \subset \mathbb{R}^d,\,d=2,3$.
However, gradient-based optimization techniques require that the density vary continuously between zero and one.
Such relaxation leads to `grey' transition regions where $0 < \rho(\V{x}) < 1$.
The transition zone can be controlled by penalization~\cite{Bendsoe1999}, projections~\cite{Lazarov2016}, or by reformulating the optimization problem to include manufacturing uncertainties~\cite{Wang2011, Schevenels2011, Lazarov2017}.
Applying a penalization, i.e., making all intermediate densities unfavorable, without incorporating additional restrictions on the density field leads to an ill-posed optimization problem, i.e., a mesh-dependent solution.
A standard remedy is to employ filtering~\cite{Bourdin2001}, where the material distribution is obtained by convolving the density field $\rho$ with a filter kernel.
Instead of an explicit convolution integral, the filtered density can be modeled as the solution of the following screened Poisson equation~\cite{Lazarov2011},
\begin{equation}\label{eqopt02}
 -\nabla\cdot r^2\nabla \tilde{\rho}+\tilde{\rho}=\rho ~~~\text{in~}
 D\,,
\end{equation}
where $r>0$ is often referred to as a length scale parameter for design.
This PDE filter is equivalent to the standard convolution filter~\cite{Lazarov2011}.
However, it offers additional benefits such as convenient enforcement of boundary conditions on the filtered density field $\tilde{\rho}$, using the same finite element discretization machinery for solving the state PDE, and an efficient parallel implementation.
Typically, the homogeneous Neumann boundary condition $\nabla \tilde{\rho}\cdot \mathbf{n} =0$ is applied along the boundary $\partial D$, with $\mathbf{n}$ denoting the outward-facing unit normal.
However, mixed Dirichlet/Neumann boundary conditions enforcing the density to take values zero or one on specified parts of the boundary, $\rho=0\lor 1$ on $\Gamma_{\mathrm{D}} \subset \partial D$, are often employed in practical applications.

The most popular penalization approach in density-based compliance topology optimization is the so-called SIMP (solid isotropic material with penalization) interpolation~\cite{Bendsoe1989}.
Taking the modulus of the elasticity as an representative example, it is written as
\begin{equation}
 \label{eqopt03}
 E=E_{\text{min}}+\tilde{\rho}^p\left(E_{\text{max}}-E_{\text{min}} \right),
\end{equation}
where $\tilde{\rho}$ is the filtered density field, $p$ is a penalization parameter, and $E_{\text{max}}$ is the modulus of elasticity of the solid material.
$E_{\text{min}}$ is a small positive constant ensuring that the state equation~\eqref{eqopt01_stateeq} is well-defined.
Depending on the application, analogous equations govern other physical coefficients in the state PDE, for instance, the material diffusion coefficient $\kappa$ in~\eqref{eq:diffusionSIMP} below.
In~\cite{ Bendsoe1989}, $E_{\text{min}}$ is set to zero, and in such cases, $\rho$ and correspondingly $\tilde{\rho}$ has to be bounded from below to remain greater than zero.
The penalization~\eqref{eqopt03} is controlled by the power parameter $p$, which, in combination with an active volume constraint, makes intermediate densities $0 < \rho(\V{x}) < 1$ unfavorable.
The penalization parameter is often set to be $p\geq 3$, with larger values resulting in a steeper transition between void and solid.
The intermediate densities cannot be removed entirely in the final optimized solution, and if necessary, these parts of the design can be realized by using spatially varying microstructures~\cite{Bendsoe1999}.

Projections were introduced in~\cite{Guest2004} to sharpen the transition from void to solid.
The original idea is to use a filter function with finite support combined with a pointwise Heaviside projection.
Applying a zero threshold to the Heaviside projection -- where all filtered densities above zero are projected to one -- ensures that the final solid design corresponds to the support of the filtered density function $\tilde \rho$, thereby guaranteeing the non-trivial size of all solid features in the design.
Conversely, setting the threshold to one in the Heaviside projection -- meaning all filtered densities below one are projected to zero -- enforces a non-zero size upon all void features in the design.
The Heaviside projection is replaced with a non-linear differentiable expression in practical implementations, with its transition sharpness is controlled by a single parameter.

The above approach does not work for the PDE filter regularization approach because the filter is a Green's function whose support coincides with the entire computational domain.
Thus, the so-called robust formulation was introduced in~\cite{Wang2011} to alleviate the lack of a length scale in either the void or the solid part of the design for the classic convolution filter with finite support size, as well as to sharpen the transition regions for problems regularized with the screened Poisson equations~\eqref{eqopt02}.
The projection in~\cite{Wang2011} is performed point-wise using the following  non-linear transformation:
\begin{equation}
 \label{eqopt04}
 \overline{\rho}=\frac{\tanh\left(\beta\eta \right)-\tanh\left(\beta\left(\tilde{\rho}-\eta\right) \right) }{\tanh\left(\beta\eta \right)-\tanh\left(\beta\left(1-\eta\right) \right)},
\end{equation}
where $\eta\in\left(\eta_d,\eta_e\right)$ is a threshold parameter, $0<\eta_d \leq \eta_e<1$, and $\beta$ controls the sharpness of the projection.
The above expression approaches the Heaviside projection as $\beta\rightarrow\infty$.
The projection alone does not enforce a length scale on the design.
The length scale is enforced by minimizing the worst possible compliance from all possible realizations for $\eta$ between $\eta_d$ and $\eta_e$, with a bound on the worst possible volume.
Thus, the optimization problem~\eqref{eqopt01} is replaced by
\begin{equation}
 \label{eqopt0r51}
 \begin{aligned}
  \text{minimize}   & \quad \max_{\eta_d \leq \eta \leq \eta_e} j\left(\overline{\rho}\left(\eta\right),\V{u}\right) \text{~over~} \left(\rho, \V{u}\right)\in Z\times V \\
  \text{subject to} & \quad  a\left(\V{u}, \V{v}; \overline{\rho}\left(\eta\right) \right)=\ell\left(\V{v}\right)\,\,  \text{for all~} \V{v}\in V                        \\
                    & \quad  a_{\mathrm{f}}\left(\tilde{\rho},{w}\right)=\ell_{\mathrm{f}}\left({w};\rho\right)\,\,  \text{for all~} {w}\in W\, ,
 \end{aligned}
\end{equation}
where the equation with the linear form $\ell_{\mathrm{f}} \colon W \times Z \to \mathbb{R}$ represents the PDE filter equation~\eqref{eqopt02}, and the design space is redefined to be
\begin{equation*}
 Z=\left\{\rho\in {L}^1\left(D\right)\,:\,0\leq \rho \leq 1\,\,{\text{a.e.
   }},\, \max_\eta \int_D \overline{\rho}\left(\rho\right) \operatorname{d}\!\V{x} \leq \gamma \left|D\right| \right\}
 \,.
\end{equation*}

The compliance objective is the most popular choice of performance measure. It is monotonic with respect to the threshold parameter $\eta$.
Thus, searching for the worst possible objective and material volume case can be avoided~\cite{Lazarov2016}.
Threshold $\eta=\eta_d$ represents the most dilated realization, $\eta=\eta_e$ the most eroded one, with all realizations in between providing a simple model for the outcomes of a production process with uniform manufacturing errors along the perimeter of the topology-optimized design.
Including these simplified production uncertainties in the optimization process avoids post-processing and engineering intervention between design and manufacturing~\cite{Andreassen2014}.

Uniform erosion or dilation of the design is a crude approximation of real-life manufacturing uncertainties.
These can vary spatially and, in many cases, introduce changes in the topology between different realizations.
Therefore, more realistic models where the design's geometry or material stiffness varies spatially were proposed in~\cite{Schevenels2011, Lazarov2012}.
For geometric uncertainties, the threshold $\eta$ is modeled as a random field, and for material uncertainties, the random field represents the modulus of elasticity.
In these models, finding the worst case is extremely expensive, and the maximum in~\eqref{eqopt0r51} is replaced by the expectation or another risk measure.

In this work, we use several topology optimization problems to demonstrate various properties and effects of the SPDE approach to modeling random fields.
The exact objectives and constraints are explained for every numerical example individually, ensuring a comprehensive understanding of the application and impact of the SPDE method in each case.
It is important to note that the threshold interpretation of density-based topology optimization introduced in~\cite{Wang2011} shares similarities with many level-set approaches~\cite{Sigmund2013, Dijk2013}, and such formulations under geometric, material, or operational uncertainties can be found in~\cite{Chen2011a, MartinezFrutos2016, Martinez-Frutos2018, MartinezFrutos2018}.
Thus, the SPDE representation of random fields applies beyond the density-based formulations utilized here and could also be employed in other topology optimization approaches.

\subsection{Heat sink}
\label{sub:heat_sink}

Optimizing thermal compliance is a textbook example of topology optimization~\cite{Bendsoe2003}.
The goal is to find a topology that extracts heat from a given domain $D$ by minimizing the thermal compliance of the design, given a distributed heat source $f$ on $D$.
The problem may be considered with a constant (deterministic) or spatially-varying (stochastic) heat source.
We first describe the physical model and the deterministic problem in~\ref{sec:deterministic_thermal_compliance}.
Afterward, we explore how GRFs influence the design in~\ref{sec:stochatic_thermal_compliance}, explain efficient symmetrization techniques based on the SPDE method in~\ref{sec:enforcing_symmetry}, and demonstrate the effect of anisotropy in~\ref{sec:effects_of_anisotropy}.
Sections~\ref{sec:thermal_compliance_sphere} and~\ref{sec:material_damage} present additional examples of optimized topologies on embedded manifolds and those subject to random material damage, respectively.

\subsubsection{Deterministic thermal compliance optimization}
\label{sec:deterministic_thermal_compliance}

We describe the topology with a zero/one bounded density field with intermediate grey values penalized using SIMP material interpolation with penalty parameter $p=3$.
We consider the domain $D=(-0.5,0.5)^2$ upon which we define a heat source represented by the function $f$.
The domain's boundary $\partial D$ is divided into two parts, $\Gamma_\mathrm{N}$ and $\Gamma_\mathrm{D}$.
The boundary $\Gamma_\mathrm{N}$ is insulated, i.e., the normal gradient of the heat flow along the boundary is zero. On $\Gamma_\mathrm{D}$, with length $1/7$ centered at the bottom of the domain, the temperature is fixed to zero.
A diffusion equation describes the physics of the heat transfer, namely,
\begin{equation}
 \label{eq_ex1_001}
 -\nabla\cdot \kappa\left(\tilde{\rho}\right) \nabla u = f\quad \text{in}\, D
 \,,
 \quad
 \nabla u\cdot \V{n}=0 ~~\text{on } \Gamma_\mathrm{N}
 \,,
 \quad
 u=0 ~~\text{on } \Gamma_\mathrm{D}
 \,,
 \quad
 \overline{\Gamma_\mathrm{N} \cup \Gamma_\mathrm{D}} = \partial D \,,
\end{equation}
where $u$ represents the temperature in the domain.
The weak formulations of~\eqref{eq_ex1_001} is given as follows,
\begin{equation}\label{eq:heat_sink_diffusion_space}
 a_1\left(u,v\right)=\ell_1\left(v\right),\quad \forall  v \in V_0:=\left\{v \in H^1\left(D\right):v = 0\,\rm{on}\,\Gamma_\mathrm{D}\right\}.
\end{equation}
with bilinear form $a_1$ and linear form $\ell_1$ defined by
\begin{equation}
 a_1\left(u,v\right)=\int_D \big(\kappa\left(\tilde{\rho}\right)\nabla u \cdot \nabla v \big)\, {\rm{d}} \mathbf{x},  \quad  \ell_1\left(v\right)= \int_D f\, v\, { \rm{d}} \mathbf{x}\,.
\end{equation}
At every point of the design domain, the diffusion coefficient is evaluated as
\begin{equation}
 \label{eq:diffusionSIMP}
 \kappa\left(\tilde{\rho}\right)=\kappa_{\min}+{\tilde{\rho}}^p\left(\kappa_{\max}-\kappa_{\min} \right)\,,
\end{equation}
with $\tilde{\rho}$ obtained by solving the finite element discretized PDE-filter equation~\eqref{eqopt02} with weak bilinear and linear forms ($W = H^1\left(D\right)$) defined as
\begin{equation}
 a_{\text f}\left(\tilde{\rho},v\right)=\int_D \big( r^2 \nabla \tilde{\rho} \cdot \nabla v + \tilde{\rho} v \big)\, {\rm{d}} \mathbf{x}, \quad  \ell_{\text f}\left(v\right)= \int_D \rho\,v\, {\rm{d}} \mathbf{x}\,.
\end{equation}

\begin{figure}
 \centering
 \includegraphics[width=0.25\textwidth]{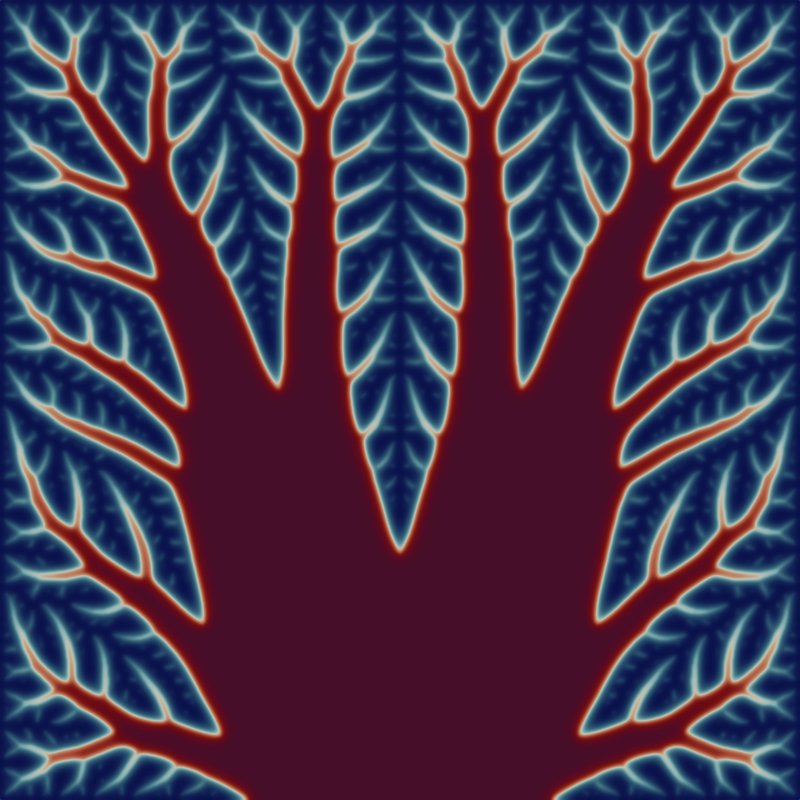}
 \caption{Optimized density $\tilde{\rho}$ for  thermal compliance problem with deterministic heat source and volume constraint 50\% (i.e., $\gamma = 0.5$).
  The length parameter in the PDE filter is set to $r=0.02$.
 }
 \label{fig:thermal_compliance_deterministic}
\end{figure}

In the deterministic setting, the heat source is a constant $f=1$, and the objective is to minimize thermal compliance given as
\begin{equation}
 \label{bb00001}
 j\left(\rho,u\right)=\int_D u f {\text{d}}\V{x} \, ,
\end{equation}
with $u$ defined via~\eqref{eq_ex1_001} and with $\rho$ subject to a volume constraint
\begin{equation}\label{eq:volume_constraint}
 \int_D \tilde{\rho}(\V{x}) {\text{d}}\V{x} \leq V
 \, .
\end{equation}
We note that (\ref{bb00001}) depends indirectly on $\rho$, i.e., $u=u\left(\tilde{\rho}\left(\rho\right)\right)$.
The obtained optimized topology minimizing the thermal compliance for the uniform deterministic heat source is shown in Figure~\ref{fig:thermal_compliance_deterministic}.
The deterministic problem is self-adjoint, and gradients of the thermal compliance with respect to the discrete representation of the original density field $\rho$ are obtained using the chain rule.
The domain is discretized using quad elements, with approximately 0.33M DOFs representing the temperature field.

\subsubsection{Stochastic thermal compliance optimization via SPDE}
\label{sec:stochatic_thermal_compliance}

In the stochastic setting, we model $f$ as a zero-mean M\'atern-type GRF with different correlation length parameters and smoothness $\nu = 1$ obtained via the SPDE method, i.e., as the solution to~\eqref{eq:SPDE_fullspace} using the same discretization as the physical problem.
The considered stochastic optimization problem is formulated akin to the deterministic one.
The objective becomes the minimization of the \textit{expected} thermal compliance defined as
\begin{equation}\label{eq:thermal-compliance-objective}
 j\left(\rho,u\right)=\mathbb{E}\left[\int_D u f {\text{d}}\V{x}\right] \approx \frac{1}{N} \sum_{i=1}^N \int_D u_i f_i {\text{d}}\V{x}\,.
\end{equation}
For the numerical examples to follow, we approximate the expectation operator with the fixed number of samples $N=300$, and each optimization step uses the same $300$ GRF samples.
We determined the number of samples with numerical experiments systematically testing the effect of the sample size on the objective and resulting designs; from 300 samples onwards, the designs and objective showed only very minor differences.
More details may be found in~\ref{apx:sample_size}.
We remark that more sophisticated, adaptive algorithms can address the approximation of the expectation operator during stochastic optimization, such as the ones presented in~\cite{bollapragada2018adaptive, Kodakkal2022, xie2023constrained, Beiser2023, Bollapragada2023,lau2024adadagrad}.
They begin with a small number of samples and adaptively increase it as the solution approaches the minimum.
This allows the algorithms to reduce the number of forward model evaluations needed to converge to the solution.
However, due to the stochastic nature of the algorithm and the small number of initial samples, these algorithms are prone to falling into a local, asymmetric minimum.
For the heat sink with a risk-neutral objective, i.e., mean compliance, an adaptive sampling algorithm is presented in the authors' earlier work~\cite{Bollapragada2023} producing results similar to the ones presented in this section.
As the presented SPDE approach applies to any optimization algorithm based on sampling, and our focus is on the effect of uncertainties on the design, the results presented here will use the most popular optimization algorithm in topology optimization, the Method of Moving Asymptotes (MMA) \cite{Svanberg1987}.
Additionally, one may use more suitable risk measures instead of the expectation in~\eqref{eq:thermal-compliance-objective}, for example, the \textit{conditional value-at-risk} \cite{rockafellar2000optimization,kouri2016risk,kouri2022primal,Kodakkal2022}.
However, such risk measures often require specialized numerical optimization algorithms lying beyond the scope of this work.

\begin{figure}
 \centering
 \begin{minipage}[b]{0.245\textwidth}\centering
  \includegraphics[width=\textwidth]{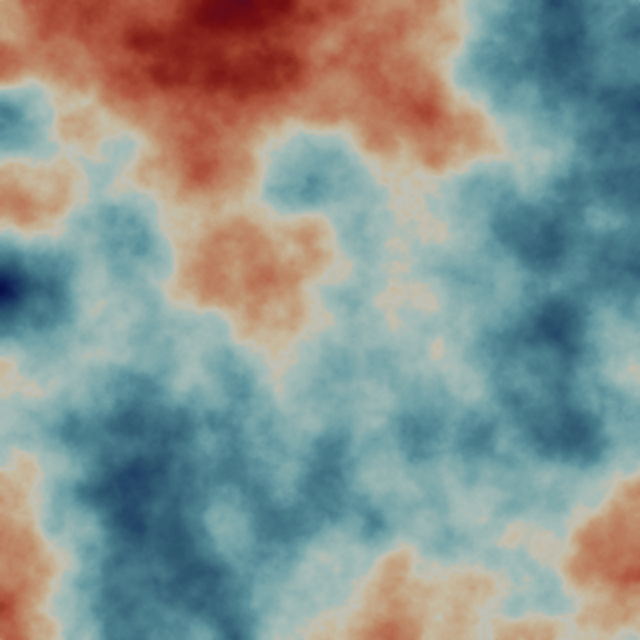}
 \end{minipage}
 \begin{minipage}[b]{0.245\textwidth}\centering
  \includegraphics[width=\textwidth]{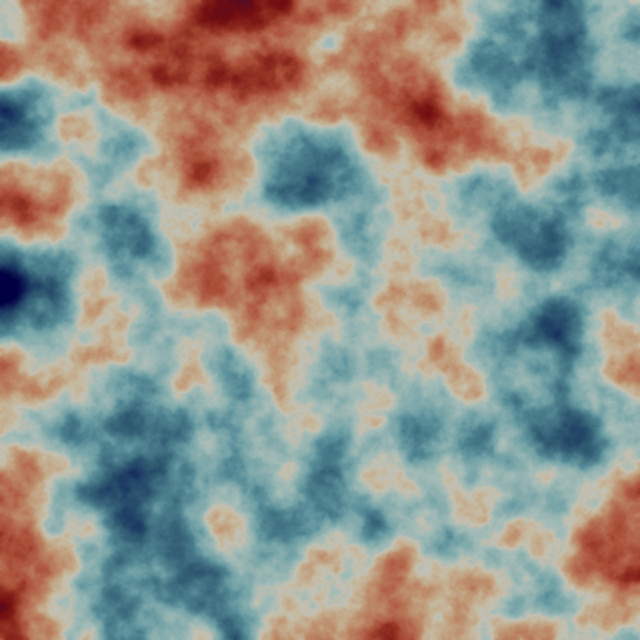}
 \end{minipage}
 \begin{minipage}[b]{0.245\textwidth}\centering
  \includegraphics[width=\textwidth]{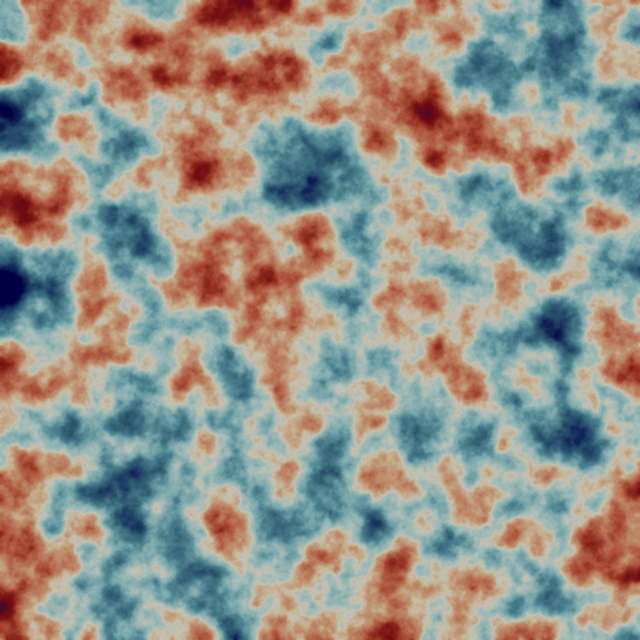}
 \end{minipage}
 \begin{minipage}[b]{0.245\textwidth}\centering
  \includegraphics[width=\textwidth]{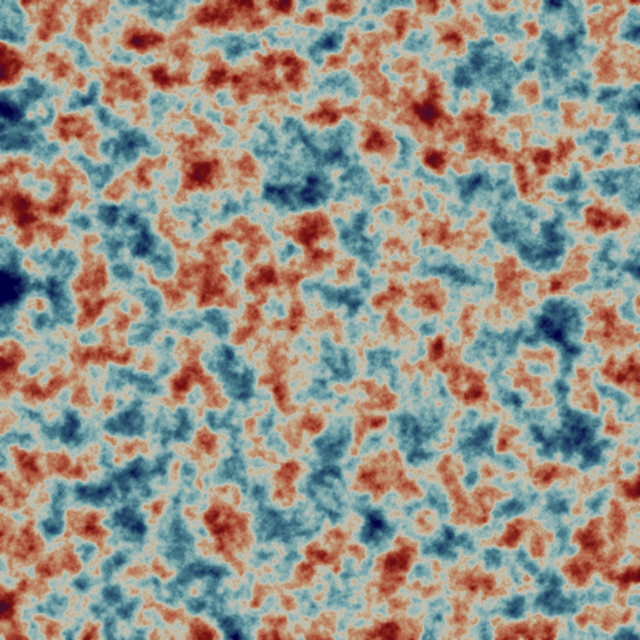}
 \end{minipage} \\
 \begin{minipage}[b]{0.245\textwidth}\centering
  \includegraphics[width=1.0\textwidth]{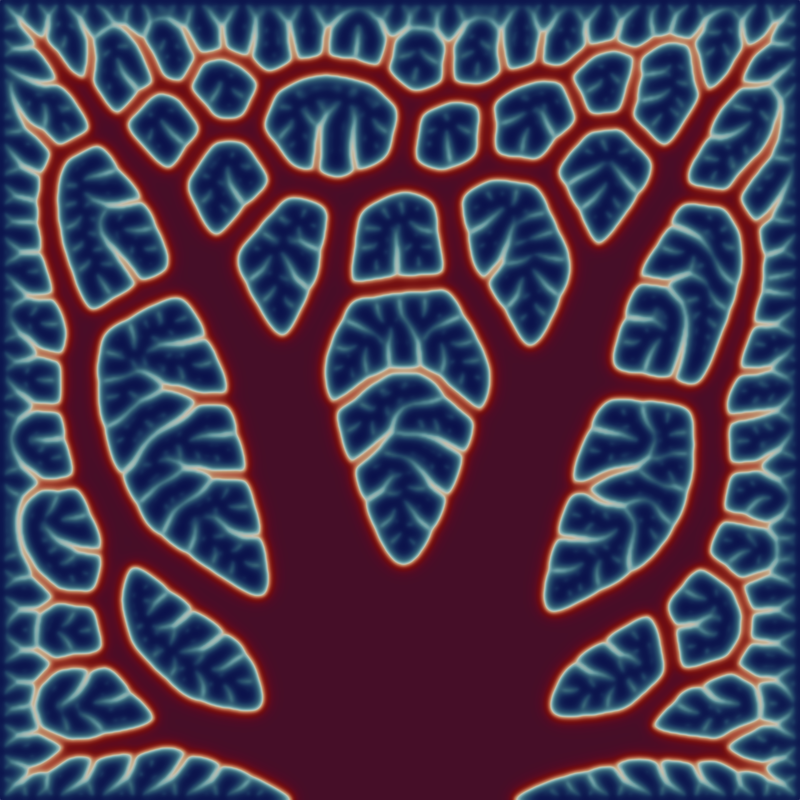}
  \caption*{(a) $l=0.2$, $j = 0.589$}
 \end{minipage}
 \begin{minipage}[b]{0.245\textwidth}\centering
  \includegraphics[width=1.0\textwidth]{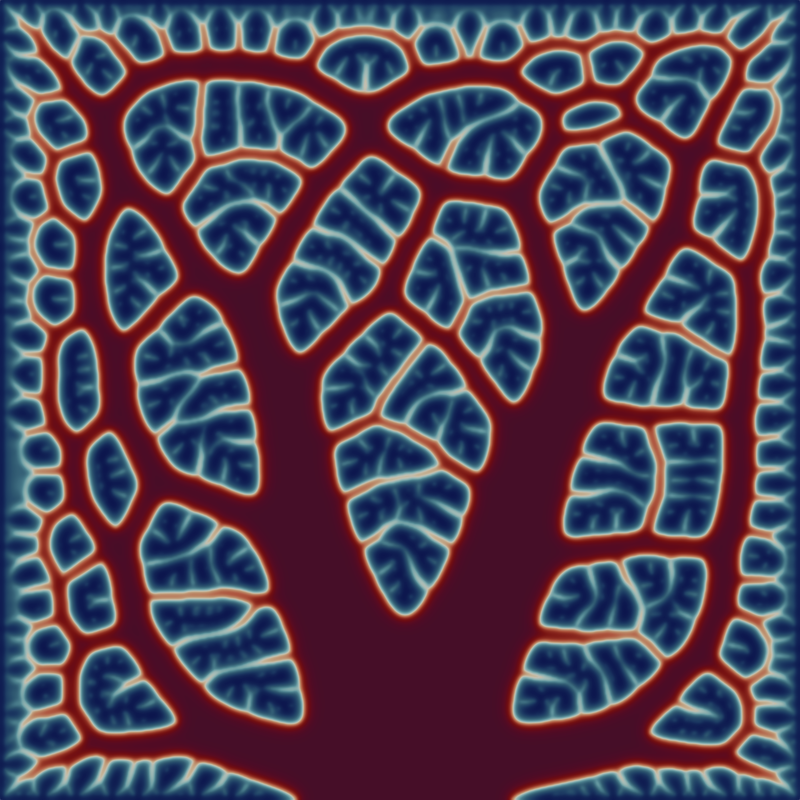}
  \caption*{(b) $l= 0.1$, $j = 0.215$}
 \end{minipage}
 \begin{minipage}[b]{0.245\textwidth}\centering
  \includegraphics[width=1.0\textwidth]{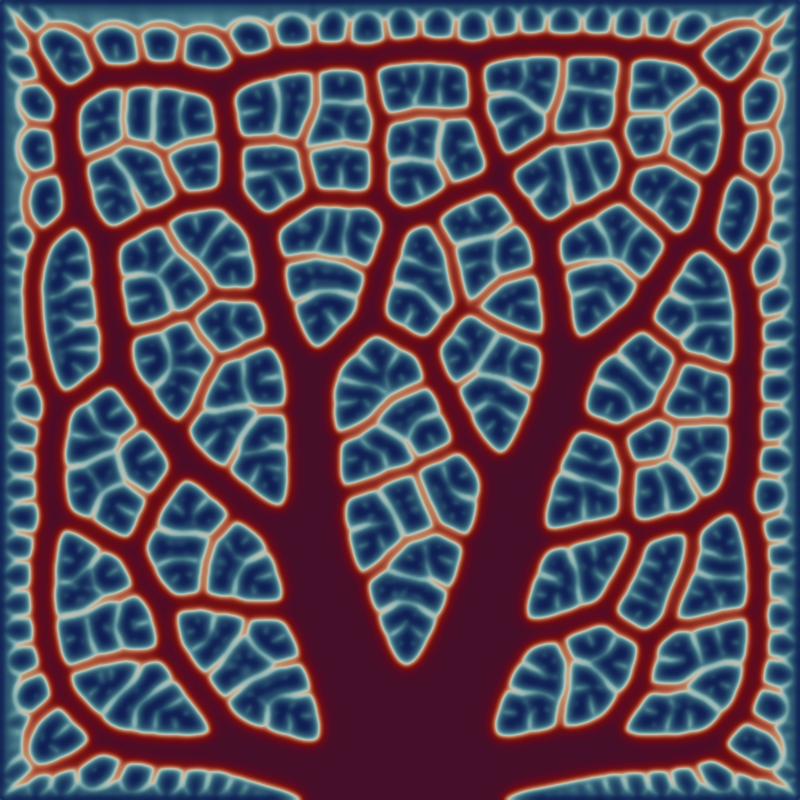}
  \caption*{(c) $l = 0.05$, $j = 0.074$}
 \end{minipage}
 \begin{minipage}[b]{0.245\textwidth}\centering
  \includegraphics[width=1.0\textwidth]{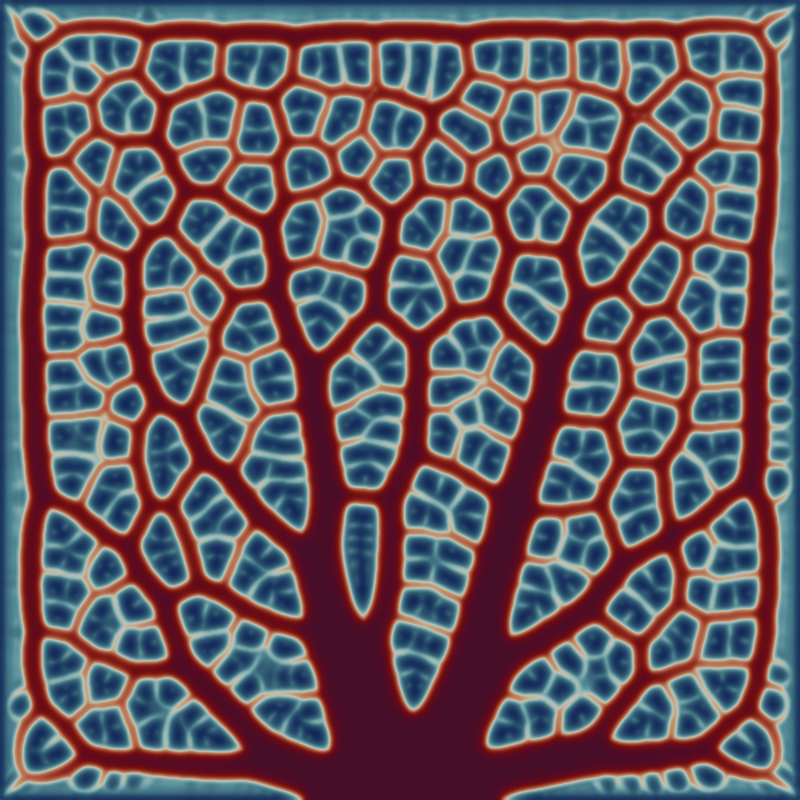}
  \caption*{(d) $l= 0.025$, $j = 0.024$}
 \end{minipage}
 \caption{
  Optimized density distributions $\tilde{\rho}$ (bottom) for thermal compliance problems with stochastic heat source of different correlation lengths (top, representative sample).
  The length parameter in the PDE filter is set to $r=0.02$, and the volume constraint is 50\% (i.e., $\gamma = 0.5$).
  The stochastic fields are isotropic, i.e., $\M{\Theta} = l^2 \, \M{I}$.
  The value of the objective function is denoted with $j$.
 }
 \label{fig:thermal_compliance_stochastic_isotropic}
\end{figure}

Comparing the results of the deterministic and stochastic problems in Figure~\ref{fig:thermal_compliance_deterministic} and~\ref{fig:thermal_compliance_stochastic_isotropic}, respectively, we observe significant differences.
The deterministic optimization problem finds a tree-like structure trying to extract the heat through shortest paths to the support without violating the volume constraint.
The stochastic formulation results in many closed loops of solid material decorated with more minor tree-like features.
The reason for this difference can be understood by realizing that the only mechanism to extract heat from a point in a domain with uniform source is to connect conductive material to the fixed temperature support.
On the other hand, when the source is a random field that provides both positive and negative values spatially varying throughout the domain, an area with negative heating can be locally neutralized by an area with positive heating by connecting them with conductive material.
Therefore, the role of the closed loops in the obtained topologies is to establish local heat equilibrium, and only the excess heat is transferred to the support by the underlying tree-like skeleton.
Decreasing the correlation length results in smaller loops reflecting the average size of the positively/negatively heated regions.

Intuitively, increasing the correlation length yields GRF samples that show fewer local fluctuations and become more locally uniform.
This behavior can be observed in Figure~\ref{fig:rf-matrix-annotated} along the correlation length axis.
Here, the areas with similar field values become larger as $l$ increases, and in the limit of $l \rightarrow \infty$, the SPDE solutions become constant in space.
Thus, with increasing lengths, the stochastic case behaves more and more similar to the deterministic case with its uniform load.
Our results support this reasoning; by lowering the correlation length from (a) to (d) in Figure~\ref{fig:thermal_compliance_stochastic_isotropic}, we observe that the resulting designs develop structures which increasingly deviate from the deterministic case.
While we observe two dominant main branches diagonally extending from the center to the top left and right corners in Panel~(a) as for the deterministic problem, Panel~(d) shows many small loops, and most branches have similar widths.
From~(a) to~(d), the main branches become less pronounced as the correlation length decreases, resulting in significantly different designs compared to the deterministic case.

Another striking difference between the deterministic and stochastic designs is the lack of symmetry of the latter.
The deterministic problem is symmetric about the $y$-axis, i.e., all operators and loads in the SPDE and physical problem are invariant under reflection.
Introducing GRFs as loads breaks this symmetry since $f(x,y) \neq f(-x,y)$  with probability equal to one for all sample realizations.
Consequently, using a finite number of samples inevitably leads to asymmetric designs.
While considering larger sample sizes $N$ yields increasingly symmetric designs (see~\ref{apx:sample_size}), this approach is computationally infeasible and we propose to enforce the natural symmetry via a modification to the algorithm.

\subsubsection{Enforcing symmetries}
\label{sec:enforcing_symmetry}

Although the individual GRFs break the system's symmetry, the expectation $\mathbb{E}\left(f(x,y) \right) = \mathbb{E}\left(f(-x,y) \right)$ preserves it.
In fact, the reflection symmetry extends to the full probability density function of $f$ at each $(x,y) \in D$.
This, together with the same reflection symmetry in the state equation~\eqref{eq_ex1_001}, boundary conditions, and other problem constraints ensures that at least a locally optimal symmetric design exists.
While this is not sufficient to conclude that a \emph{globally optimal} design must obey any symmetry, it shows that it is of substantial interest to investigate optimized designs exhibiting the same reflection symmetry found in the deterministic problem.
The following changes to the algorithm ensure the design's symmetry while simultaneously lowering the computational cost.
We describe these changes in detail because their complexity is greater than in the deterministic optimal design setting.

As before, we begin with a symmetric initial guess for the density field satisfying the volume constraint~\eqref{eq:volume_constraint}.
The gradients and the corresponding design updates preserve symmetry if we consider mirrored pairs of GRF realizations ($f(x,y)$ and $\bar{f}(x,y) = f(-x,y)$) instead of randomly chosen ones.
Conveniently, the SPDE allows for constructing such pairs efficiently, only considering half of the domain $D_{1/2} = (-0.5,0)\times(-0.5,0.5)$.
Moreover, the solution of the state equation~\eqref{eq_ex1_001} can also be computed on $D_{1/2}$ since it exhibits the same symmetry.

We consider a GRF $f$ generated via the SPDE method for a given, random white noise realization $\mathcal{W}$ and decompose it into its symmetric and antisymmetric components, i.e., $f = (f_s + f_a) / \sqrt{2}$ and $\bar{f} = (f_s - f_a) / \sqrt{2}$.
Linearity of the SPDE ensures that both $f_s$ and $f_a$ satisfy~\eqref{eq:SPDE_fullspace} with white noise realizations $\mathcal{W}_s$ and $\mathcal{W}_a$ that are proportional to the symmetric and antisymmetric components of $\mathcal{W}$.
Symmetry and anti-symmetry imply $\partial_x f_s(x=0,y) = 0$ and $f_a(x=0,y)=0$.
In other words, $f_s$ and $f_a$ obey homogeneous Neumann and Dirichlet boundary conditions, respectively, along the centerline of the domain $D$, defined $\Gamma_\mathrm{C} = \{(x,y) \in D  \mid x=0\}$.
Well-posedness of the SPDE~\eqref{eq:SPDE_fullspace} implies that we can solve the equation on $D_{1/2}$ with homogeneous Dirichlet boundary conditions on $\Gamma_\mathrm{C}$ and white noise $\mathcal{W}_a|_{D_{1/2}}$ to obtain $f_a|_{D_{1/2}}$.
Likewise, using homogeneous Dirichlet boundary conditions and white noise $\mathcal{W}_s|_{D_{1/2}}$ gives $f_s|_{D_{1/2}}$.
Together, $f_s|_{D_{1/2}}$ and $f_a|_{D_{1/2}}$ allow us to construct $f|_{D_{1/2}}$ and $\bar{f}|_{D_{1/2}}$, which can be reflected to construct $f$
and $\bar{f}$ at any point in $D$, if necessary.
In practice, we do not derive $\mathcal{W}_s$ and $\mathcal{W}_a$ from $\mathcal{W}$, instead we simply generate two independent Gaussian white noise samples on $D_{1/2}$ using the procedure explained in Section~\ref{sub:sampling_the_white_noise_vector}.

The GRFs $f$ serve as loads in the state equation of this section~\eqref{eq_ex1_001}.
Given that the physical problem is linear and also exhibits a reflection symmetry, we may restrict the entire optimization workflow to $D_{1/2}$ by considering the symmetric and antisymmetric solutions of~\eqref{eq_ex1_001}, $u_s$ and $u_a$, respectively.
The boundary conditions are analogous to the restricted SPDEs discussed in the previous paragraph, meanwhile, $f_s$ and $f_a$ become the loads for the symmetric and antisymmetric restricted state equations, respectively.
Finally, we may recover the final temperature fields $u$ and $\bar{u}$ corresponding to $f$ and $\bar{f}$ from the equations
\begin{equation}
 u=\frac{1}{\sqrt{2}}\left(u_s+u_a\right),\quad \bar{u}=\frac{1}{\sqrt{2}}\left(u_s-u_a\right) \, .
\end{equation}
Although the above modification requires twice as many PDE solves, it reduces the overall computational cost because each solve requires only half of the original domain.

\begin{figure}
 \centering
 \begin{minipage}[b]{0.245\textwidth}\centering
  \includegraphics[width=1.0\textwidth]{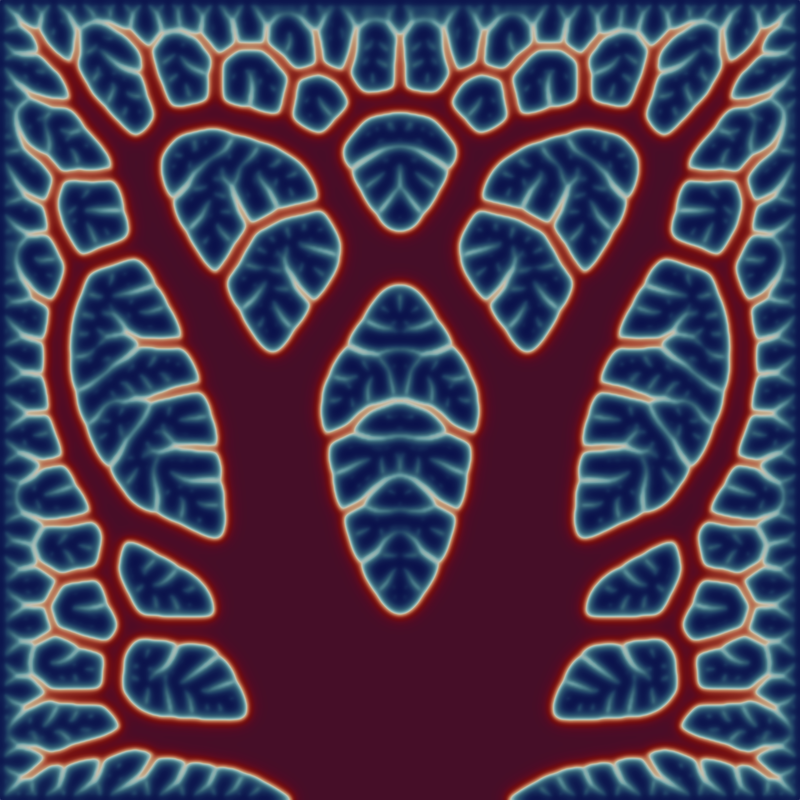}
  \caption*{(a) $l = 0.2$, $j = 0.589$}
 \end{minipage}
 \begin{minipage}[b]{0.245\textwidth}\centering
  \includegraphics[width=1.0\textwidth]{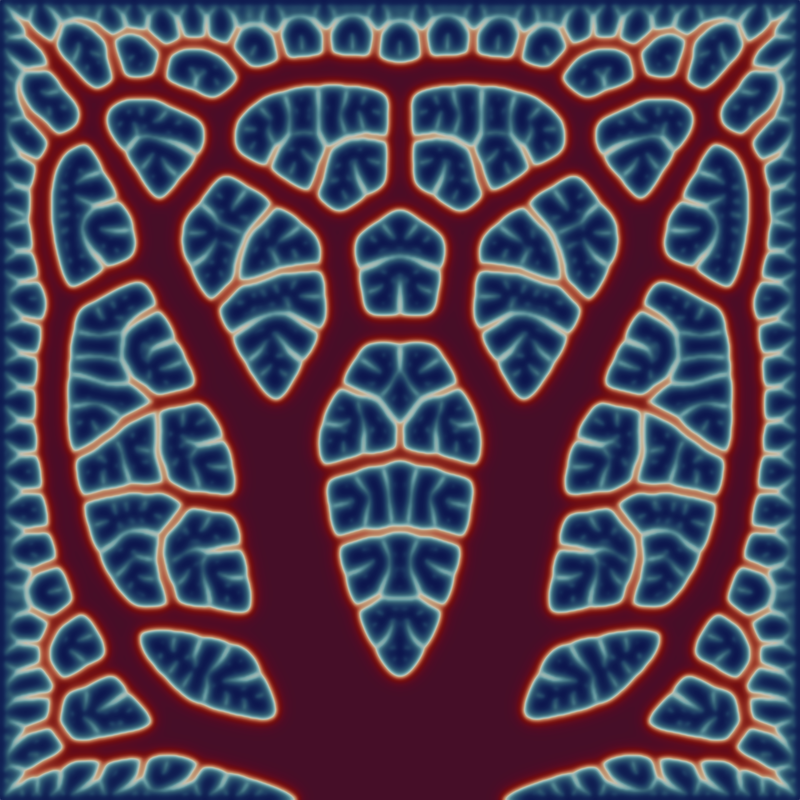}
  \caption*{(b) $l = 0.1$, $j = 0.214$}
 \end{minipage}
 \begin{minipage}[b]{0.245\textwidth}\centering
  \includegraphics[width=1.0\textwidth]{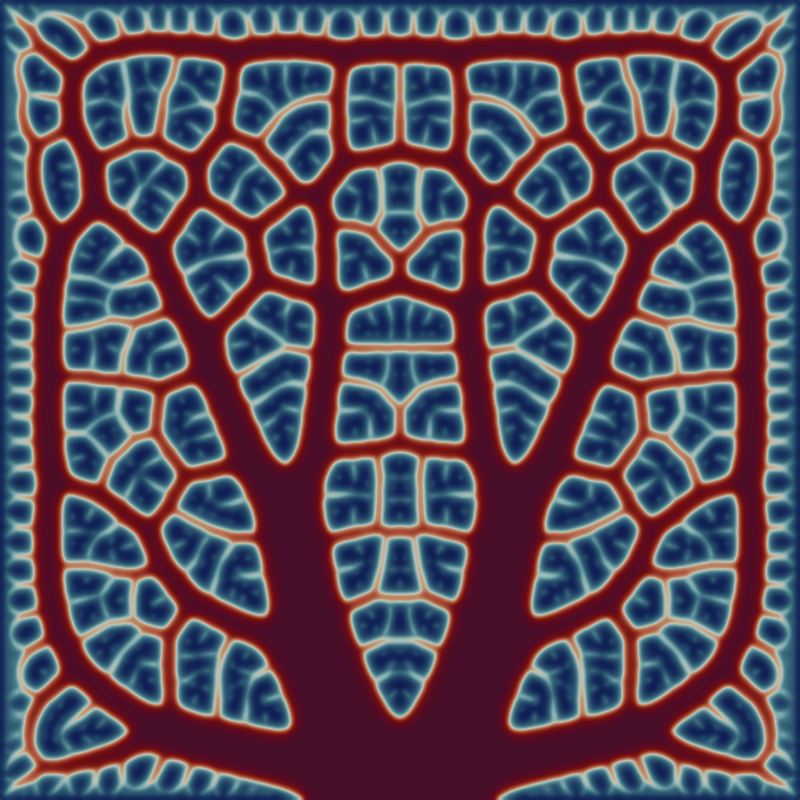}
  \caption*{(c) $l = 0.05$, $j = 0.074$}
 \end{minipage}
 \begin{minipage}[b]{0.245\textwidth}\centering
  \includegraphics[width=1.0\textwidth]{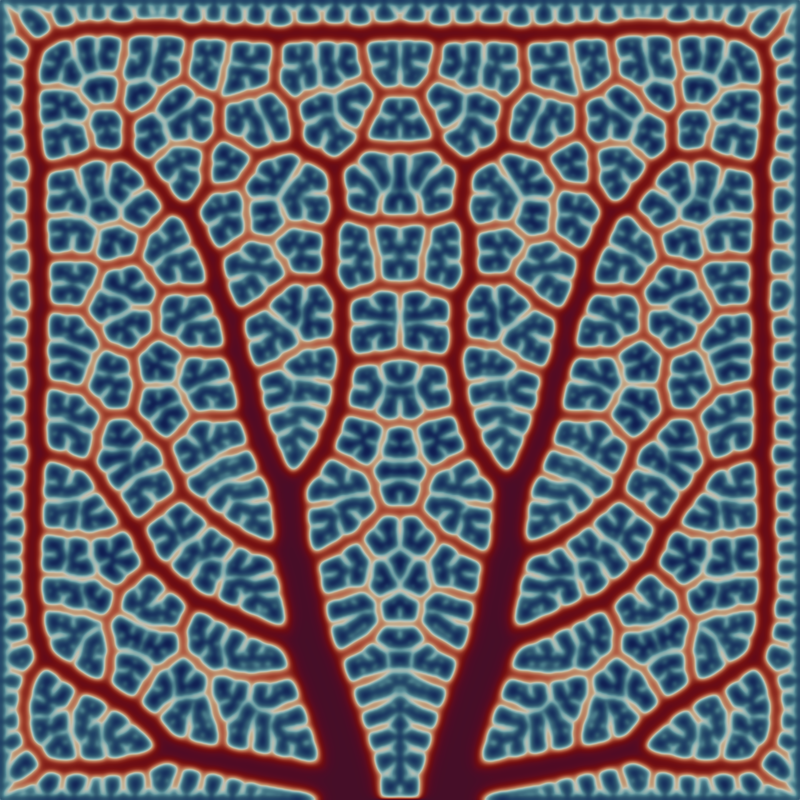}
  \caption*{(d) $l = 0.025$, $j = 0.024$}
 \end{minipage}
 \caption{
  Optimized symmetric density distributions $\tilde{\rho}$ for thermal compliance problems with stochastic heat source, different correlation lengths, and volume constraint 50\%  (i.e., $\gamma = 0.5$).
  The length parameter in the PDE filter is set to $r=0.02$.
  The stochastic fields are isotropic, i.e., $\M{\Theta} = l^2\, \M{I}$.
  Symmetry is enforced across the vertical axis $x=0$ of the design domain.
  The optimized value of the objective function is denoted by $j$.
 }
 \label{fig:thermal_compliance_stochastic_symmetric_isotropic}
\end{figure}

Figure~\ref{fig:thermal_compliance_stochastic_symmetric_isotropic} shows the stochastic heat sink designs with enforced symmetry.
The proposed modification produces symmetric designs with topological features similar to Figure~\ref{fig:thermal_compliance_stochastic_isotropic} for a lower computational cost.
Comparing Figures~\ref{fig:thermal_compliance_stochastic_isotropic} and~\ref{fig:thermal_compliance_stochastic_symmetric_isotropic} shows that similar correlation lengths yield similar structures.
For instance, the main branches in (a), (b), and (c) show strong similarities.
Moreover, the characteristic loop sizes match.
We observe that the point where the main branches split above $x=0$ moves closer to the bottom boundary with decreasing correlation length.
This point seems to be consistently lower for the symmetrized designs.
The most significant differences between symmetrized and non-symmetrized designs appear for short correlation lengths, i.e., in panels~(d).
Here, the main branches in the symmetrized design are completely separated, while the non-symmetrized branches are still connected.
Overall, the latter branches appear thicker than the symmetrized ones.

The optimization of the symmetrized examples was carried out entirely on $D_{1/2}$.
The images in Figure~\ref{fig:thermal_compliance_stochastic_symmetric_isotropic} are post-processed such that the second half is added by reflection.
While every optimization on $D_{1/2}$ with subsequent reflection will produce symmetric results, we emphasize that the presented algorithm is equivalent to running the optimization on $D$ with mirrored sample pairs.
However, it bypasses the need for a symmetric mesh and reduces computational costs.

\subsubsection{Effects of anisotropy}
\label{sec:effects_of_anisotropy}

\begin{figure}
 \centering
 \begin{minipage}[b]{0.245\textwidth}\centering
  \includegraphics[width=\textwidth]{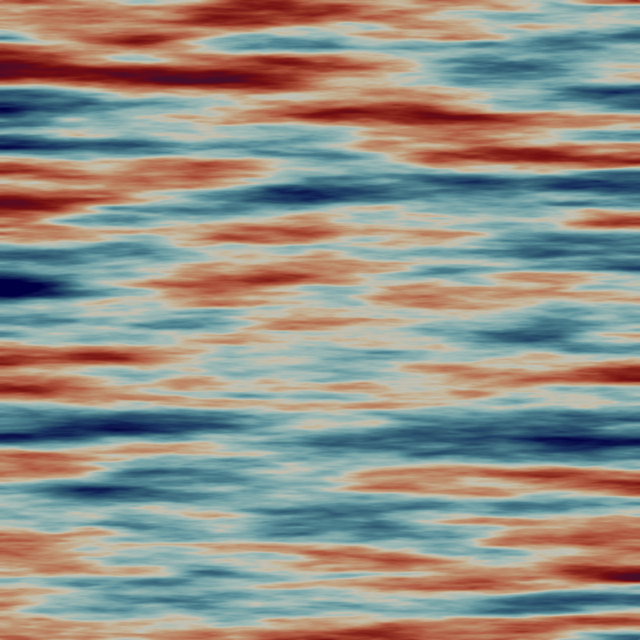}
 \end{minipage}
 \begin{minipage}[b]{0.245\textwidth}\centering
  \includegraphics[width=\textwidth]{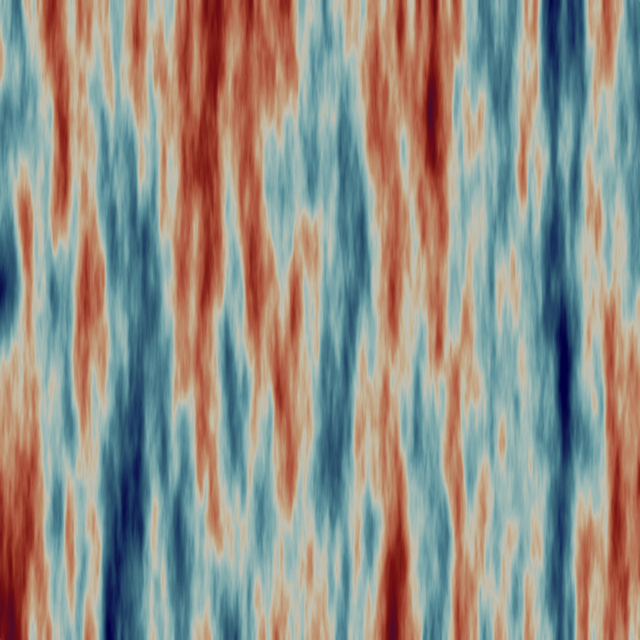}
 \end{minipage}
 \begin{minipage}[b]{0.245\textwidth}\centering
  \includegraphics[width=\textwidth]{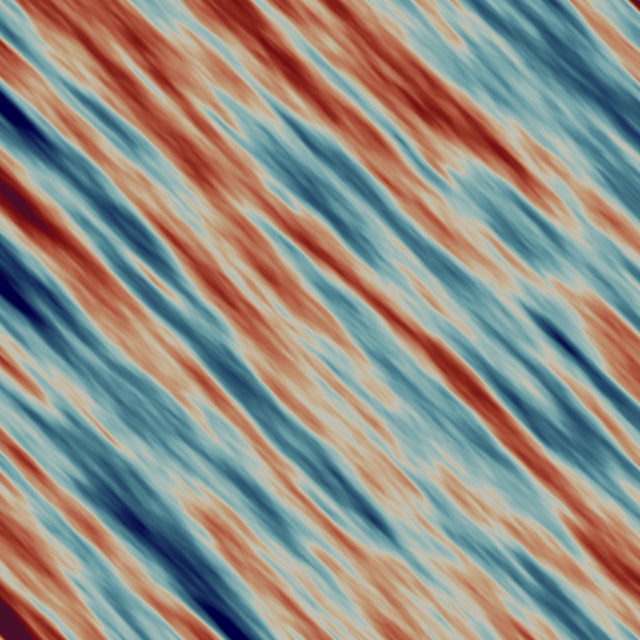}
 \end{minipage}
 \begin{minipage}[b]{0.245\textwidth}\centering
  \includegraphics[width=\textwidth]{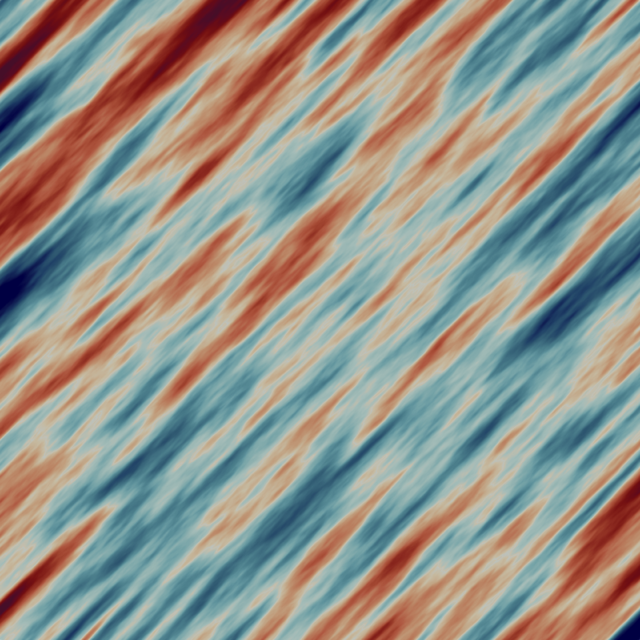}
 \end{minipage} \\
 \begin{minipage}[b]{0.245\textwidth}\centering
  \includegraphics[width=1.0\textwidth]{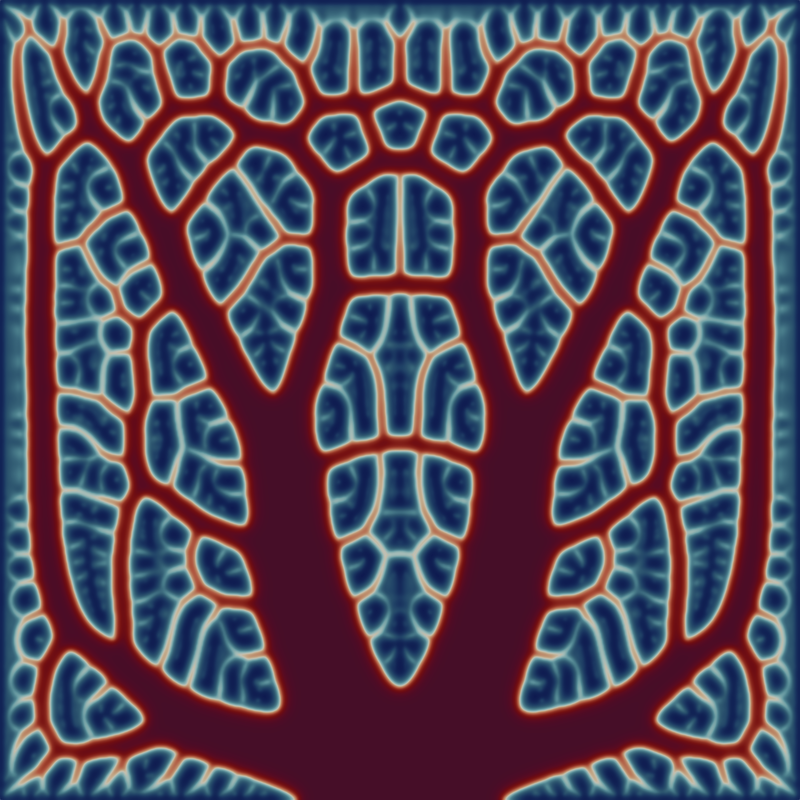}
  \caption*{(a) $l_x = 0.2, l_y = 0.025$, $\vartheta = 0$,\\ $j = 0.109$}
 \end{minipage}
 \begin{minipage}[b]{0.245\textwidth}\centering
  \includegraphics[width=1.0\textwidth]{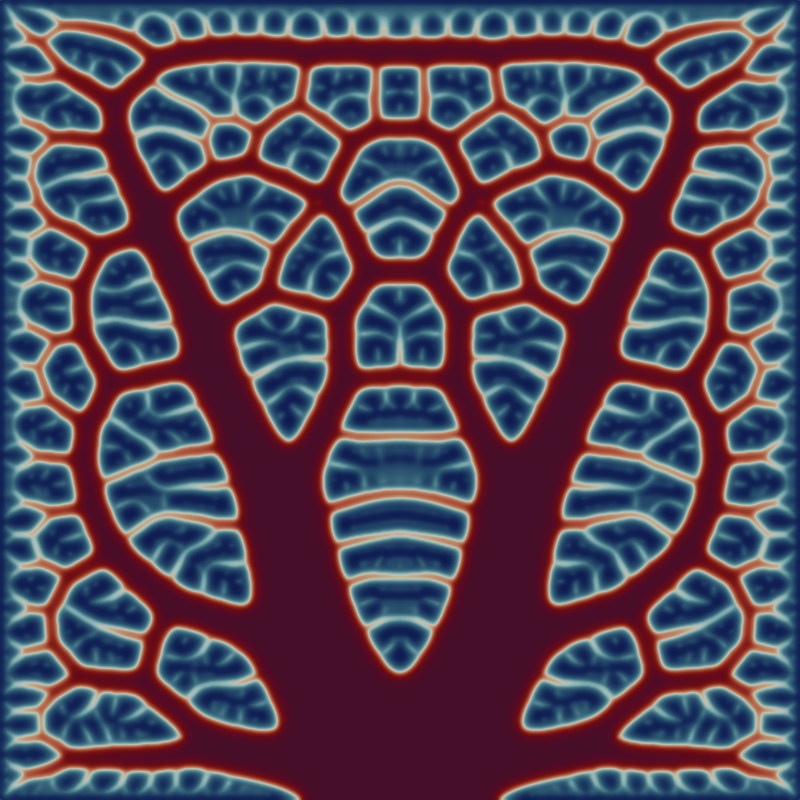}
  \caption*{(b) $l_x = 0.025, l_y = 0.2$, $\vartheta = 0$,\\ $j = 0.110$}
 \end{minipage}
 \begin{minipage}[b]{0.245\textwidth}\centering
  \includegraphics[width=1.0\textwidth]{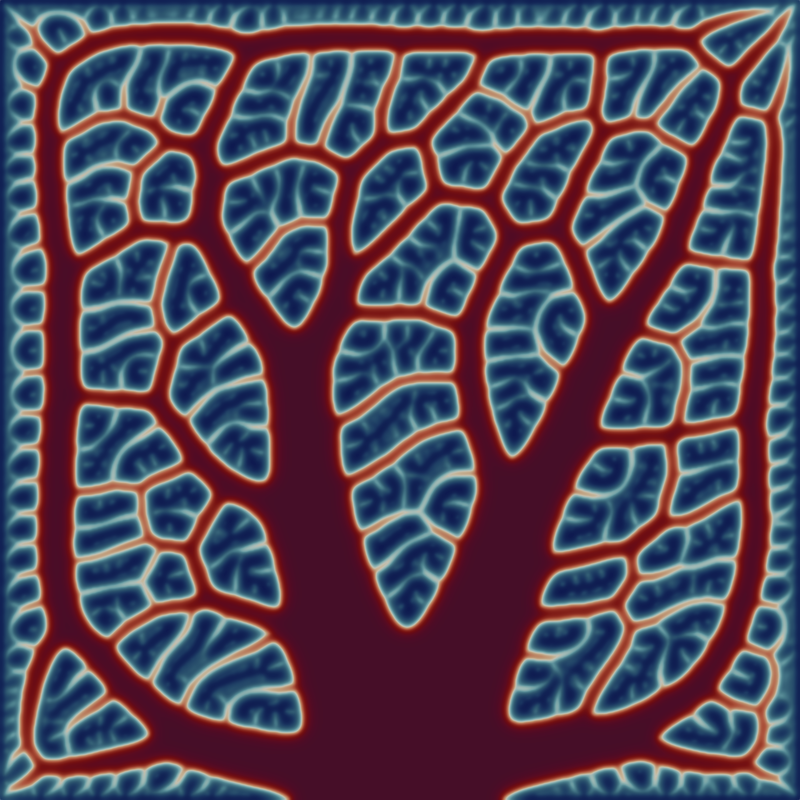}
  \caption*{(c) $l_x = 0.2, l_y = 0.025, \vartheta = \frac{\pi}{4}$,\\ $j = 0.114$}
 \end{minipage}
 \begin{minipage}[b]{0.245\textwidth}\centering
  \includegraphics[width=1.0\textwidth]{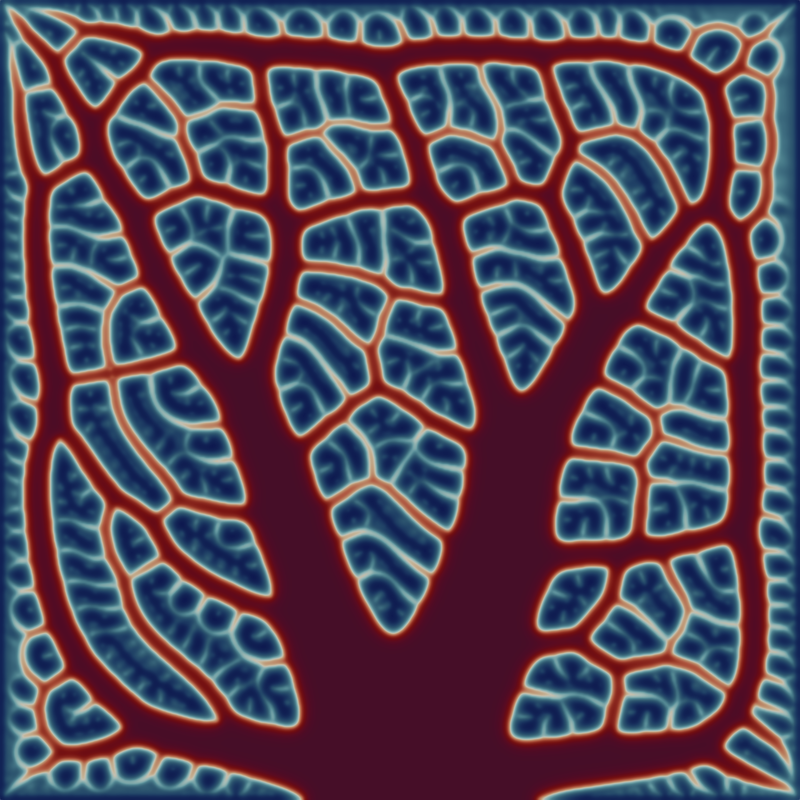}
  \caption*{(d) $l_x = 0.2, l_y = 0.025, \vartheta = \frac{-\pi}{4}$,\\ $j = 0.114$}
 \end{minipage}
 \caption{
  Optimized density distributions $\tilde{\rho}$ (bottom) for thermal compliance problems with stochastic heat source of different correlation lengths (top, representative sample).
  The length parameter in the PDE filter is set to $r=0.02$, the volume constraint is 50\% (i.e., $\gamma = 0.5$)and the stochastic fields are anisotropic..
  For (a,c), symmetry is enforced across a vertical axis $x=0$.
  The value of the objective function is denoted with $j$.
 }
 \label{fig:thermal_compliance_stochastic_anisotropic}
\end{figure}

Introducing different correlation lengths, $l_i \neq l_j$ for $i\neq j$, when constructing the matrix $\M{\Theta}$ in~\eqref{eq:SPDE_fullspace} allows the SPDE method to generate anisotropic random fields.
The effect of such anisotropy on topology-optimized heat sinks is demonstrated in Figure~\ref{fig:thermal_compliance_stochastic_anisotropic}.
Panel~(a) shows the designs resulting from using long correlation lengths in the horizontal $x$-direction and short correlation lengths in the vertical $y$-direction, and vice-versa for panel~(b).
In both cases, the anisotropy does not break the mirror symmetry of the problem, and we employ the symmetrization technique proposed in Section~\ref{sec:enforcing_symmetry} above.
The two designs show significant differences that reflect the implied directionality of the different random heat sources.
Additionally, we may rotate $\M{\Theta}$ with a standard 2-dimensional rotation matrix; cf.\ $\M{R}$ in~\eqref{eq:SPDE_fullspace}.
Introducing this rotation breaks the symmetry, and we solve the problem without symmetrization.
Panels~(c) and~(d) depict the designs given two different rotation angles $\vartheta = \pm \pi/4$.
Overall, we find that anisotropy significantly biases the topologies, demonstrating the modeling freedom of the SPDE method.
While not explicitly considered here, we remark that the method generalizes to spatially-varying coefficients $\M{\Theta} = \M{\Theta}(\V{x})$, providing great flexibility in the construction of inhomogeneous fields.

\subsubsection{Optimization on embedded surfaces}
\label{sec:thermal_compliance_sphere}

\begin{figure}[t]
 \centering
 \begin{minipage}[b]{0.07\textwidth}\centering
  \scriptsize{(a) $l = 1.2$}
  \vspace{1.9cm}
 \end{minipage}
 \begin{minipage}[b]{0.3\textwidth}\centering
  \includegraphics[width=0.8\textwidth]{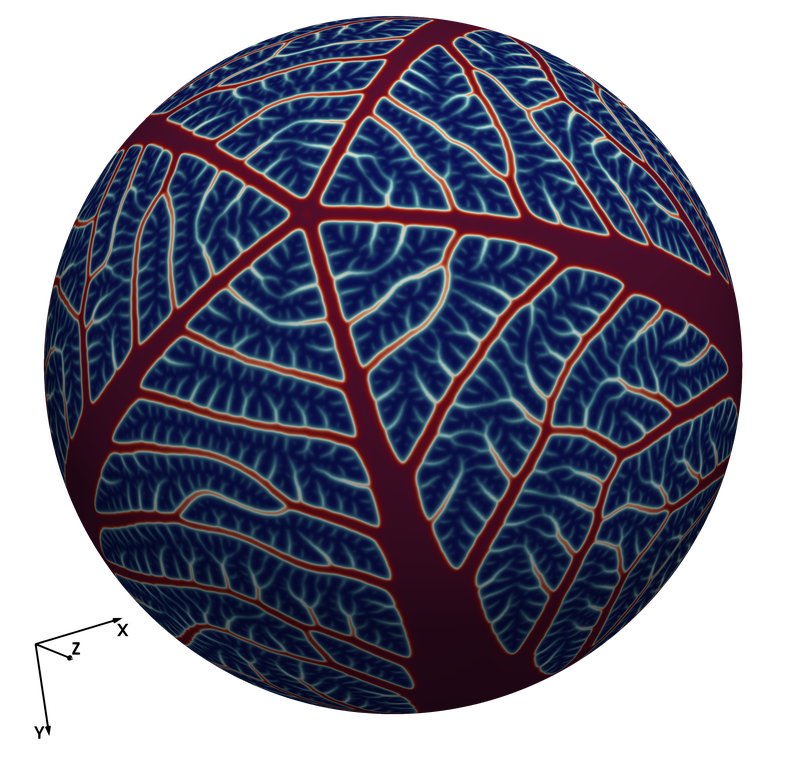}
 \end{minipage}
 \begin{minipage}[b]{0.3\textwidth}\centering
  \includegraphics[width=0.8\textwidth]{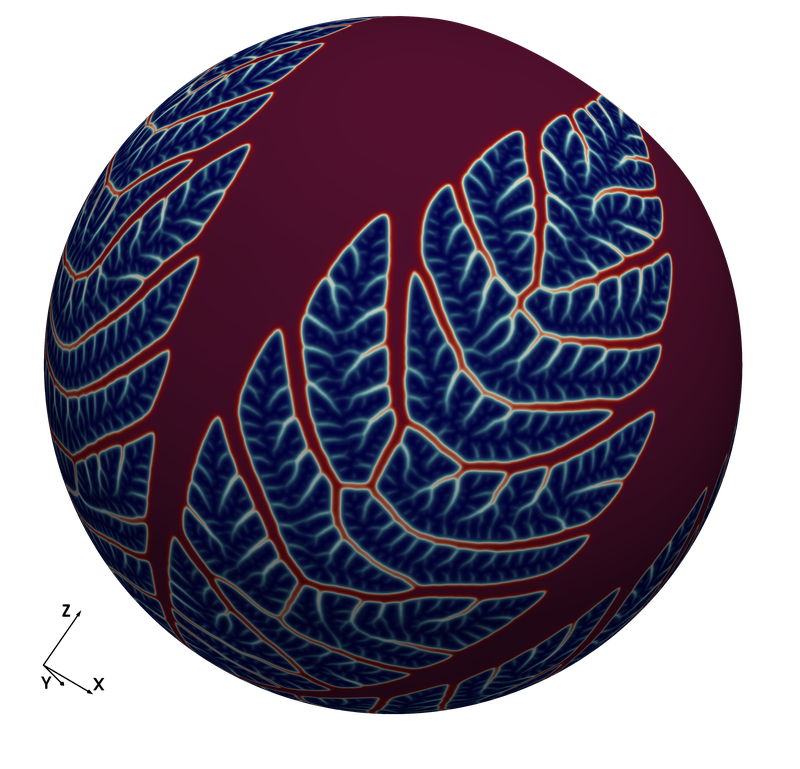}
 \end{minipage}
 \begin{minipage}[b]{0.3\textwidth}\centering
  \includegraphics[width=0.8\textwidth]{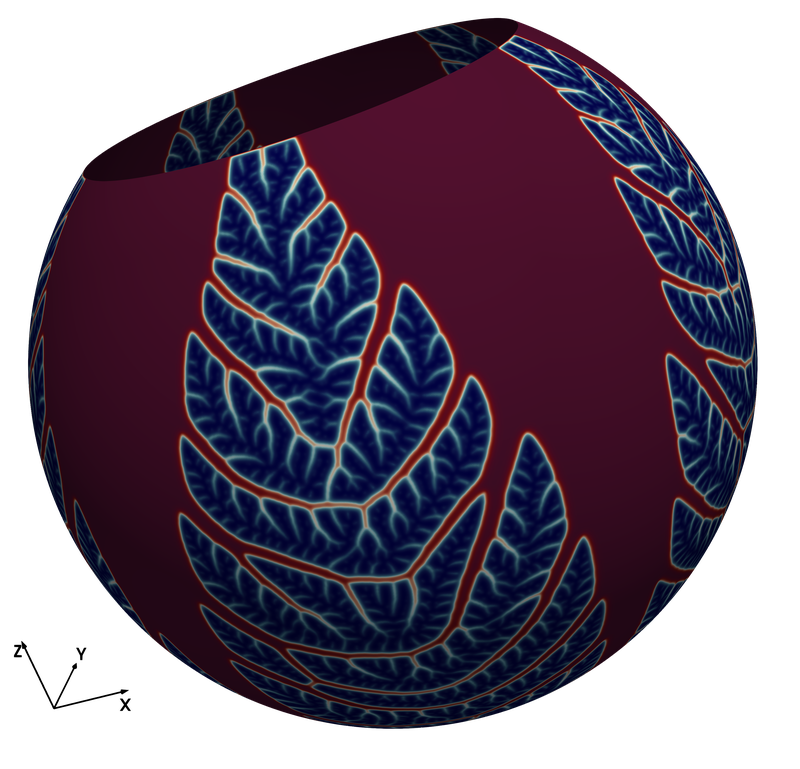}
 \end{minipage}
 \begin{minipage}[b]{0.07\textwidth}\centering
  \scriptsize{(b) $l = 0.6$}
  \vspace{1.9cm}
 \end{minipage}
 \begin{minipage}[b]{0.3\textwidth}\centering
  \includegraphics[width=0.8\textwidth]{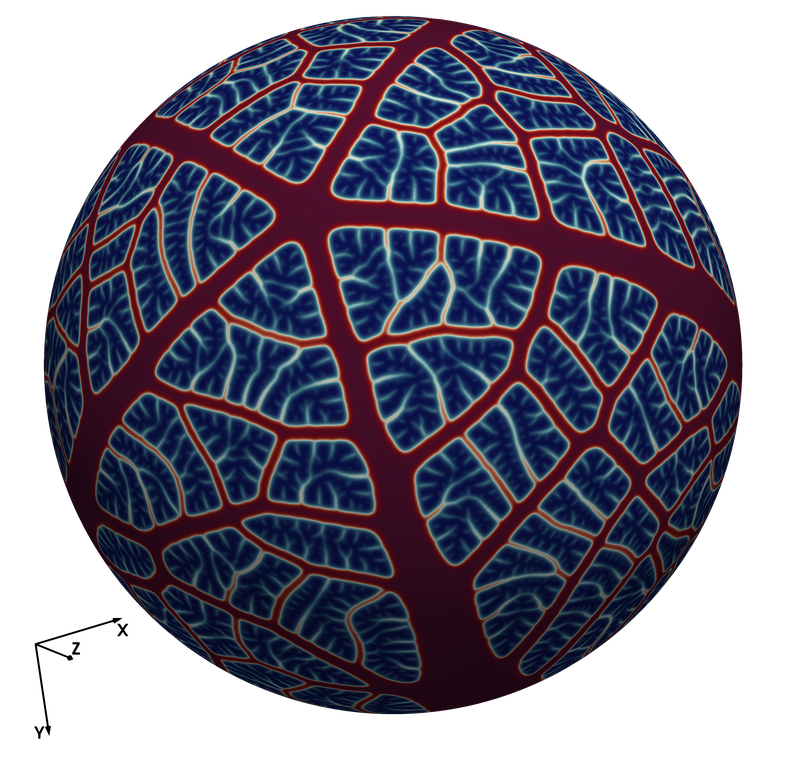}
 \end{minipage}
 \begin{minipage}[b]{0.3\textwidth}\centering
  \includegraphics[width=0.8\textwidth]{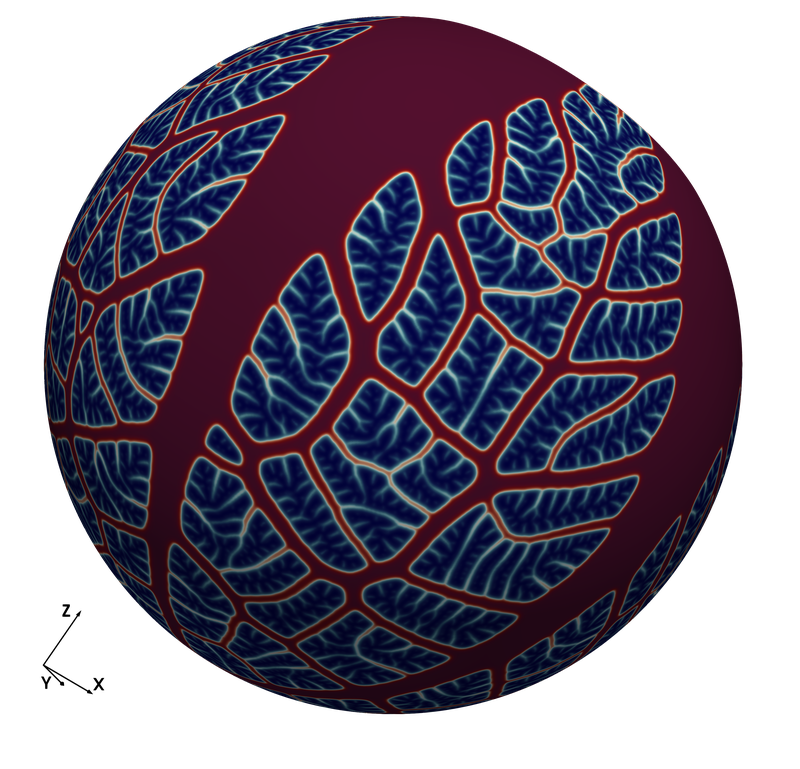}
 \end{minipage}
 \begin{minipage}[b]{0.3\textwidth}\centering
  \includegraphics[width=0.8\textwidth]{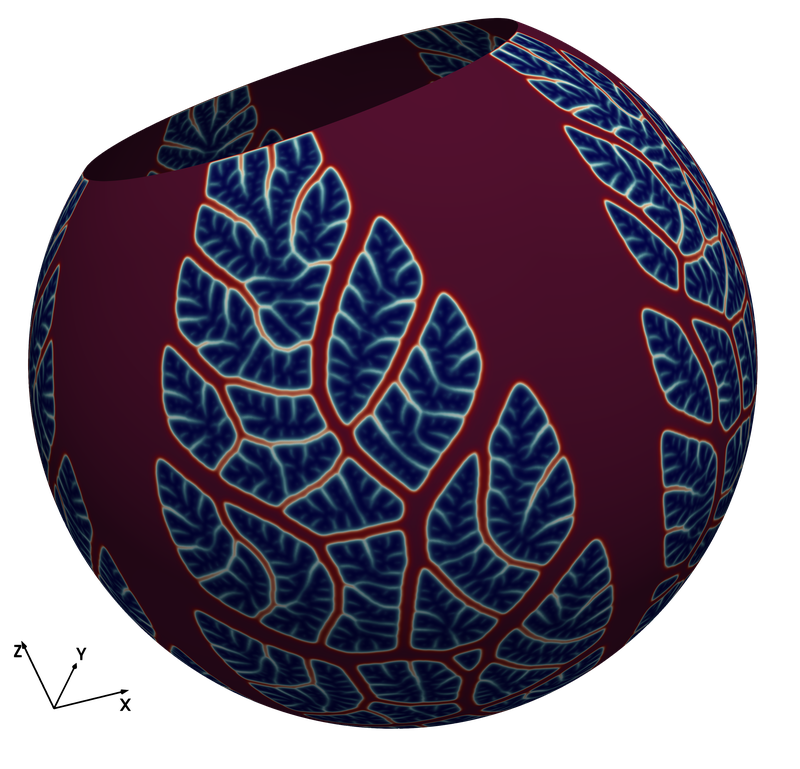}
 \end{minipage}
 \begin{minipage}[b]{0.07\textwidth}\centering
  \scriptsize{(c) $l = 0.3$}
  \vspace{2.6cm}
 \end{minipage}
 \begin{minipage}[b]{0.3\textwidth}\centering
  \includegraphics[width=0.8\textwidth]{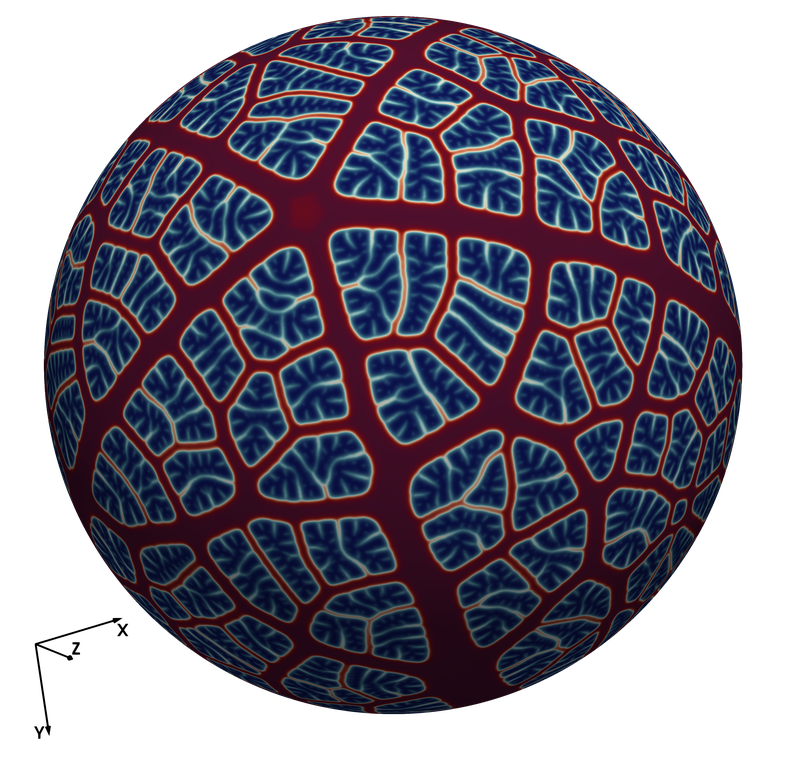}
  \caption*{View 1}
 \end{minipage}
 \begin{minipage}[b]{0.3\textwidth}\centering
  \includegraphics[width=0.8\textwidth]{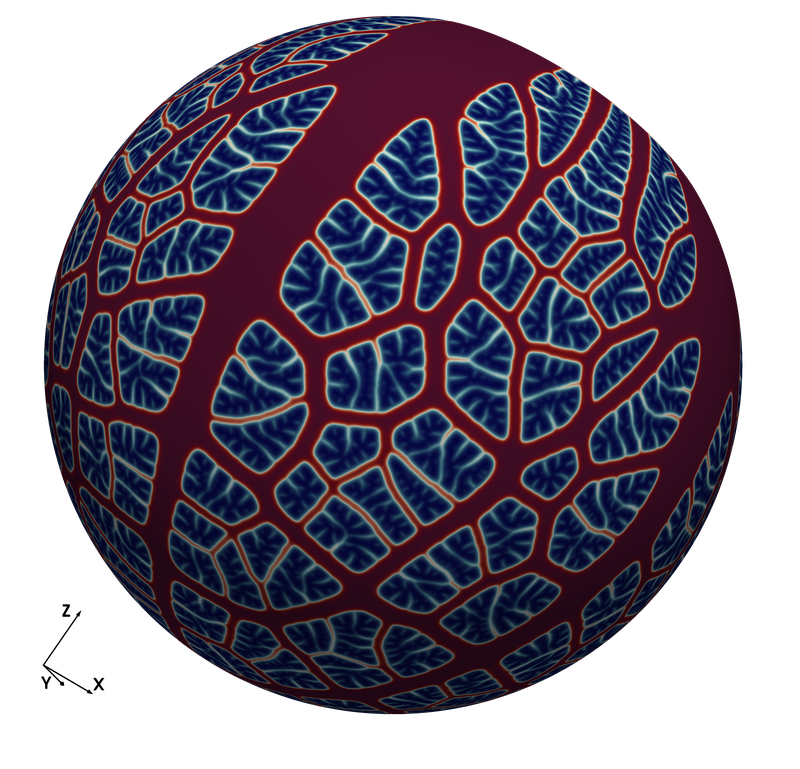}
  \caption*{View 2}
 \end{minipage}
 \begin{minipage}[b]{0.3\textwidth}\centering
  \includegraphics[width=0.8\textwidth]{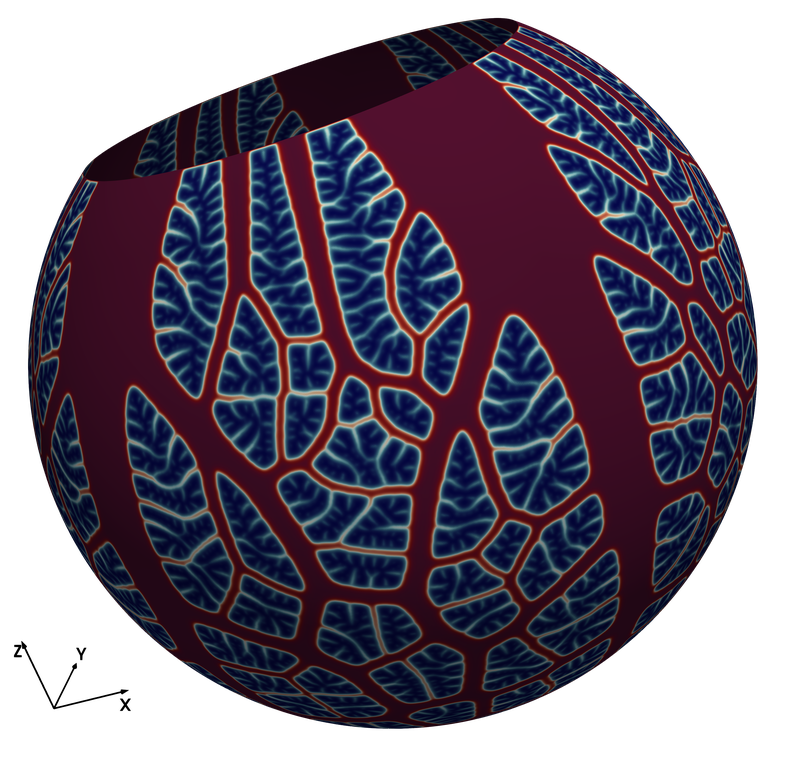}
  \caption*{View 3}
 \end{minipage}
 \caption{
  Optimized density distributions $\tilde{\rho}$ for thermal compliance problems with stochastic heat source, different correlation lengths, and volume constraint 50\% on a cut sphere.
  The length parameter in the PDE filter is set to $r=0.02$.
  The stochastic fields are isotropic.
  The values of the objective function are $j_{(a)}=39.74$, $j_{(b)}=17.75$, and $j_{(c)}=7.61$.
 }
 \label{fig:thermal_compliance_sphere}
\end{figure}

We continue investigating stochastic thermal compliance optimization and consider a surface embedded in 3D.
In this case, $D$ is now a manifold equal to the surface of the unit-radius sphere intersected with a half-space passing through the point $(0,0,4/5)$ and parallel to the $x$-$y$-plane; see Figure~\ref{fig:thermal_compliance_sphere}.
More specifically, $D$ is defined to be
\begin{equation}\label{eq:domain-cut-sphere}
 D = \{ (x,y,z) \in \mathbb{R}^3 \, | \, x^2 + y^2 + z^2 = 1 \text{ and } z < 4/5 \} \, .
\end{equation}
The problem formulation is identical to the one posed on the unit square; however, the boundary $\partial D$ is now the circle $\{ (x,y,z) \in \mathbb{R}^3 \, | \, x^2 + y^2 + z^2 = 1 \text{ and } z = 4/5 \}$.
We impose a homogeneous Dirichlet boundary condition along the entire boundary $\Gamma_\mathrm{D} = \partial D$.
Moreover, all Laplace-type operators in the problem formulation~\eqref{eq_ex1_001} are now interpreted as Laplace--Beltrami operators.
For a detailed discussion of the covariance of GRFs on the sphere, including the Mat\'ern-type covariance, we refer to~\cite{Lang2015, Jansson2022,lindgren2023diffusionbased}.
We initialize the density to $\rho_0 = \left(\sin (5 \varphi) + 1 \right) / 2$ with $\varphi \in (0,2\pi)$ denoting the azimuthal angle.
This prescription satisfies the volume constraint~\eqref{eq:volume_constraint} with $\gamma = 0.5$ and biases the topology towards a periodic pattern.
Besides these minor differences, the optimization proceeds akin to the previous cases.
Figure~\ref{fig:thermal_compliance_sphere} shows the density fields $\tilde{\rho}$ resulting from random heating modeled by isotropic Mat\'ern-type GRFs of different correlation lengths.
Similar to optimization on the unit square, decreasing the correlation length results in fast oscillatory fields, causing the optimization to form smaller closed loops of solid material to equilibriate the heat distribution locally.
The tree-like skeleton transfers the excess heat to the support.
Such capability extends to arbitrary surfaces embedded in $\mathbb{R}^3$, enabling heat sink designs on complex surfaces such as battery packs, tablets, and laptops~\cite{MartinezMaradiaga2019}.

\subsubsection{Effects of random material damage}
\label{sec:material_damage}

Apart from operational uncertainties like external excitations or time variations in boundary conditions, material properties are another common source of randomness.
Examples of such formulations can be found in~\cite{Lazarov2012,Lazarov2012a,Guilleminot2019}.
The experience from these studies shows that differences between design topologies are barely noticeable for slowly varying uncertain material properties, modeled by random fields with long correlation lengths compared to the characteristic size of the design domain.
On the other hand,~\cite{Jansen2013c} presents a very different outcome by modeling the localized damage caused by a small random projectile in the design domain.
This form of uncertainty leads to significant differences in the optimized topologies.
Indeed, redundant members are clearly observable in~\cite{Jansen2013c}, providing numerous possibilities to transfer the load to the supports in case neighboring members fail.

\begin{figure}
 \centering
 \begin{minipage}[b]{0.18\textwidth}\centering
  \includegraphics[width=1.0\textwidth]{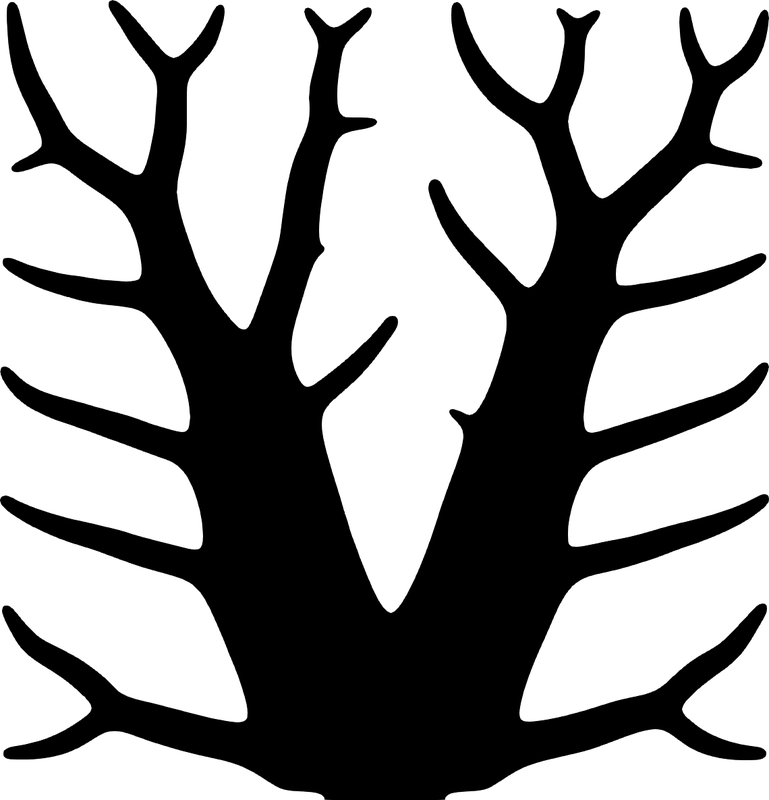}
 \end{minipage}
 \begin{minipage}[b]{0.18\textwidth}\centering
  \includegraphics[width=1.0\textwidth]{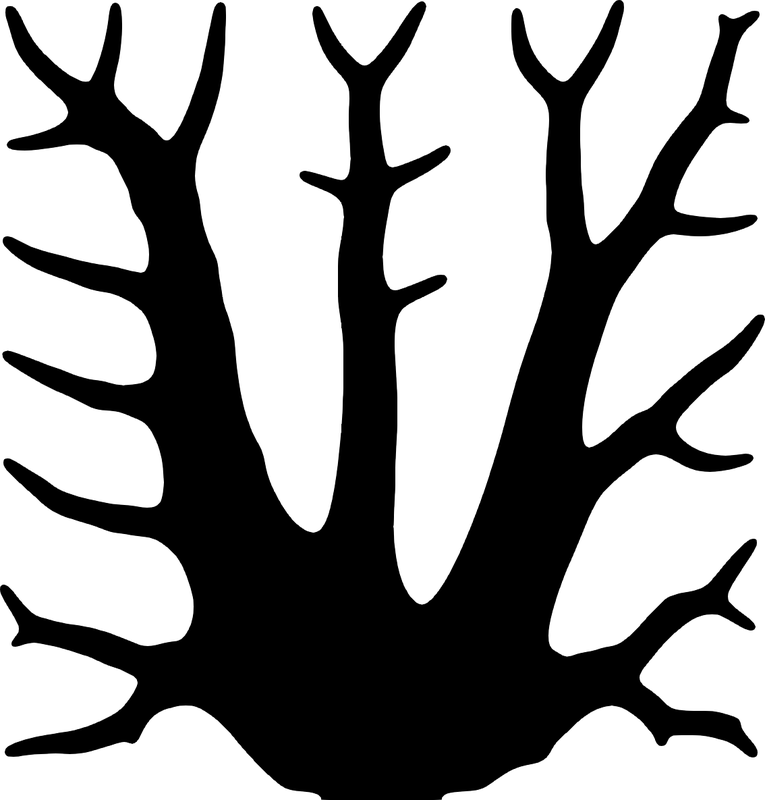}
 \end{minipage}
 \begin{minipage}[b]{0.18\textwidth}\centering
  \includegraphics[width=1.0\textwidth]{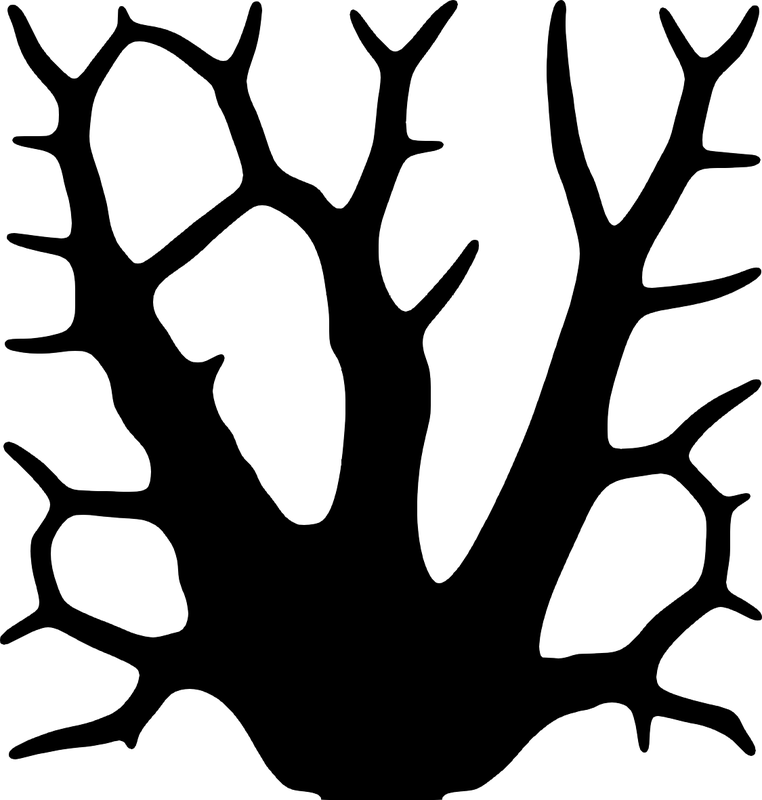}
 \end{minipage}\\
 \begin{minipage}[b]{0.18\textwidth}\centering
  \includegraphics[width=1.0\textwidth]{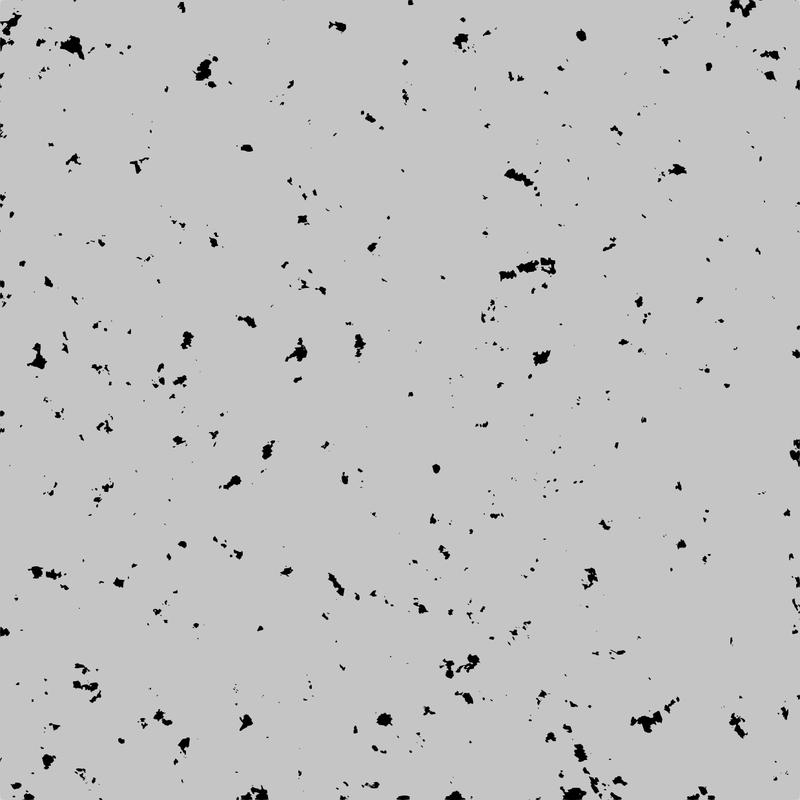}
  \caption*{(a) $l_x = 0.01, l_y = 0.01$,\\ $j = 2.23$}
 \end{minipage}
 \begin{minipage}[b]{0.18\textwidth}\centering
  \includegraphics[width=1.0\textwidth]{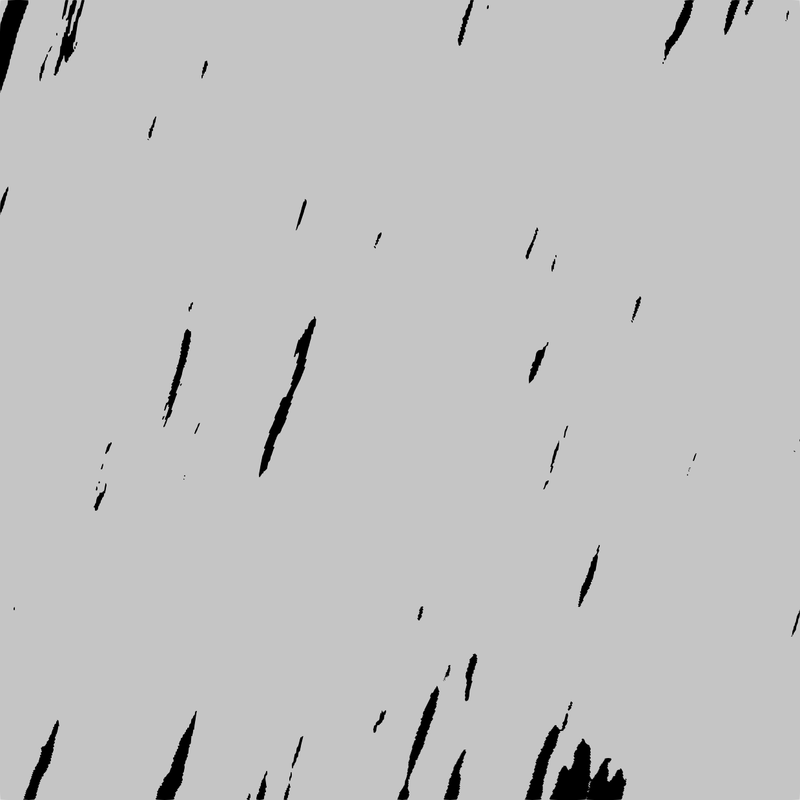}
  \caption*{(b) $l_x = 0.01, l_y = 0.10$,\\ $j = 2.52$}
 \end{minipage}
 \begin{minipage}[b]{0.18\textwidth}\centering
  \includegraphics[width=1.0\textwidth]{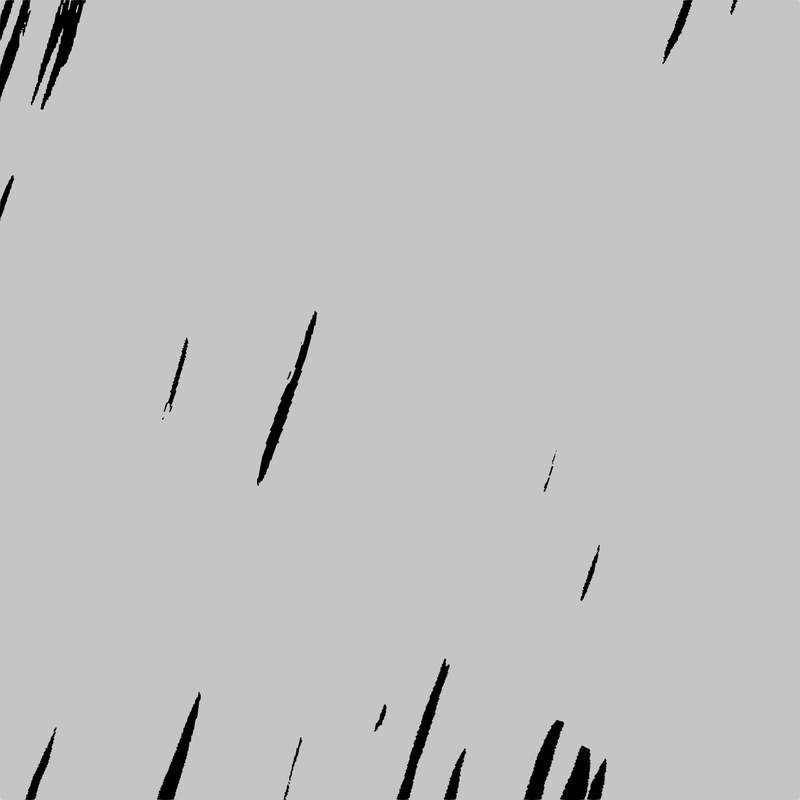}
  \caption*{(c) $l_x = 0.01, l_y = 0.20$,\\ $j = 2.76$}
 \end{minipage}
 \caption{The figure shows optimized topologies (top) for heat sinks obtained using the robust formulation with materials uncertainties.
  Samples of the material distributions are shown in the bottom row and correspond to $\tau$ in~\eqref{eq:uncertain_kappa_robust}.
  In this row, the color grey indicates $\tau = 1$, while black indicates regions with $\tau = 0$.
  The random damage fields are rotated 0.3 radians counterclockwise with correlation lengths specified above for each optimization case.
 }
 \label{fig:thermal_compliance_robust}
\end{figure}

These summarized examples from the literature represent two idealized corner cases that are relatively easy to model with limited computational resources but avoid the full range of spatial variations necessary for more realistic damage and material models.
The SPDE approach allows for representing intermediate cases as demonstrated in Figure~\ref{fig:thermal_compliance_robust}.
Instead of using SIMP interpolation~\eqref{eq:diffusionSIMP} as before, we employ the robust formulation of topology optimization~\cite{Wang2011} with thresholds $\eta_d = 0.3$ for the dilation and $\eta = 0.7$ for the erosion parameters.
The dilated volume is restricted to be less than 35\% of the total volume of the design domain, and the filter parameter is selected to be $r = 0.05$.
The projection parameter is chosen to be $\beta=8$.
In this example, the uncertainty is encompassed solely within the diffusion coefficient and the load is deterministic.
More specifically, we use a uniform heat input $f$ and model the random diffusion coefficient in~\eqref{eq_ex1_001} as
\begin{equation}\label{eq:uncertain_kappa_robust}
 \kappa\left(\overline{\rho}\right)=\kappa_{\min}+\tau{\overline{\rho}}\left(\kappa_{\max}-\kappa_{\min} \right)\,,
\end{equation}
where $\tau$ is a binary (zero/one) field obtained by a pointwise transformation of the SPDE solution.
This transformation projects all values of above 2.5 to zero, and maps the remaining values to one.
The zero values model local damage or material inclusions with very low conductivity.

Similar to the examples shown in Figure~\ref{fig:thresholded_fields}, the shape of the damage regions, shown in the bottom row of Figure~\ref{fig:thermal_compliance_robust}, and their scale and relative positions can be controlled easily.
The impact of such material variations can be seen on the top row.
Transforming an isotropic Gaussian random field with a small correlation length cannot impact the optimized topology since it results in a model with small, well-distributed damage regions and characteristic sizes much smaller than the length scale enforced on the design.
Increasing the correlation length in one of the directions and rotating the random field results in elongated damage regions following the prescribed orientation.
Furthermore, the small damage regions are replaced by elongated damage zones with larger lengths than the length scale of the design.
Therefore, the optimization provides alternative paths to transfer the heat to the support and creates closed loops of high-conductivity material.
It should be pointed out that the selected material model visually affects the topology of an optimization formulation with mean value (risk-neutral) objective.
This stands in contrast to all previously mentioned cases in the literature, where topology changes are only observed for risk-averse objective functions like the sum of mean and standard deviation or the min-max formulation.

\subsection{Bridge design}

\label{sub:bridge}

We are now concerned with finding the optimal topology of a bridge to support a given load.
We use the SPDE method to impose random fluctuations on the boundaries of a design problem, mimicking random horizontal loads.
For all 2D and 3D solid mechanics examples, the system's response is governed by the Navier--Cauchy PDE, given as
\begin{align}
 \label{eqne0001}
 \begin{split}
  -\operatorname{div}\mathbf{\sigma}\left(\mathbf{u}\right) & =\mathbf{f},                                                        \\
  \mathbf{\sigma}\left(\mathbf{u}\right)                    & =\mathbf{C}:\mathbf{\varepsilon}\left(\mathbf{u}\right)~~\text{in }
  D \, ,\end{split}
\end{align}
where $\mathbf{\sigma}$ is the so-called stress tensor, and $\mathbf{\varepsilon}$ is the linearized strain tensor defined as $\mathbf{\varepsilon}=\left(\nabla \mathbf{u}+\nabla\mathbf{u}^{\sf{T}}\right)/2$.
$\mathbf{C}$ is an elastic material properties tensor, $\mathbf{u}$ denotes the displacement field, and $\mathbf{f}$ is a distributed input supplied to the system.
The boundary $\partial D =\overline{\Gamma_\mathrm{D_i} \cup \Gamma_\mathrm{N_i}}, i=1,.,d$, is decomposed into disjoint subsets for each component $i=1,.,d$, $\Gamma_\mathrm{D_i}$ with prescribed displacements $u_i=0$, and $\Gamma_\mathrm{N_i}$ with prescribed traction $t_i$.
The material properties tensor is isotropic and has the following form $\mathbf{C}\left(\mathbf{x}\right)=E\left(\mathbf{x}\right) \mathbf{C}_0$, where $\mathbf{C}_0$ is a constant isotropic elasticity tensor obtained for a predefined Poisson ratio $\nu$ smaller than 0.5 and modulus of elasticity one. $E\left(\mathbf{x}\right)$ is a spatially varying modulus of elasticity $ E\left(x\right) \in \left[ E_{\min}, E_{\max}\right]$; cf.~\eqref{eqopt03}. Here, the Poisson's ratio is kept constant; however, if necessary, it can also be modeled as a random field.
\begin{subequations}
 The variational formulation of~\eqref{eqne0001} is given as
 \begin{equation}
  a\left(\mathbf{u},\mathbf{v}\right)=\ell\left(\mathbf{v}\right) ~~\text{for all } \mathbf{v}\in V_0 \, ,
 \end{equation}
 with $V_0=\left\{\mathbf{v}\in\left[H^1\left(D\right)\right]^d:v_i=0\,\rm{on}\,\Gamma_\mathrm{D_i},i=1,.,d\right\}\subset V=\left[H^1\left(D\right)\right]^d$ and bilinear form $a$ and linear form $\ell$ defined by
 \begin{align}
  a\left(\mathbf{u},\mathbf{v}\right) & =\int_D \left(\mathbf{C}:\mathbf{\varepsilon}\left(\mathbf{u}\right)\right):\mathbf{\varepsilon}\left(\mathbf{v}\right) {\rm{d}} \mathbf{x} \, ,        \\
  \ell\left(\mathbf{v}\right)         & = \int_D \left(\mathbf{f}\cdot\mathbf{v}\right) {\rm{d}} \mathbf{x} +  \int_{\Gamma_\mathrm{N}} \left(\mathbf{t}\cdot\mathbf{v} \right) {\rm{d}} s \, .
 \end{align}
\end{subequations}

\subsubsection{Two-dimensional bridge design}

\begin{figure}
 \centering
 \includegraphics[width=.25\textwidth]{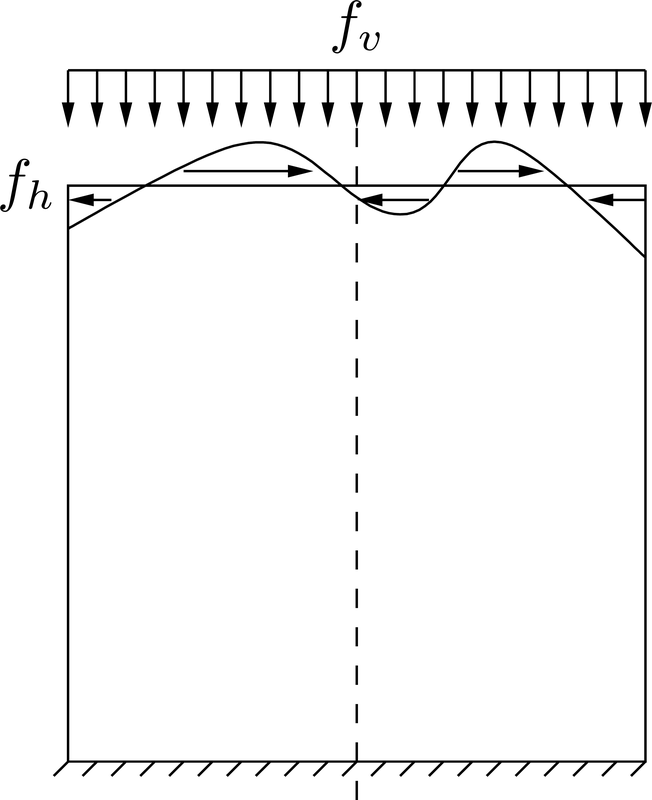}
 \caption{
  2D bridge - design domain and boundary conditions.
  The vertical and horizontal loads are denoted by $f_v$ and $f_h$, respectively.
 }
 \label{fig:fig_geo_2D}
\end{figure}

The first set of examples is inspired by~\cite{Zhao2014a, Martinez-Frutos2018} and consists of 2D bridge design problems.
The geometry of the 2D problem is shown in Figure~\ref{fig:fig_geo_2D}.
We choose the design domain to be the unit square.
Two distributed loads are applied on top of the design domain: a uniform vertical load $f_v = 1$ and a spatially varying random horizontal distributed force $f_h$.
The vertical and horizontal displacements at the bottom of the design domain are set to zero.
We use the so-called robust formulation~\cite{Wang2011, Lazarov2016} discussed earlier to obtain black-and-white designs with a prescribed length scale.
We optimize the mean compliance for the most eroded design obtained with $\eta_e=0.7$, and the volume constraint is enforced on the most dilated design with $\eta_d=0.3$.
The maximum amount of solid material is limited to 30\% of the total area of the design domain.
The projection parameter in~\eqref{eqopt04} is set to $\beta=8$, and the length parameter in the PDE filter~\eqref{eqopt02} is set to $r=0.02/\sqrt{12}$.
The interested readers are referred to~\cite{Lazarov2011} for more details on the length scale relation.
For the PDE filter, we enforce the Dirichlet boundary condition $\tilde{\rho}=1$ on the bridge deck, i.e., on top of the design domain, $\tilde{\rho}=0$ on the vertical sides, and zero Neumann on the remaining boundary.
The domain is discretized using second-order quad elements, with approximately 1M DOFs representing the displacement field and 0.5M DOFs representing the filtered density.
The density field is optimized using the PETSc implementation~\cite{Aage2013} of the Method of Moving Asymptotes (MMA)~\cite{Svanberg1987}, and the total number of iterations is set to 350.
The optimization uses a fixed number of 400 realizations for $f_h$.
The realizations do not change between the optimization iterations, resulting in a deterministic optimization problem similar to the one for thermal compliance.
Symmetrized designs are obtained as in Section~\ref{sec:enforcing_symmetry}.

\begin{figure}
 \centering
 \begin{minipage}[b]{0.195\textwidth}\centering
  \includegraphics[width=0.9\textwidth]{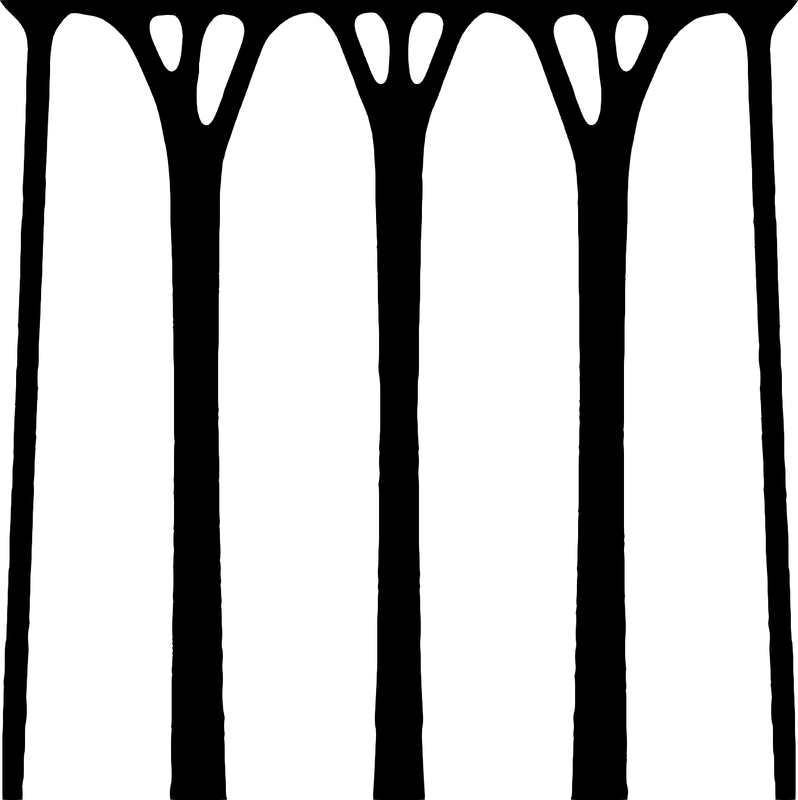}\\
  \caption*{(a) $l=0.01$, $j=5.28$, \\ symmetrized}
 \end{minipage}
 \begin{minipage}[b]{0.195\textwidth}\centering
  \includegraphics[width=0.9\textwidth]{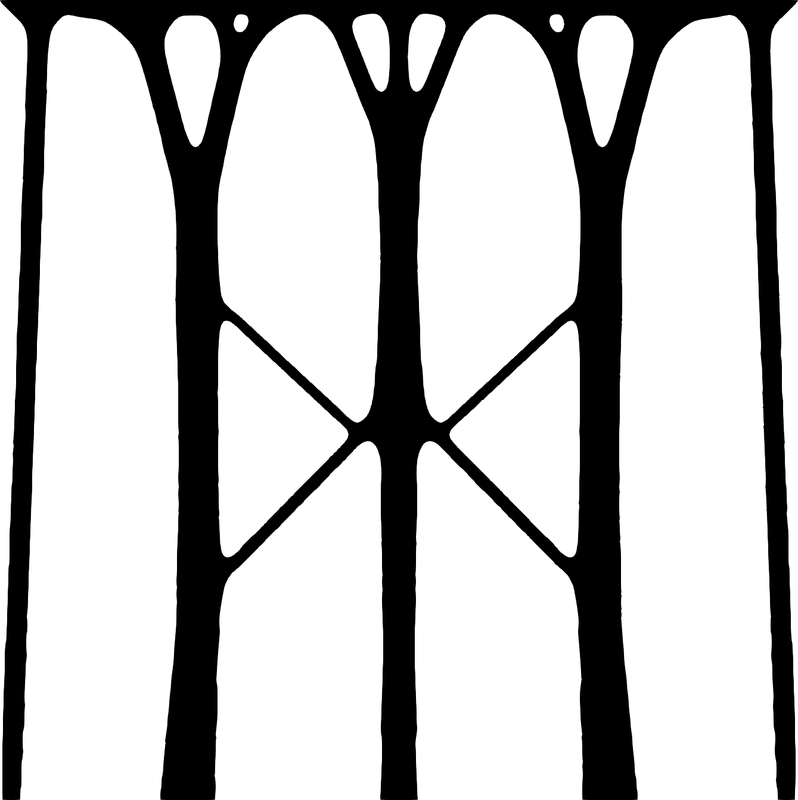}\\
  \caption*{(b) $l=0.05$, $j=5.97$, \\ symmetrized}
 \end{minipage}
 \begin{minipage}[b]{0.195\textwidth}\centering
  \includegraphics[width=0.9\textwidth]{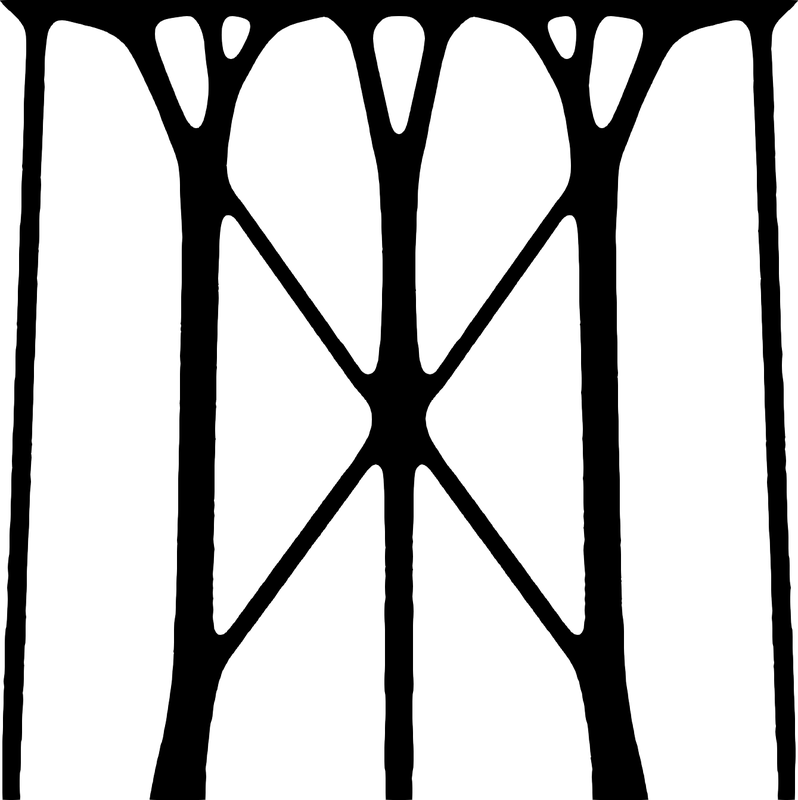}\\
  \caption*{(c) $l=0.20$, $j=6.74$, \\ symmetrized}
 \end{minipage}
 \begin{minipage}[b]{0.195\textwidth}\centering
  \includegraphics[width=0.9\textwidth]{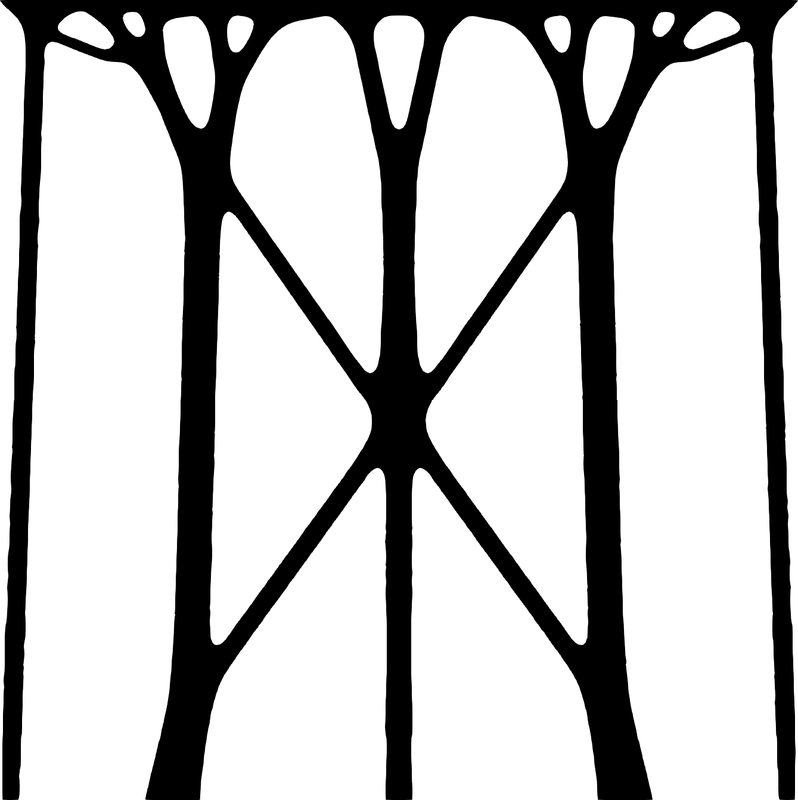}\\
  \caption*{(d) $l=0.30$, $j=7.16$, \\ symmetrized}
 \end{minipage}
 \begin{minipage}[b]{0.195\textwidth}\centering
  \includegraphics[width=0.9\textwidth]{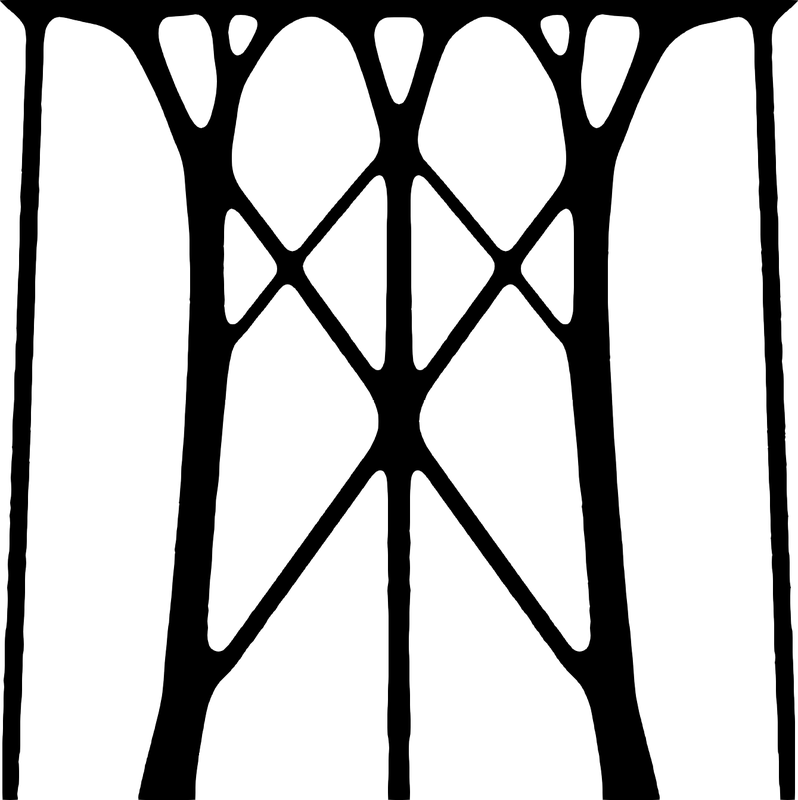}\\
  \caption*{(e) $l=0.40$, $j=7.76$, \\ symmetrized}
 \end{minipage}\\ \vspace{0.25cm}
 \begin{minipage}[b]{0.195\textwidth}\centering
  \includegraphics[width=0.9\textwidth]{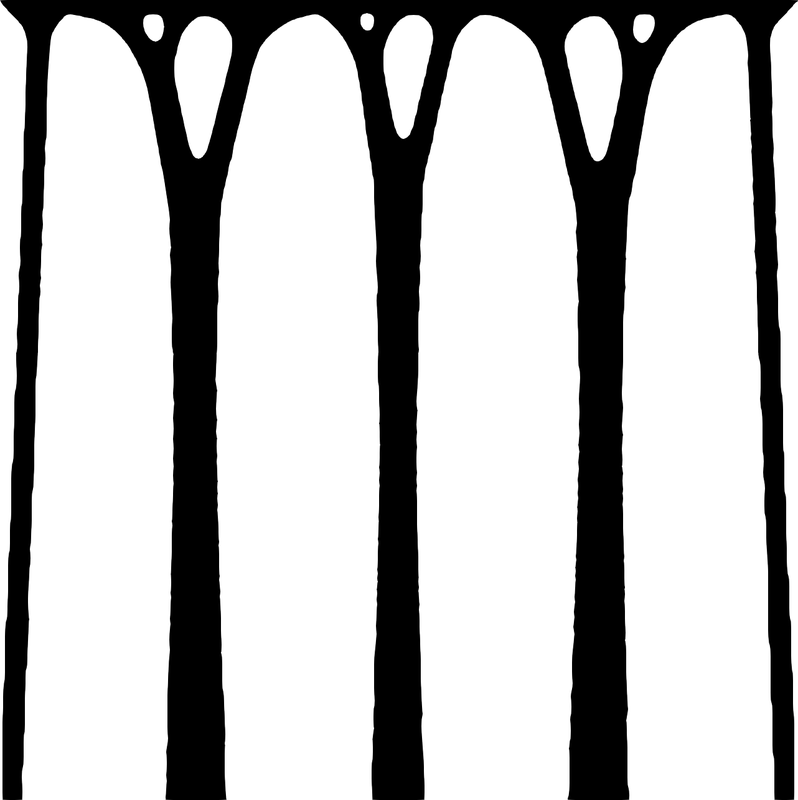}\\
  \caption*{(f) $l=0.01$, $j=5.56$}
 \end{minipage}
 \begin{minipage}[b]{0.195\textwidth}\centering
  \includegraphics[width=0.9\textwidth]{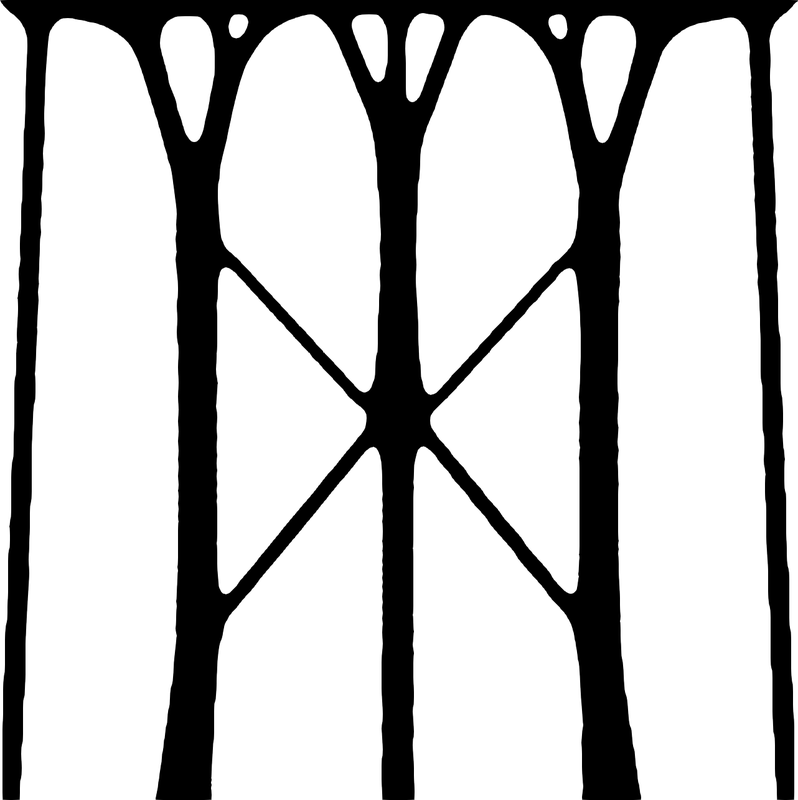}\\
  \caption*{(h) $l=0.05$, $j=6.26$}
 \end{minipage}
 \begin{minipage}[b]{0.195\textwidth}\centering
  \includegraphics[width=0.9\textwidth]{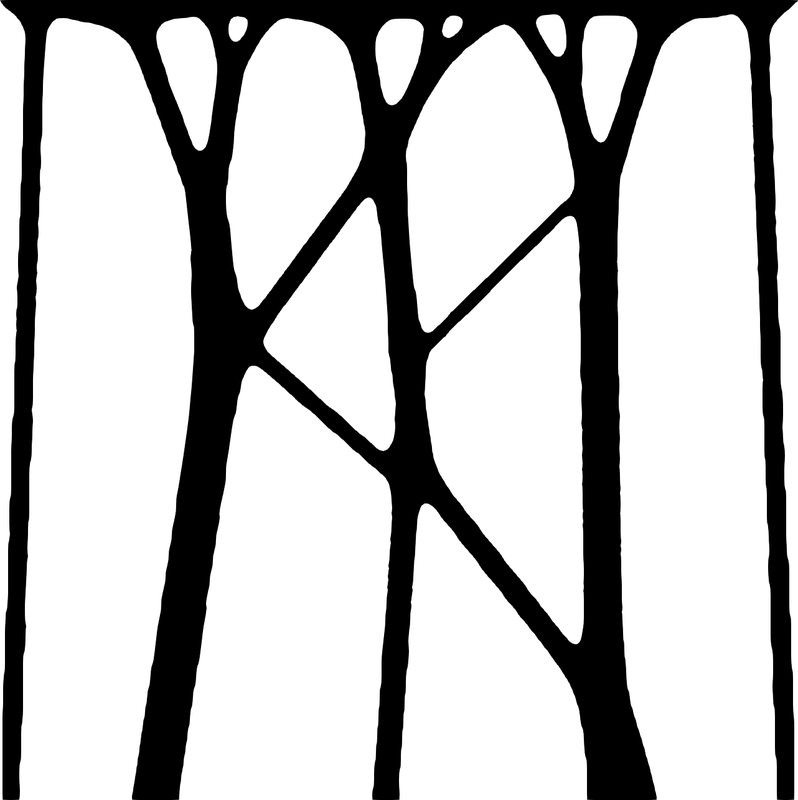}\\
  \caption*{(h) $l=0.20$, $j=7.09$}
 \end{minipage}
 \begin{minipage}[b]{0.195\textwidth}\centering
  \includegraphics[width=0.9\textwidth]{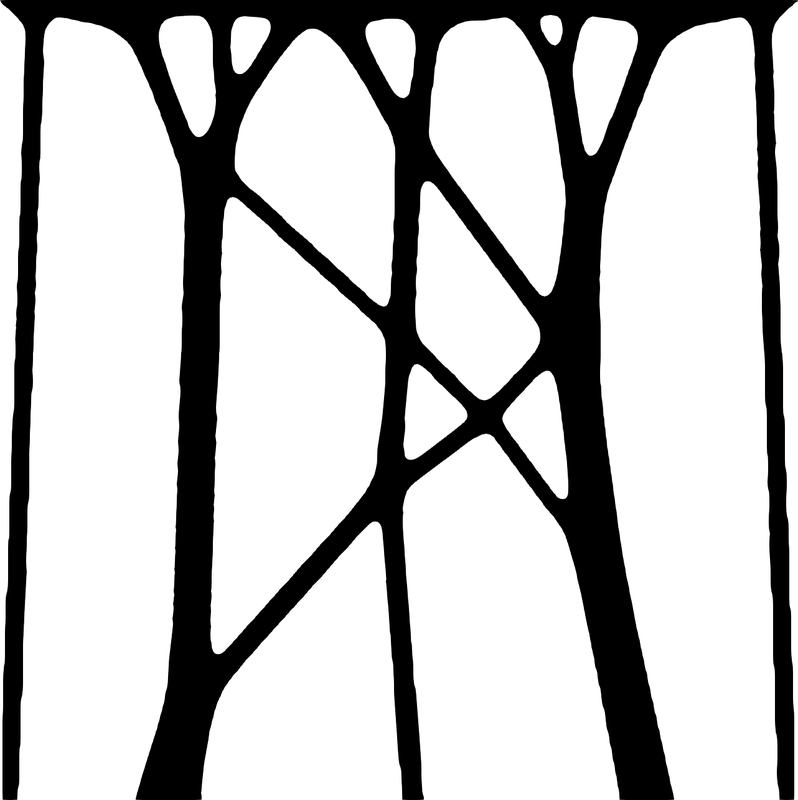}\\
  \caption*{(i) $l=0.30$, $j=7.89$}
 \end{minipage}
 \begin{minipage}[b]{0.195\textwidth}\centering
  \includegraphics[width=0.9\textwidth]{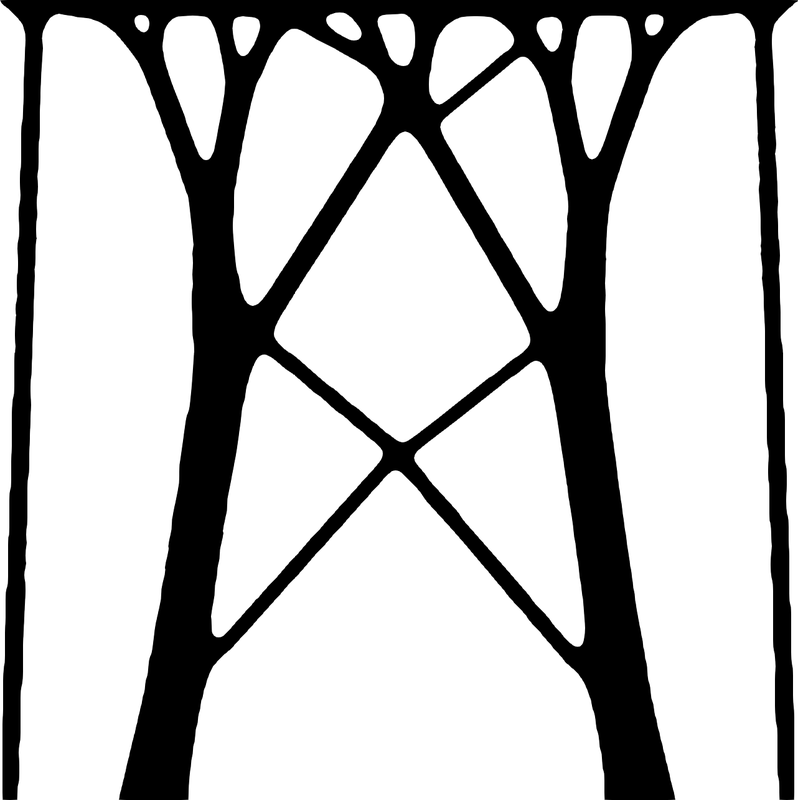}\\
  \caption*{(j) $l=0.40$, $j=8.00$}
 \end{minipage}
 \caption{
  Compliance-optimized design of a bridge structure with uniformly distributed unit vertical load and random horizontal Gaussian noise with a standard deviation 0.2 and different correlation lengths $l$.
  The correlation length $l$ increases from left to right.
  The top (a-e) and bottom (f-j) rows display the results with and without enforced symmetry, respectively.
  The initial distribution, a scaled cosine wave, aids the appearance of five supporting legs in all designs.
 }
 \label{fig:bridge_compliance_cos}
\end{figure}

\begin{figure}
 \centering
 \begin{minipage}[b]{0.195\textwidth}\centering
  \includegraphics[width=0.9\textwidth]{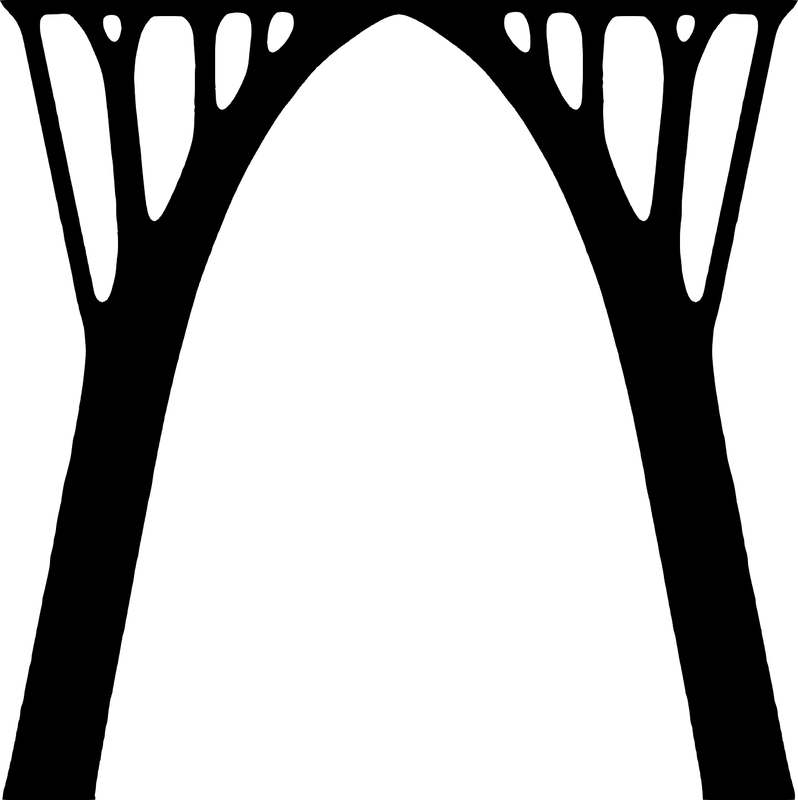}\\
  \caption*{(a) $l=0.01$, $j=5.02$, \\ symmetrized}
 \end{minipage}
 \begin{minipage}[b]{0.195\textwidth}\centering
  \includegraphics[width=0.9\textwidth]{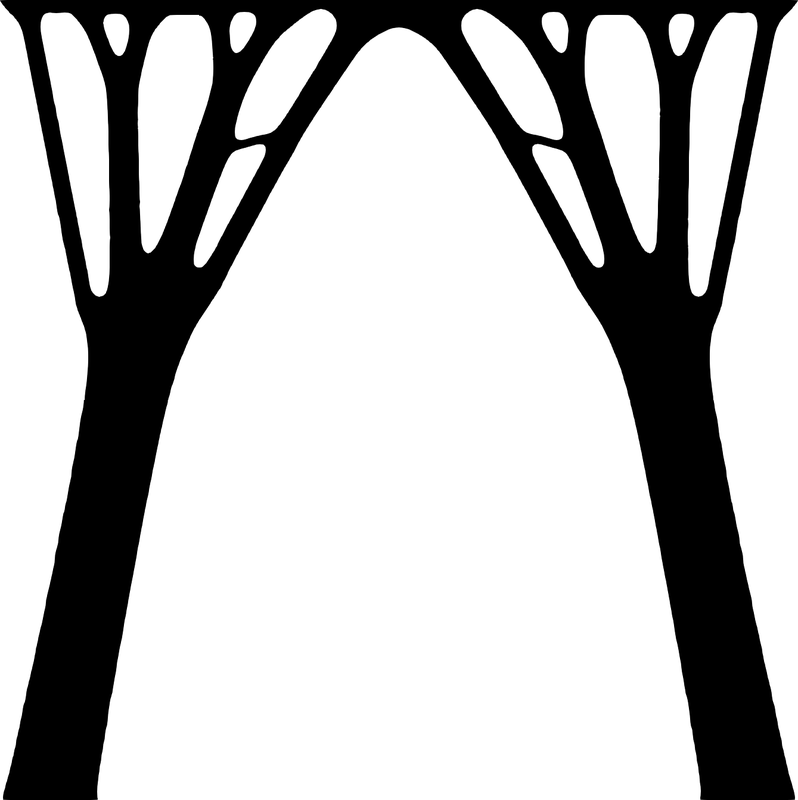}\\
  \caption*{(b) $l=0.05$, $j=5.39$, \\ symmetrized}
 \end{minipage}
 \begin{minipage}[b]{0.195\textwidth}\centering
  \includegraphics[width=0.9\textwidth]{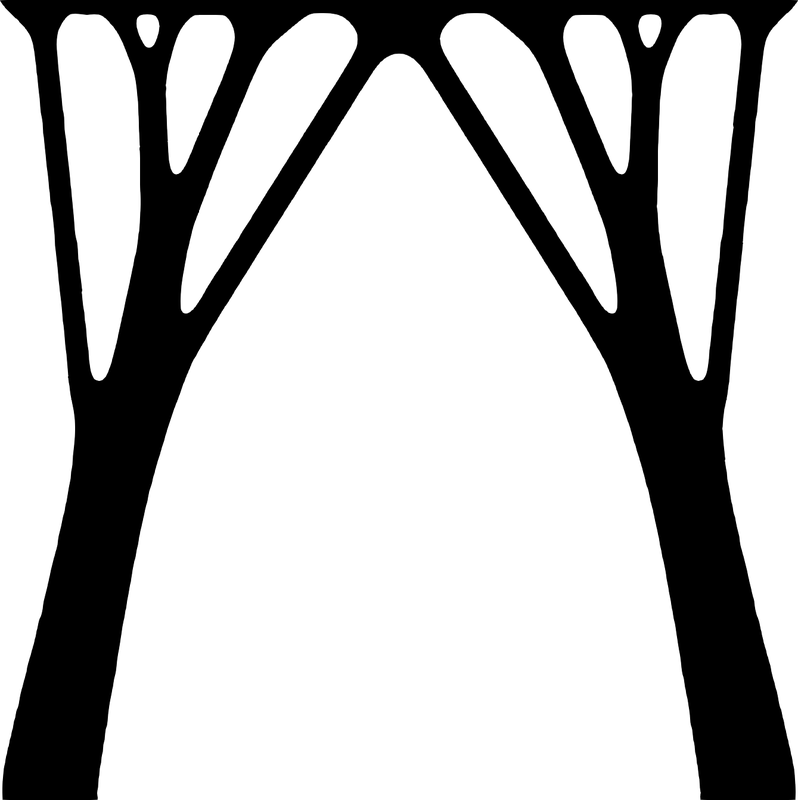}\\
  \caption*{(c) $l=0.20$, $j=5.77$, \\ symmetrized}
 \end{minipage}
 \begin{minipage}[b]{0.195\textwidth}\centering
  \includegraphics[width=0.9\textwidth]{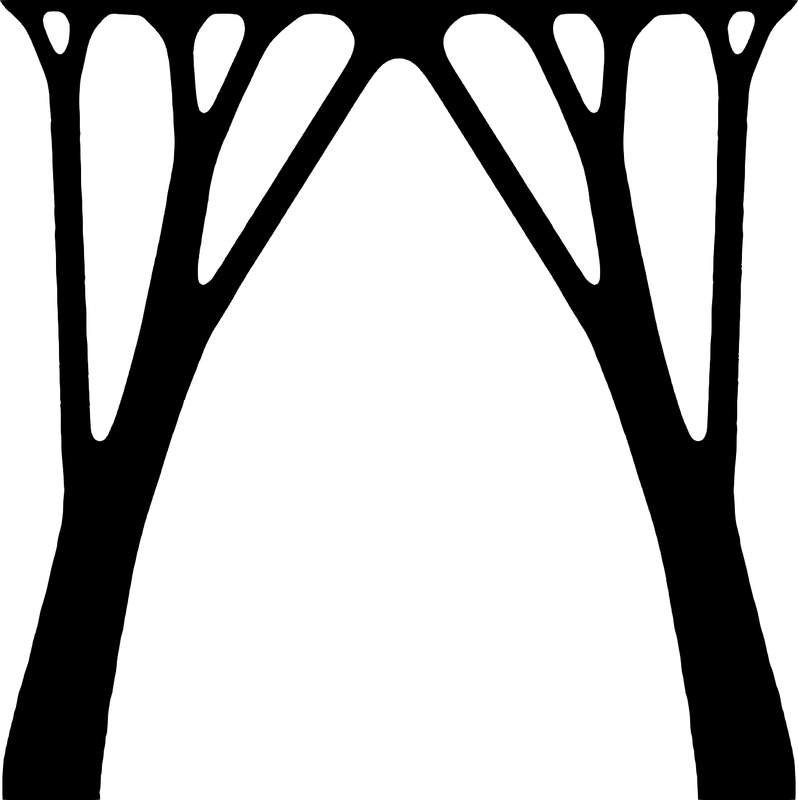}\\
  \caption*{(d) $l=0.30$, $j=5.91$, \\ symmetrized}
 \end{minipage}
 \begin{minipage}[b]{0.195\textwidth}\centering
  \includegraphics[width=0.9\textwidth]{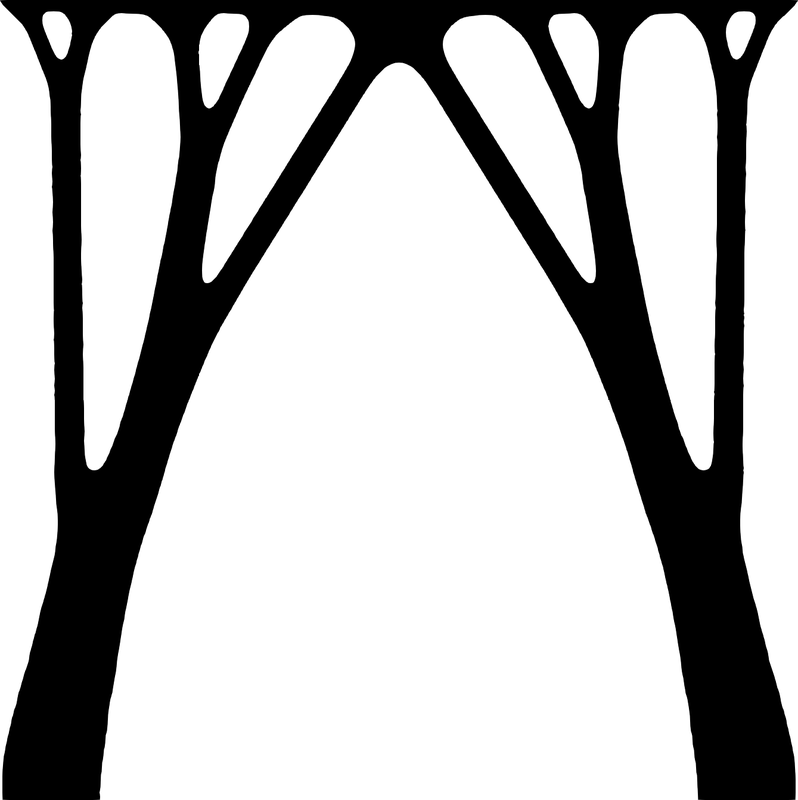}\\
  \caption*{(e) $l=0.40$, $j=6.14$, \\ symmetrized}
 \end{minipage} \\ \vspace{0.25cm}
 \begin{minipage}[b]{0.195\textwidth}\centering
  \includegraphics[width=0.9\textwidth]{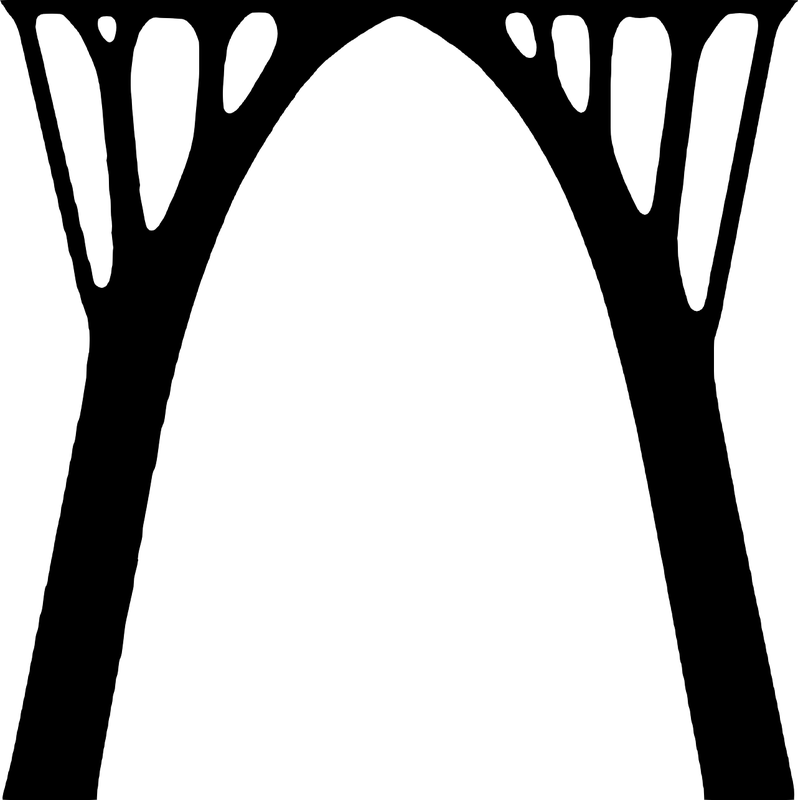}\\
  \caption*{(f) $l=0.01$, $j=5.13$}
 \end{minipage}
 \begin{minipage}[b]{0.195\textwidth}\centering
  \includegraphics[width=0.9\textwidth]{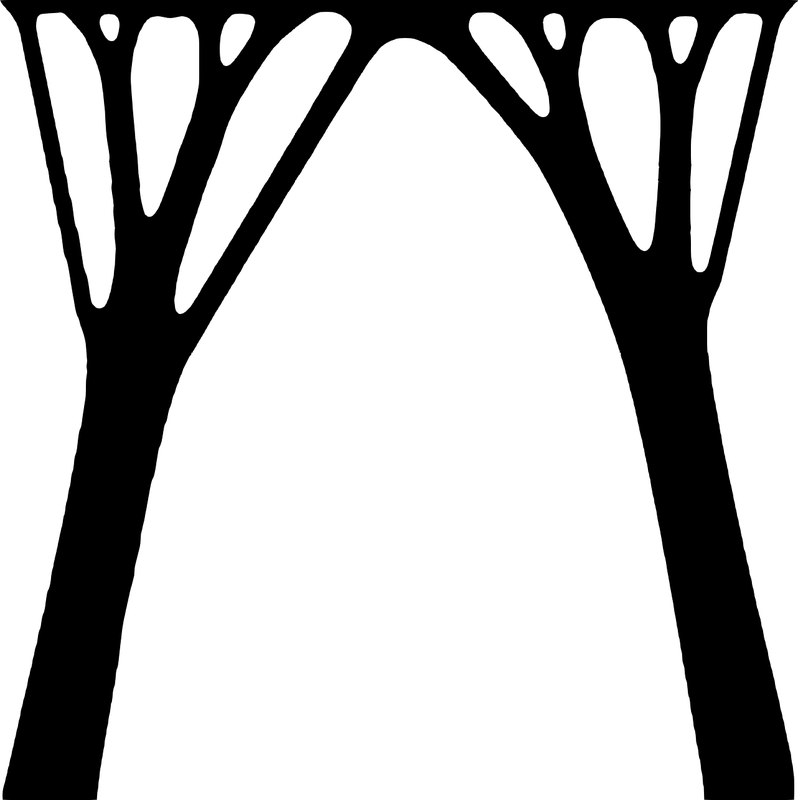}\\
  \caption*{(g) $l=0.05$, $j=5.56$}
 \end{minipage}
 \begin{minipage}[b]{0.195\textwidth}\centering
  \includegraphics[width=0.9\textwidth]{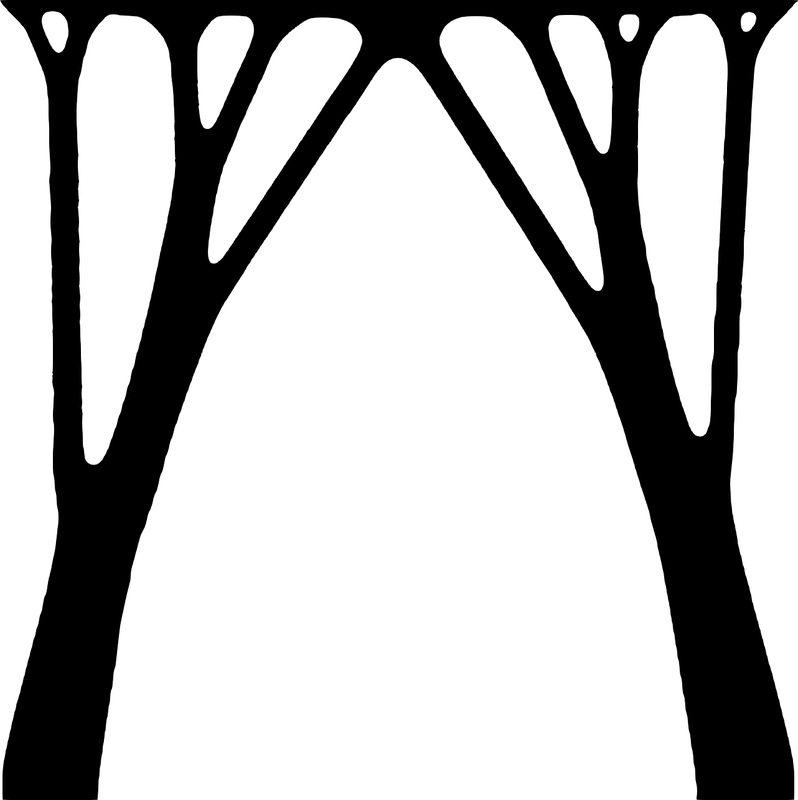}\\
  \caption*{(h) $l=0.20$, $j=6.03$}
 \end{minipage}
 \begin{minipage}[b]{0.195\textwidth}\centering
  \includegraphics[width=0.9\textwidth]{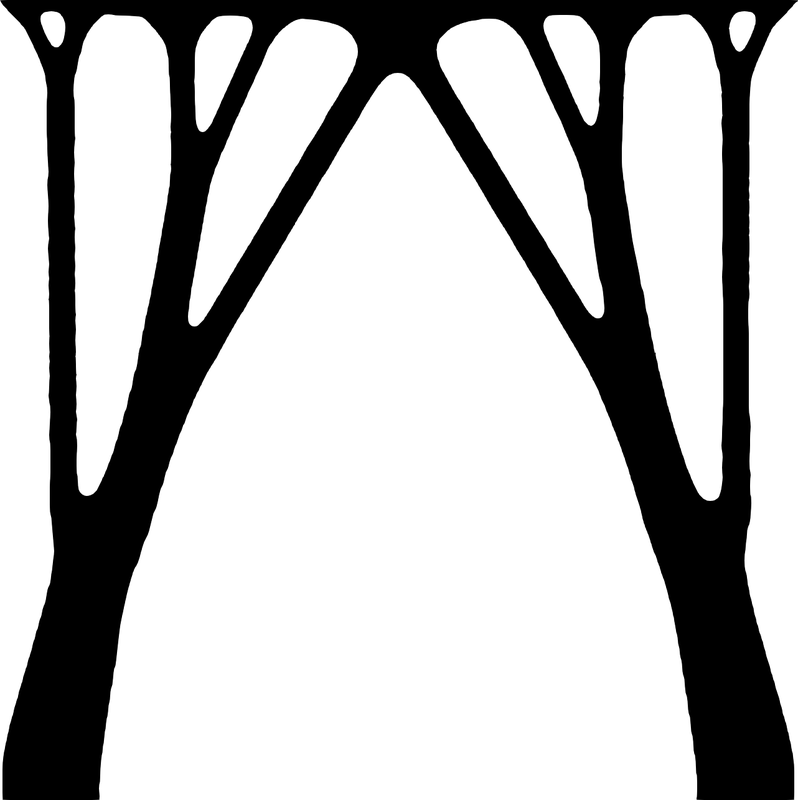}\\
  \caption*{(i) $l=0.30$, $j=6.54$}
 \end{minipage}
 \begin{minipage}[b]{0.195\textwidth}\centering
  \includegraphics[width=0.9\textwidth]{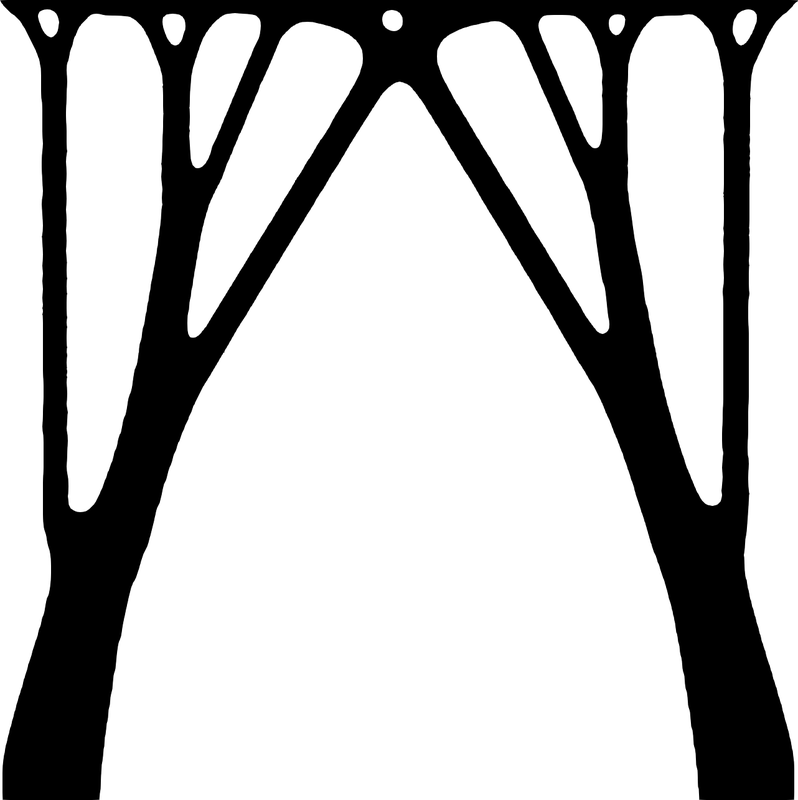}\\
  \caption*{(j) $l=0.40$, $j=6.81$}
 \end{minipage}
 \caption{
  Compliance-optimized design of a bridge structure with uniformly distributed unit vertical load and random horizontal Gaussian noise with a standard deviation 0.2 and different correlation lengths $l$.
  The correlation length $l$ increases from left to right.
  The top (a-e) and bottom (f-j) rows display the results with and without enforced symmetry, respectively.
  The initial density distribution is uniform.
 }
 \label{fig:bridge_compliance_uniform}
\end{figure}

The results from the topology optimization for different initial density distributions and correlation lengths for the random spatially varying horizontal load $f_h$ are presented in Figure~\ref{fig:bridge_compliance_cos} and~\ref{fig:bridge_compliance_uniform}.
The min-max optimization problem~\eqref{eqopt0r51} is not convex, and the iterative optimization process will likely end up in a local minimum.
Furthermore, the final topology is influenced significantly by the initial density distribution, which can be observed by comparing Figures~\ref{fig:bridge_compliance_cos} and~\ref{fig:bridge_compliance_uniform}.
The designs in the former figure are obtained from an initial density field proportional to a cosine function with a period of $1/4$, and the designs presented in the latter are obtained from uniform initial density distribution.
The non-uniform initial density distribution supports the appearance of five legs for transferring the vertical load to the support.
On the other hand, the uniform initial guess results in two inclined main legs that branch to support the bridge deck.

As mentioned earlier, the optimization problem is not convex, and even though we have symmetry in the design domain with respect to the dashed line in the middle, we cannot expect a symmetric solution as the optimization process will likely end up in a local minimum.
Furthermore, during the optimization, we use a limited number of realizations for the horizontal load $f_h$ without guaranteeing the symmetry of the realized distribution concerning the dashed axis.
Therefore, for optimization runs executed on the entire optimization domain, we clearly observe a lack of symmetry in the obtained topologies in Figures~\ref{fig:bridge_compliance_cos} (f)-(j) and~\ref{fig:bridge_compliance_uniform} (f)-(j).
Since we enforce $\tilde{\rho}=1$ for the boundary condition of the PDE filter on the deck, the solid material locally balances the negative and positive oscillations of the load, and only the unbalanced part of the horizontal force is transferred to the supports, forcing the topology optimization algorithm to develop diagonal support members.

The optimized topologies are significantly affected by the correlation length of the horizontal stochastic load $f_h$.
Shorter correlation lengths result in a fast oscillatory load applied on the bridge's deck.
In these cases, the horizontal forces are balanced locally for the non-uniform initial distribution experiments, resulting in topologies without diagonal members required for stiffening the structure against global horizontal loads.
The topologies obtained for correlation length parameter $l=0.01$ do not differ visually from those obtained for the pure vertical load without horizontal disturbance.
Therefore, to avoid redundancy, designs for deterministic cases with $f_h=0$ are omitted in Figures~\ref{fig:bridge_compliance_cos} and~\ref{fig:bridge_compliance_uniform}.

On the other hand, for longer correlation lengths, the ergodic average along the bridge deck significantly differs from zero for most realizations, forcing the optimization to develop diagonal supporting structural members for transferring the unbalanced forces to the bottom of the design domain.
For a uniform initial density distribution, the deck loads are transferred by only two legs to the bottom of the design domain.
In this case, the optimization process develops small loops close to the deck for shorter correlation lengths versus larger loops for longer lengths.
Like the multi-leg case above, the structural loops balance the forces as close as possible to the bridge deck, and the main two legs transfer only the excess, unbalanced load to the supports.

\subsubsection{Three-dimensional bridge design}

\begin{figure}
 \centering
 \begin{minipage}{0.49\textwidth}
  \centering
  \includegraphics[height=.5\textwidth]{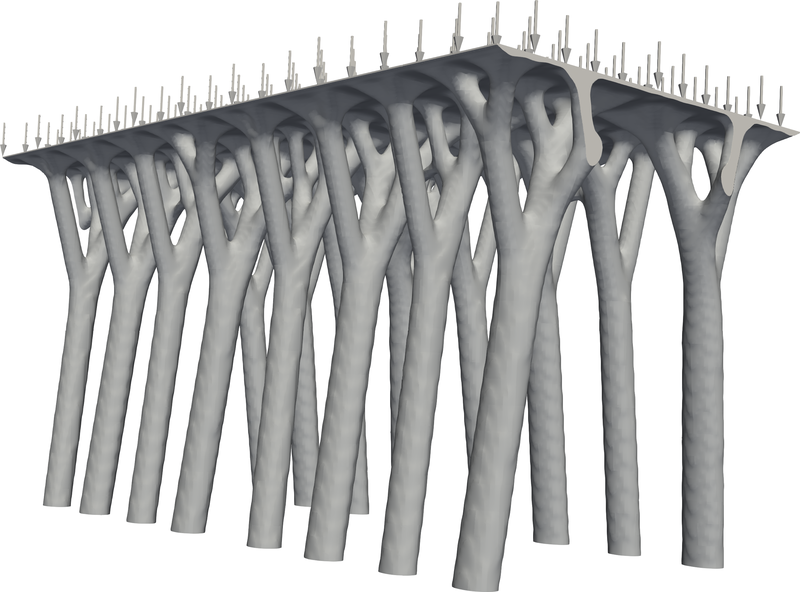}
  \caption*{(a) Optimized design from a uniform initial distribution, $j=10.80$}
 \end{minipage}
 \quad
 \begin{minipage}{0.4\textwidth}
  \includegraphics[height=.30\textwidth]{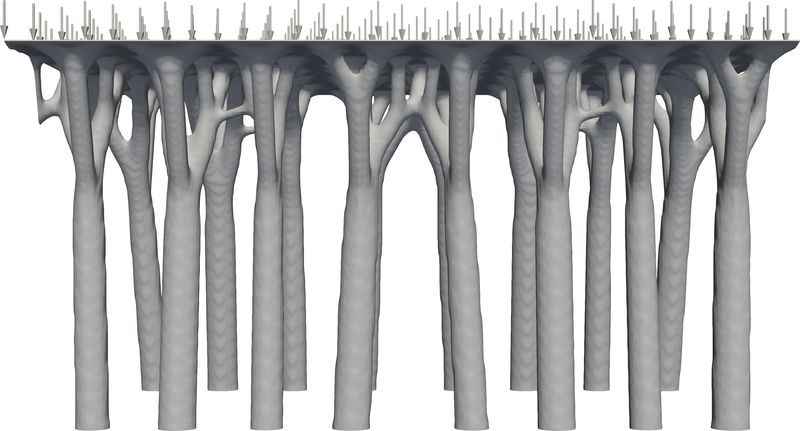}
  \includegraphics[height=.30\textwidth]{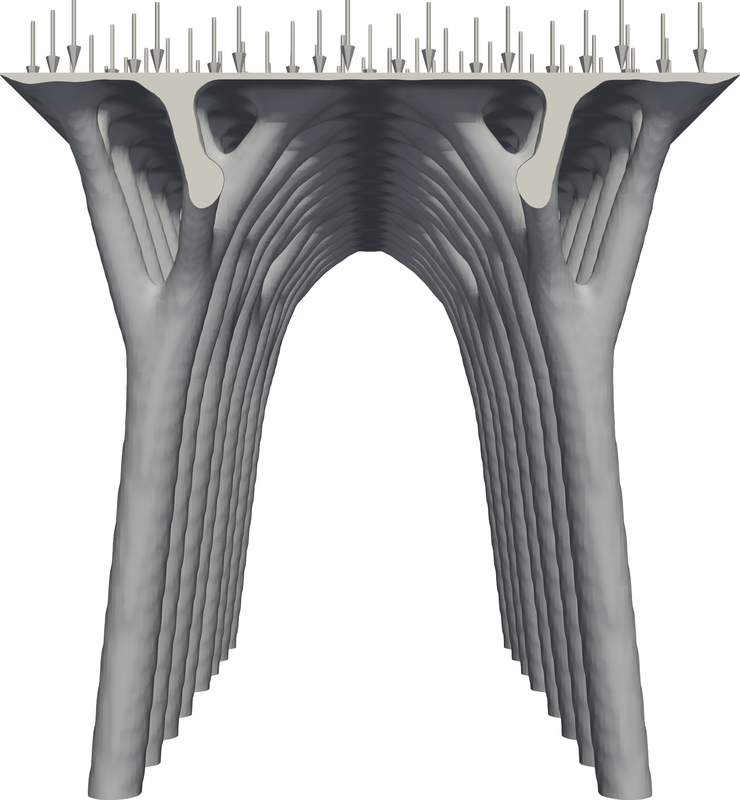}\\
  \includegraphics[height=.60\textwidth,angle=-90]{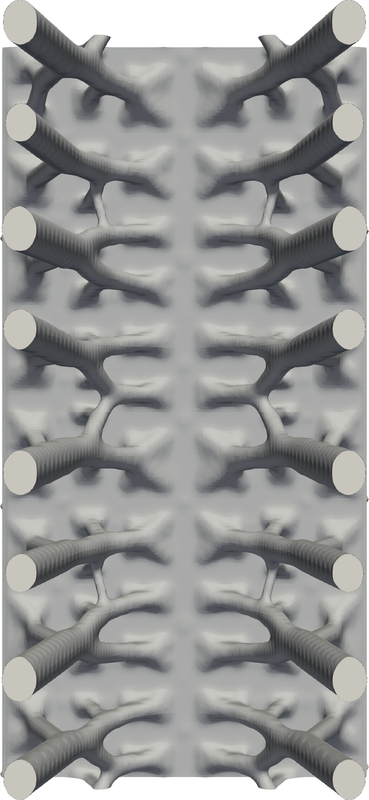}
 \end{minipage}

 \medskip

 \begin{minipage}{0.49\textwidth}
  \centering
  \includegraphics[height=.5\textwidth]{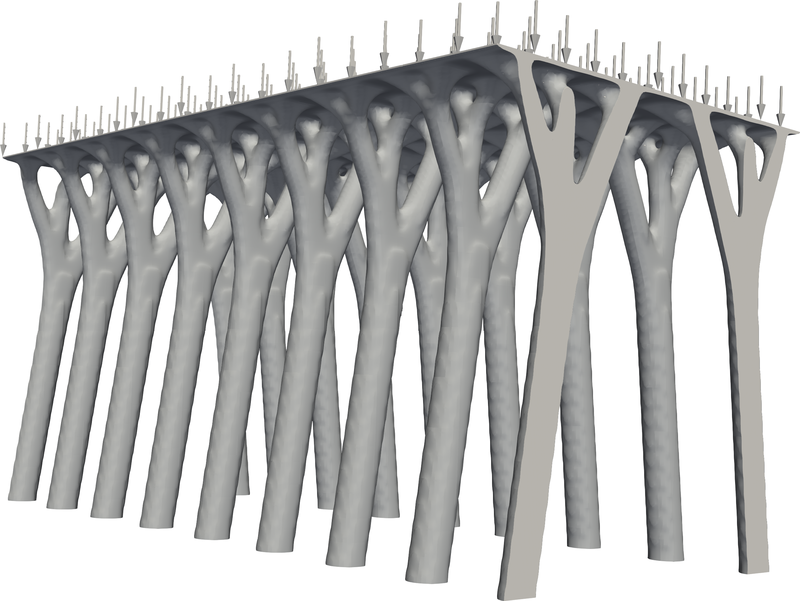}
  \caption*{(b) Optimized design from a prescribed initial distribution, $j=10.70$}
 \end{minipage}
 \quad
 \begin{minipage}{0.4\textwidth}
  \includegraphics[height=.30\textwidth]{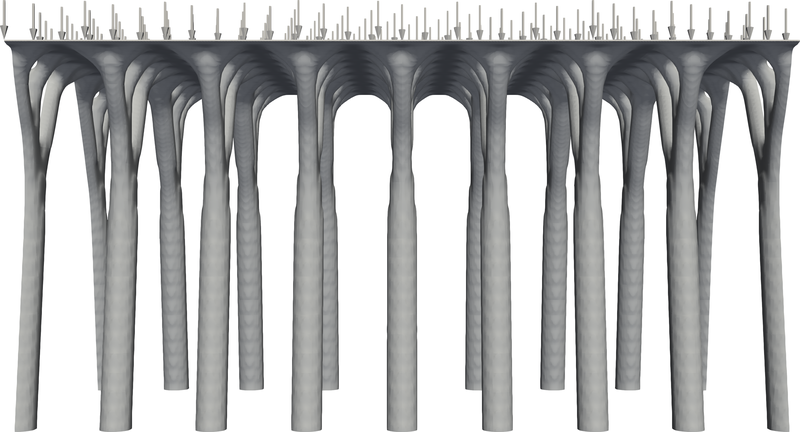}
  \includegraphics[height=.30\textwidth]{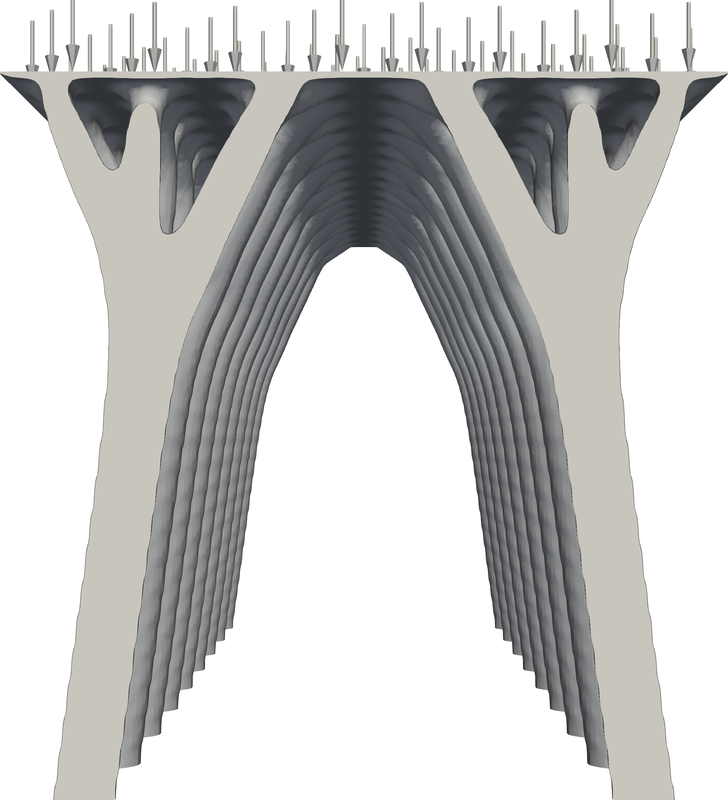}\\
  \includegraphics[height=.60\textwidth,angle=-90]{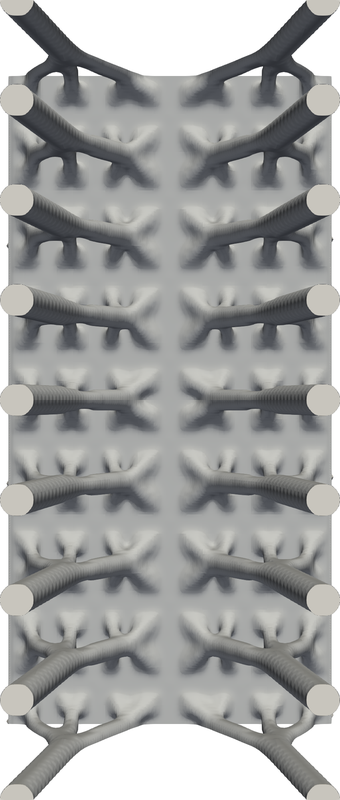}
 \end{minipage}

 \caption{Deterministic compliance optimization of a 3D bridge structure with uniform (top) and prescribed (bottom) initial density distribution. }
 \label{fig:bridge3D_full_compliance_deterministic}
\end{figure}

\begin{figure}
 \vspace*{-1em}
 \centering
 \begin{minipage}{0.45\textwidth}
  \centering
  \includegraphics[width=.9\textwidth]{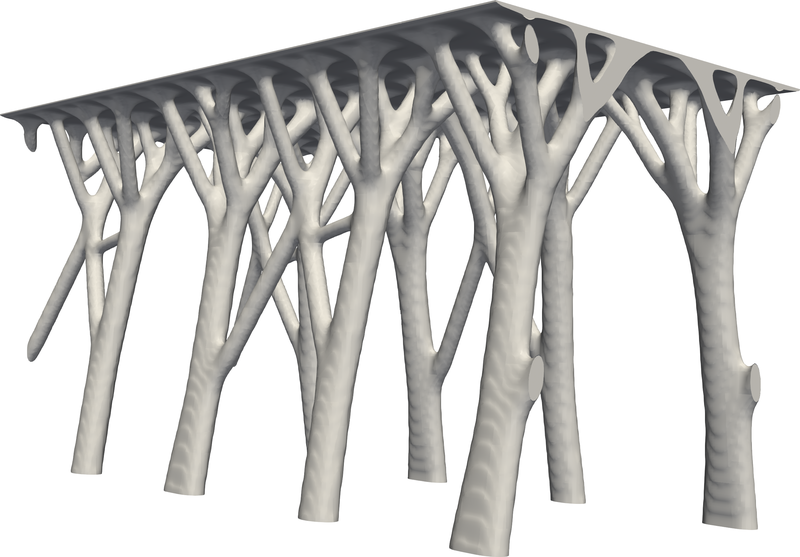}
  \caption*{(a) $l=0.01$, $j=12.21$}
 \end{minipage}
 \quad
 \begin{minipage}{0.4\textwidth}
  \includegraphics[height=.30\textwidth]{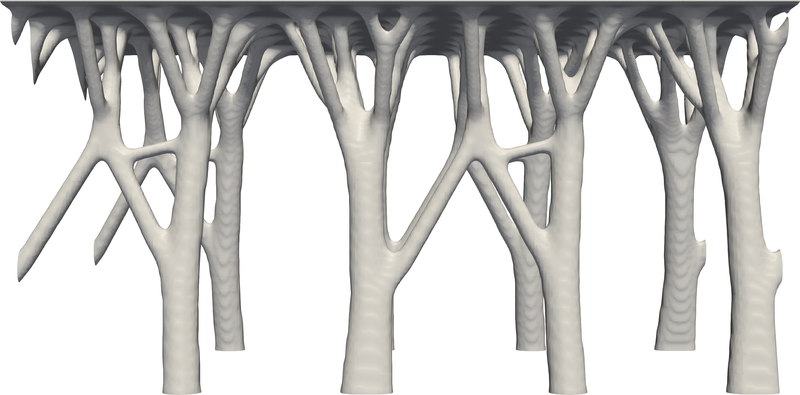}
  \includegraphics[height=.30\textwidth]{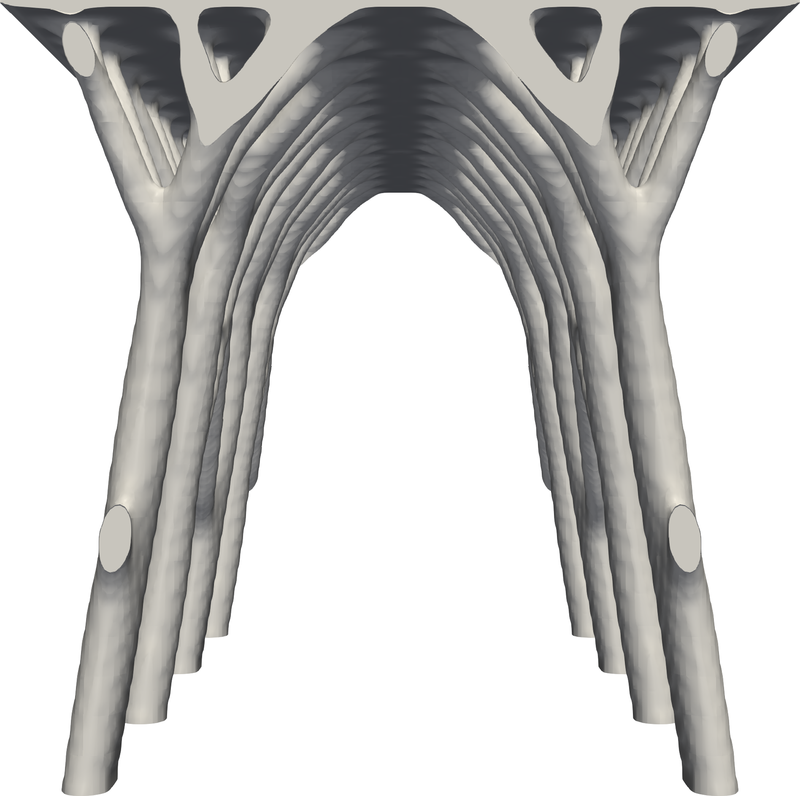}\\
  \includegraphics[height=.60\textwidth,angle=-90]{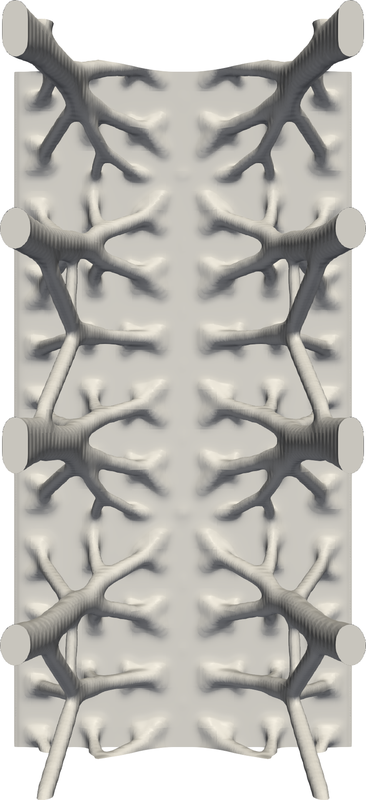}
 \end{minipage}

 \smallskip

 \begin{minipage}{0.45\textwidth}
  \centering
  \includegraphics[width=.9\textwidth]{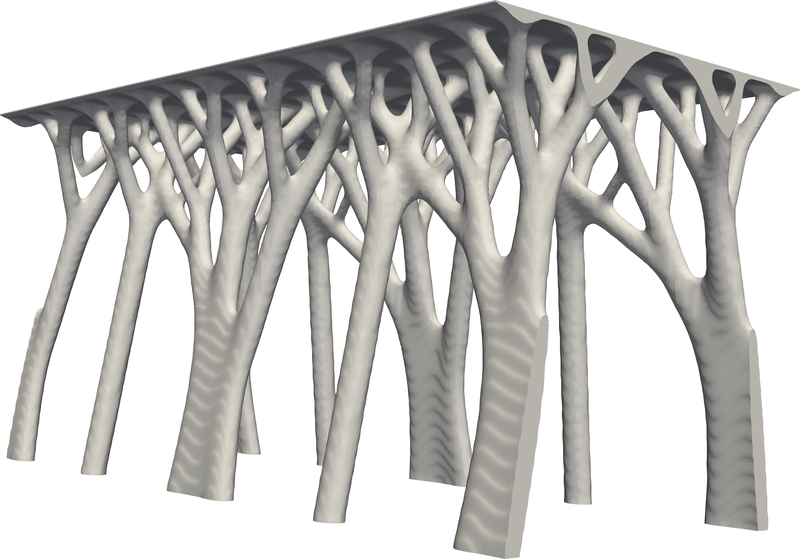}
  \caption*{(b) $l=0.02$, $j=12.43$}
 \end{minipage}
 \quad
 \begin{minipage}{0.4\textwidth}
  \includegraphics[height=.30\textwidth]{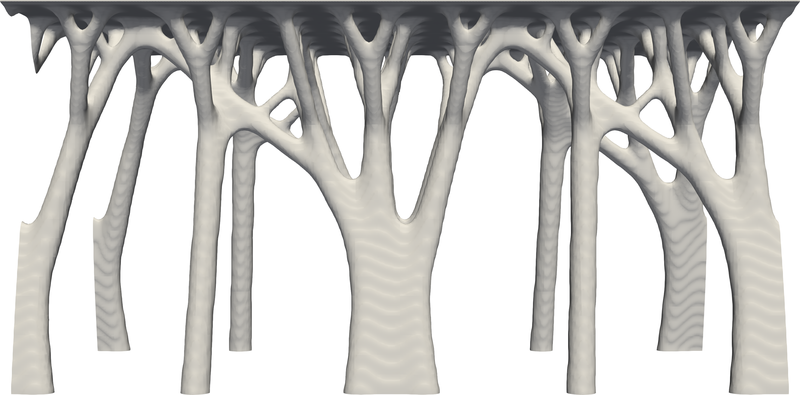}
  \includegraphics[height=.30\textwidth]{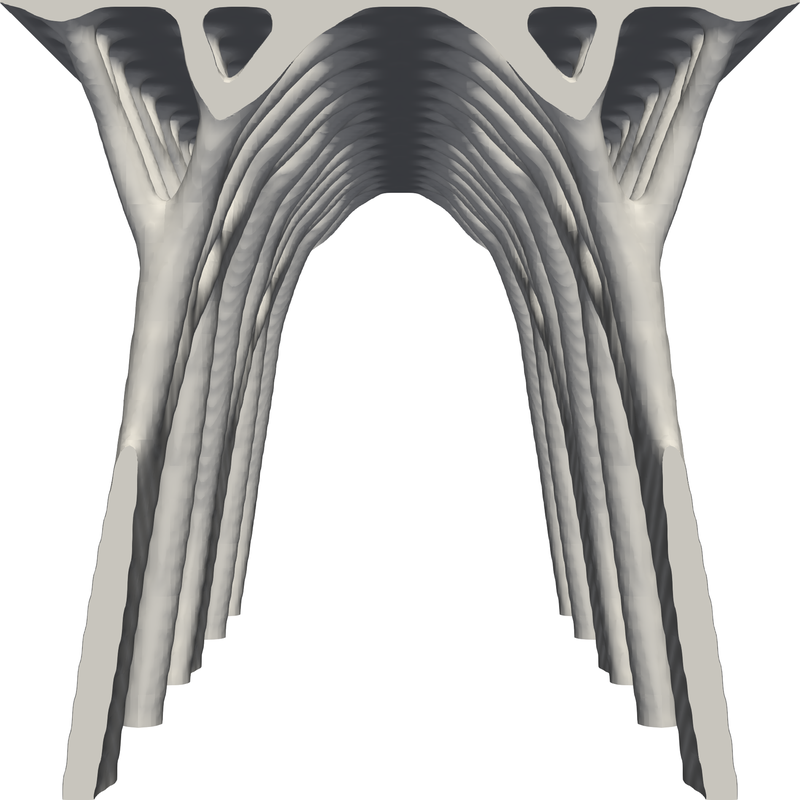}\\
  \includegraphics[height=.60\textwidth,angle=-90]{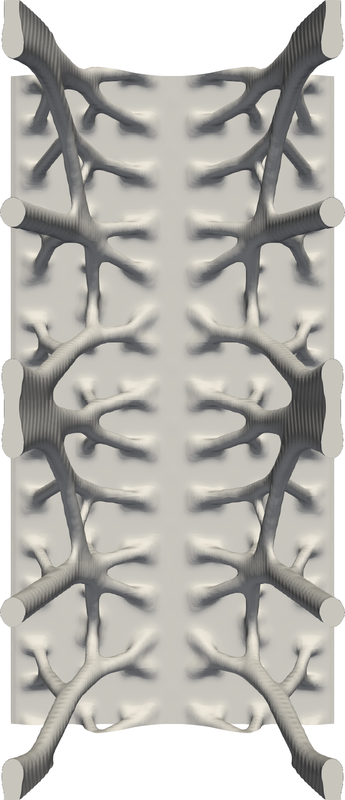}
 \end{minipage}

 \smallskip

 \begin{minipage}{0.45\textwidth}
  \centering
  \includegraphics[width=.9\textwidth]{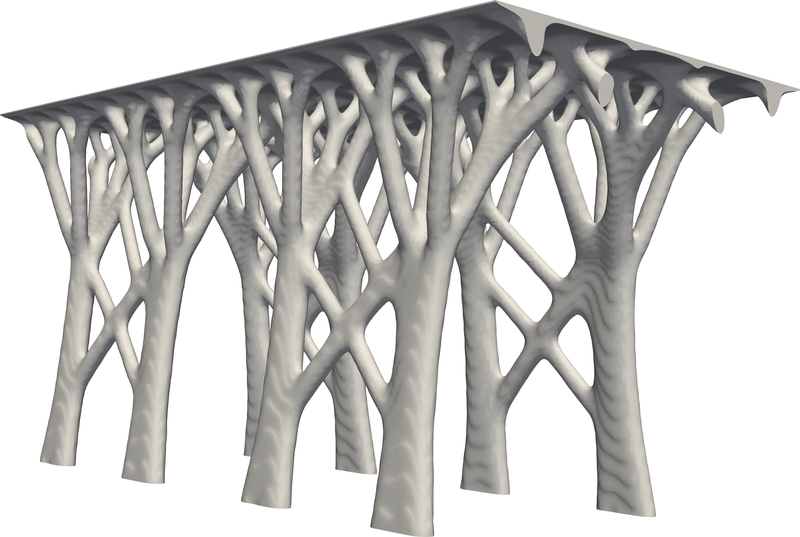}
  \caption*{(c) $l=0.03$, $j=12.59$}
 \end{minipage}
 \quad
 \begin{minipage}{0.4\textwidth}
  \includegraphics[height=.30\textwidth]{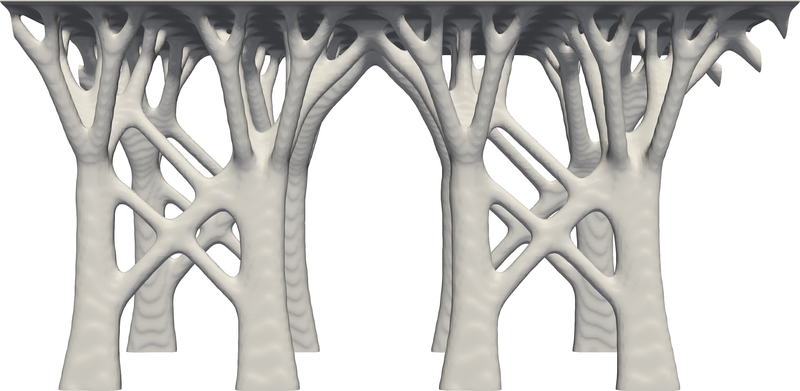}
  \includegraphics[height=.30\textwidth]{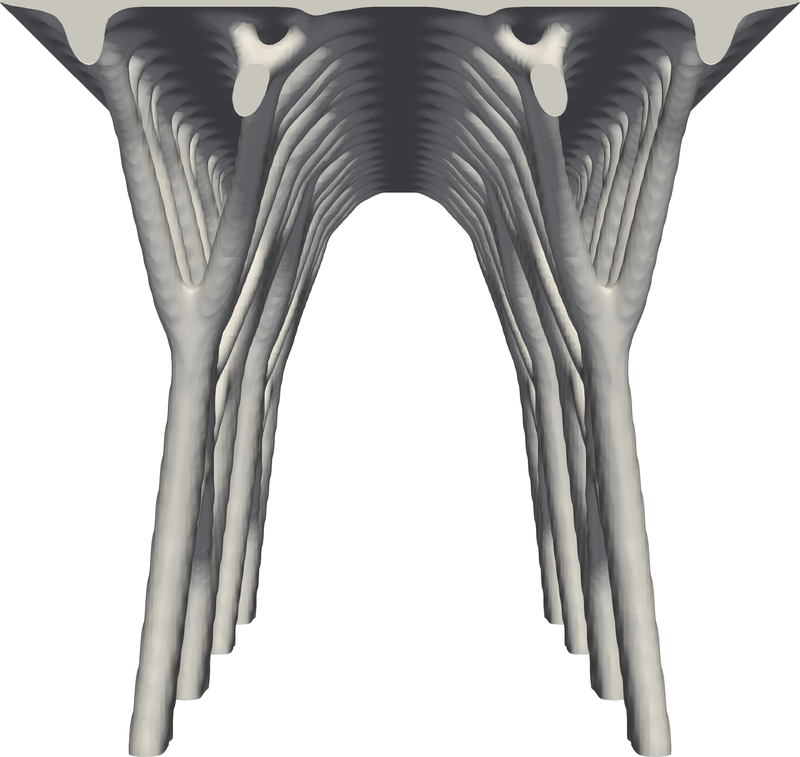}\\
  \includegraphics[height=.60\textwidth,angle=-90]{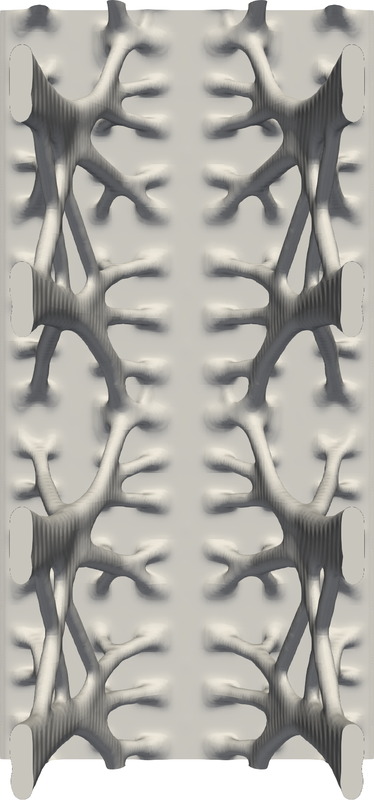}
 \end{minipage}

 \smallskip

 \begin{minipage}{0.45\textwidth}
  \centering
  \includegraphics[width=.9\textwidth]{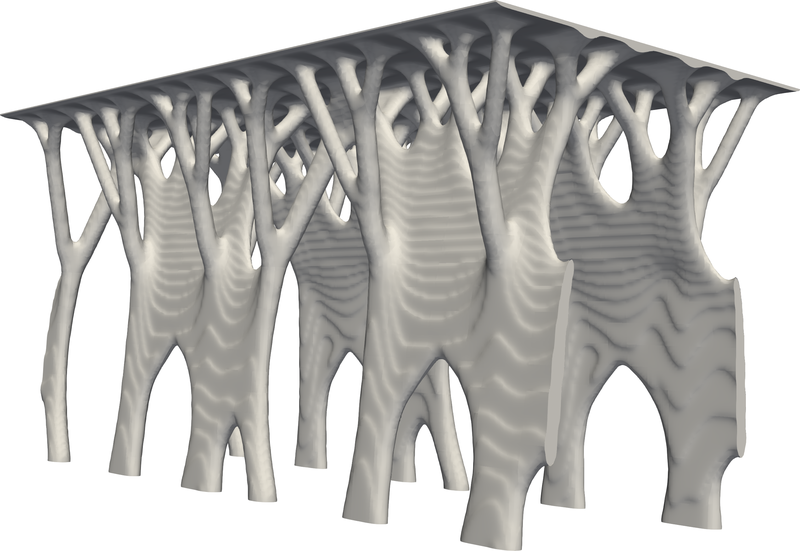}
  \caption*{(d) $l=0.04$, $j=12.39$}
 \end{minipage}
 \quad
 \begin{minipage}{0.4\textwidth}
  \includegraphics[height=.30\textwidth]{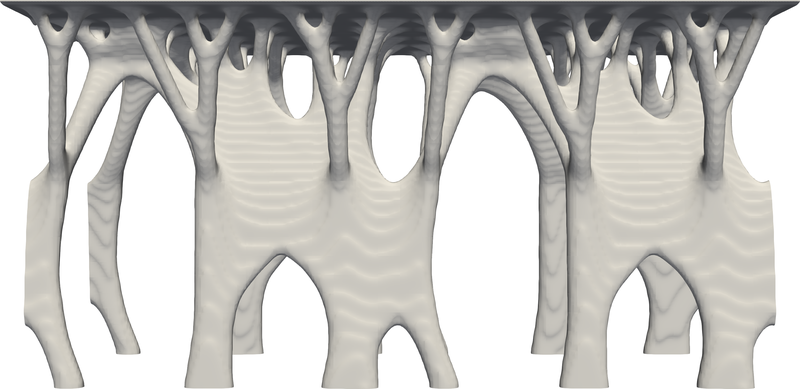}
  \includegraphics[height=.30\textwidth]{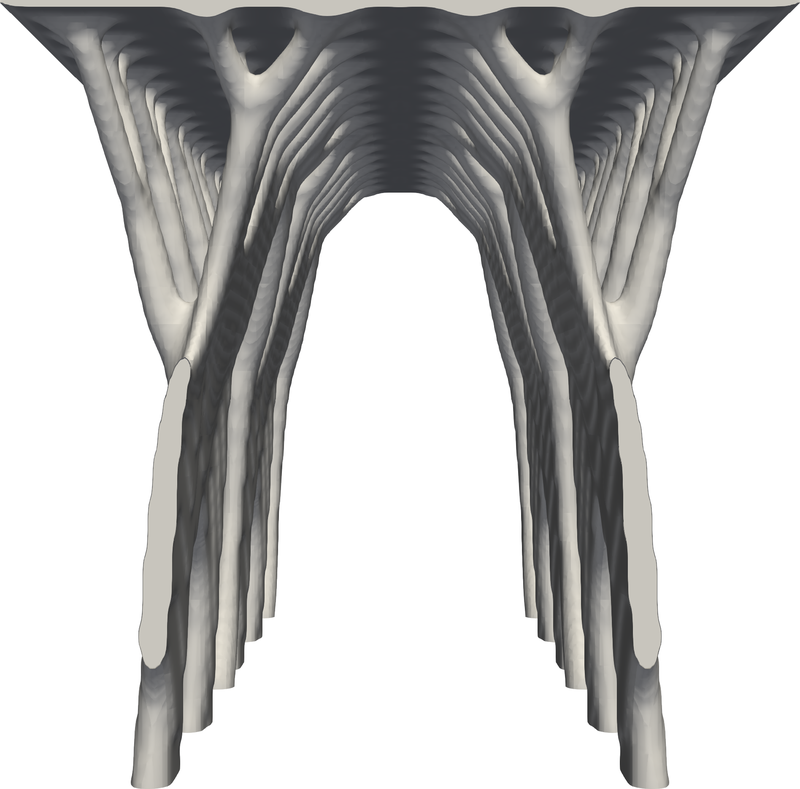}\\
  \includegraphics[height=.60\textwidth,angle=-90]{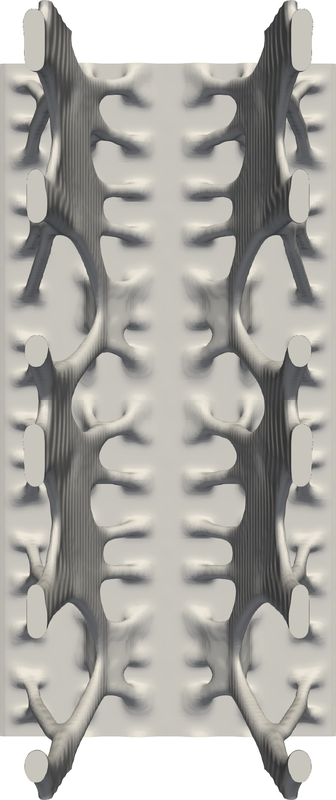}
 \end{minipage}
 \caption{Stochastic compliance optimization of a 3D bridge structure from a uniform initial density distribution.}
 \label{fig:bridge3D_full_compliance_uniform}
\end{figure}

\begin{figure}
 \centering
 \begin{minipage}{0.49\textwidth}
  \centering
  \includegraphics[width=.9\textwidth]{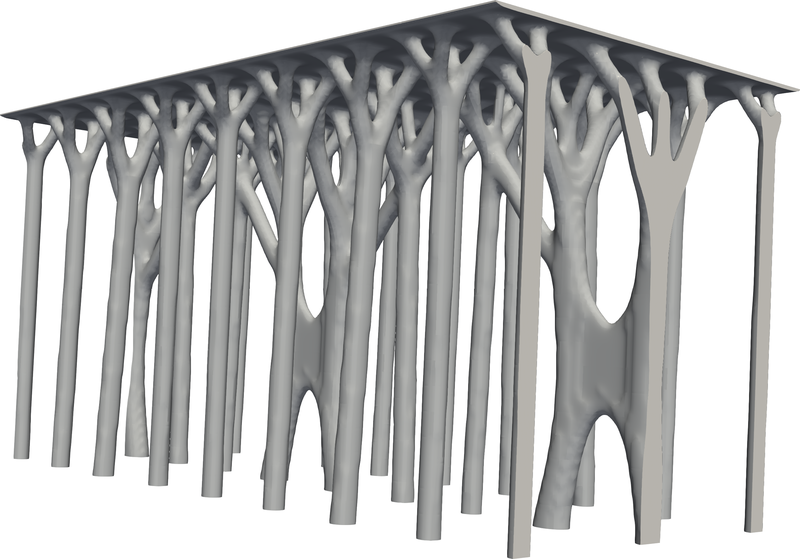}
  \caption*{(a) $l=0.01$, $j=12.10$}
 \end{minipage}
 \quad
 \begin{minipage}{0.4\textwidth}
  \includegraphics[height=.30\textwidth]{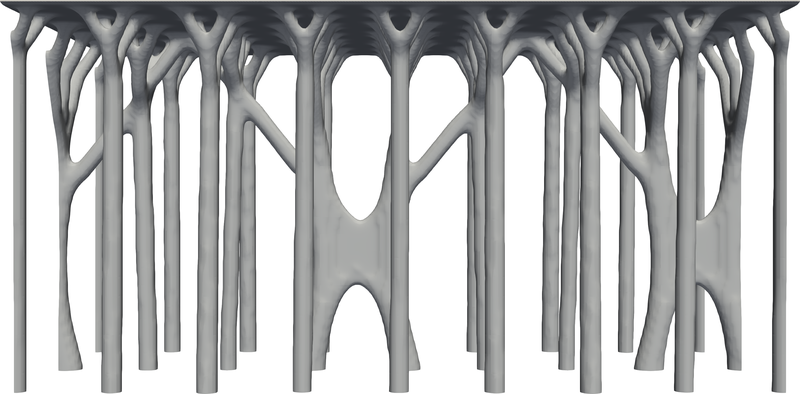}
  \includegraphics[height=.30\textwidth]{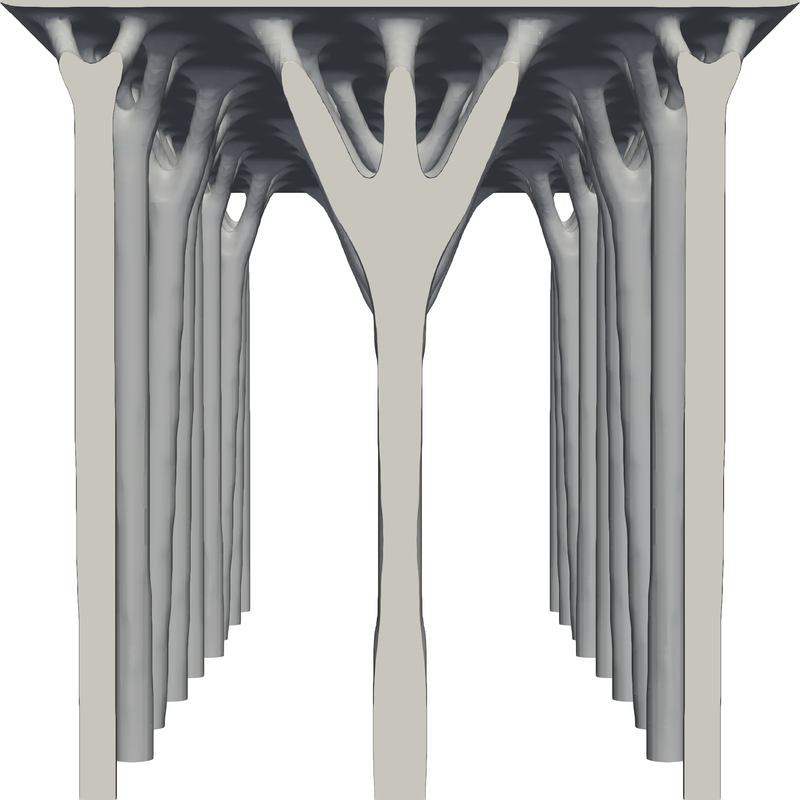}\\
  \includegraphics[height=.60\textwidth,angle=-90]{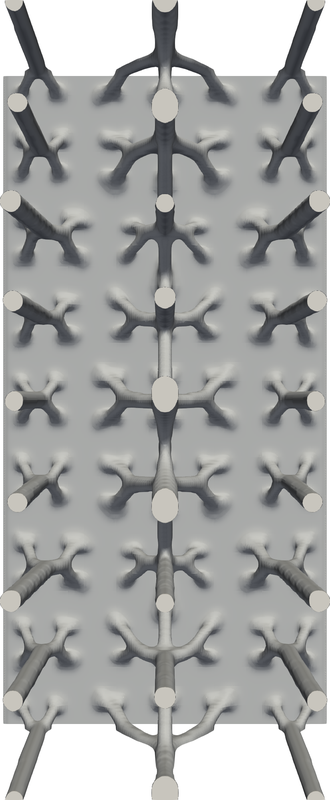}
 \end{minipage}

 \medskip

 \begin{minipage}{0.49\textwidth}
  \centering
  \includegraphics[width=.9\textwidth]{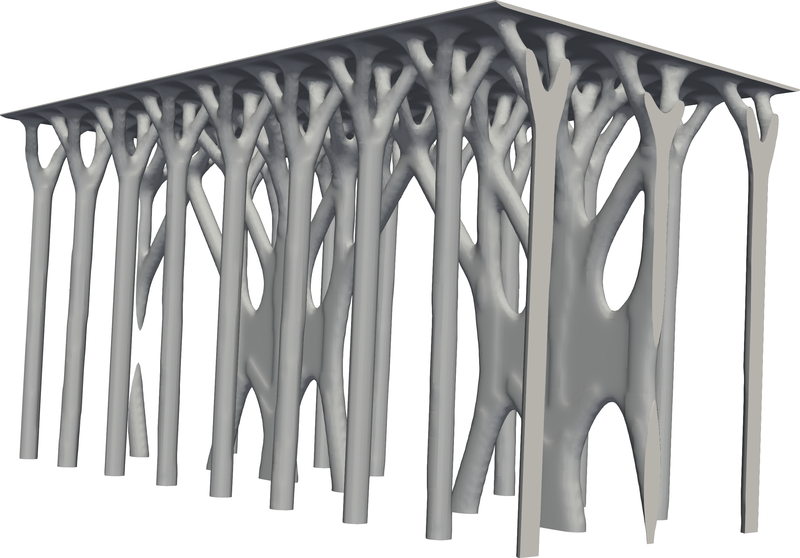}
  \caption*{(b) $l=0.02$, $j=12.46$}
 \end{minipage}
 \quad
 \begin{minipage}{0.4\textwidth}
  \includegraphics[height=.30\textwidth]{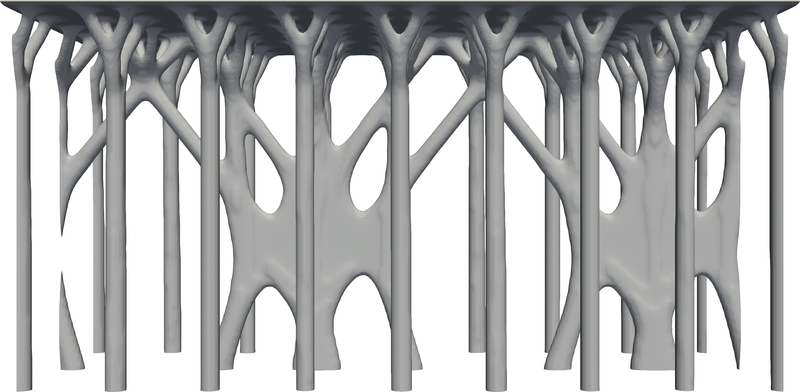}
  \includegraphics[height=.30\textwidth]{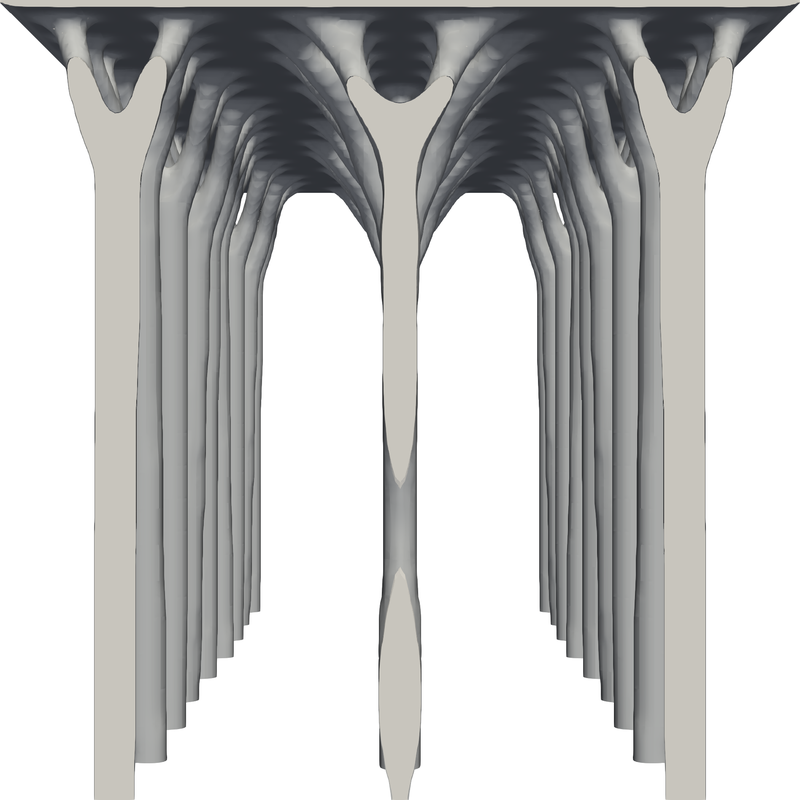}\\
  \includegraphics[height=.60\textwidth,angle=-90]{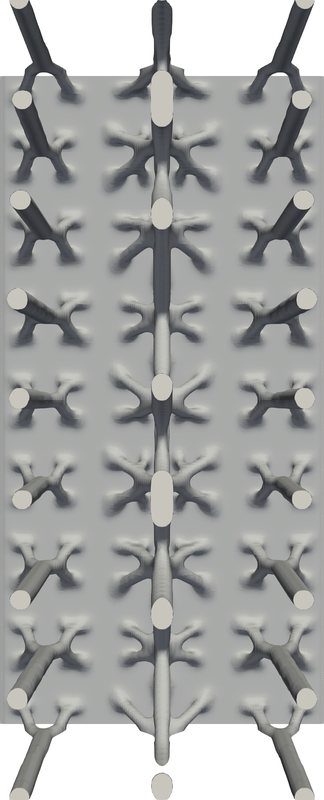}
 \end{minipage}

 \medskip

 \begin{minipage}{0.49\textwidth}
  \centering
  \includegraphics[width=.9\textwidth]{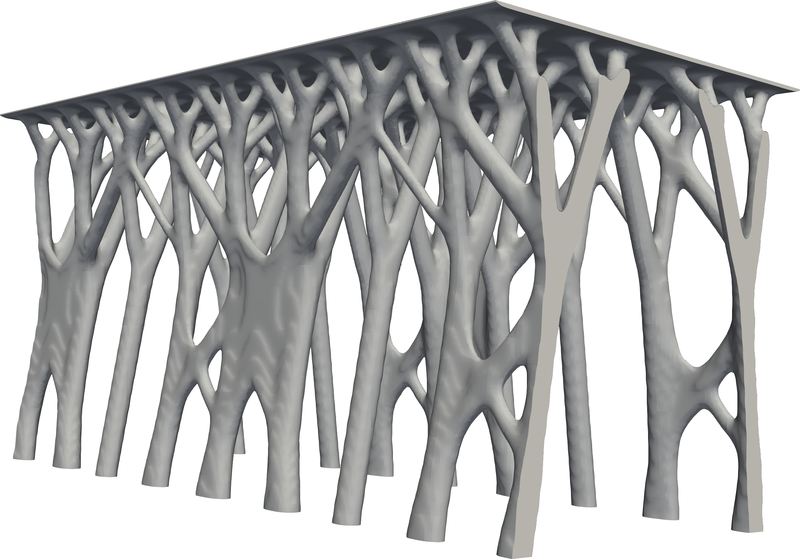}
  \caption*{(c) $l=0.03$, $j=12.91$}
 \end{minipage}
 \quad
 \begin{minipage}{0.4\textwidth}
  \includegraphics[height=.30\textwidth]{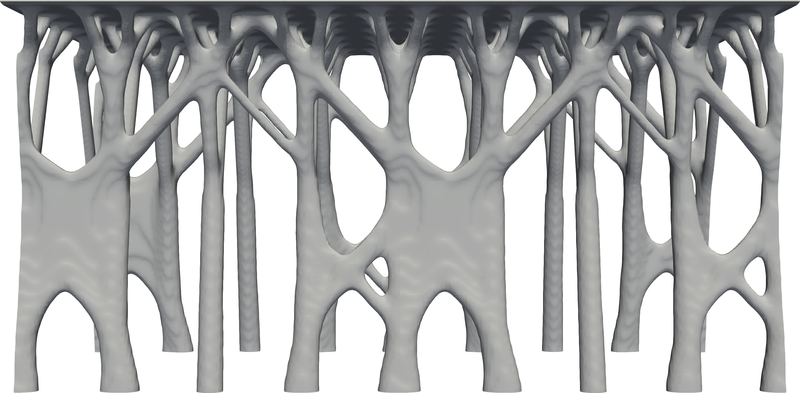}
  \includegraphics[height=.30\textwidth]{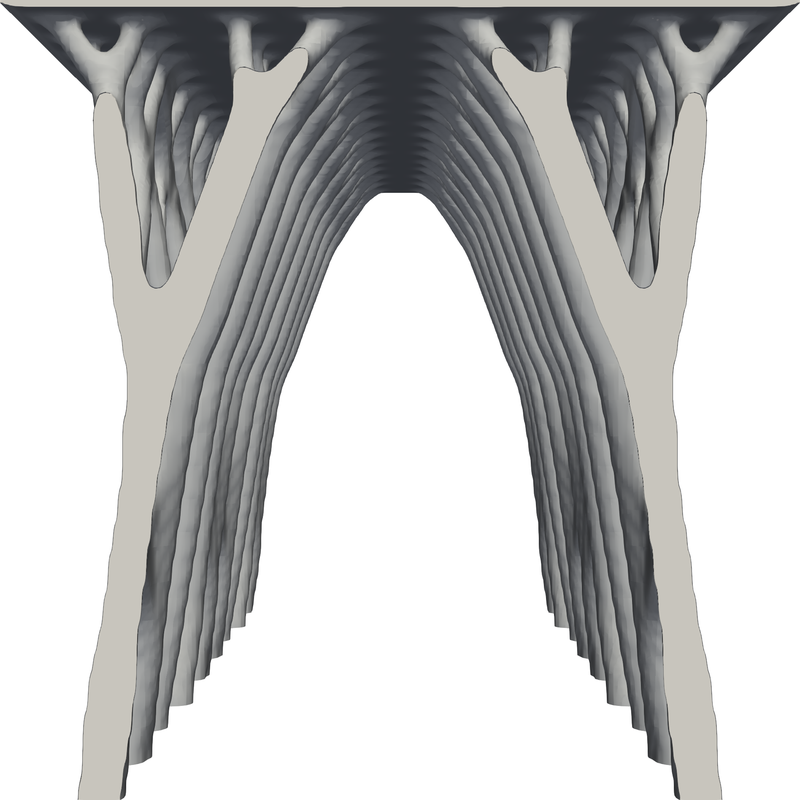}\\
  \includegraphics[height=.60\textwidth,angle=-90]{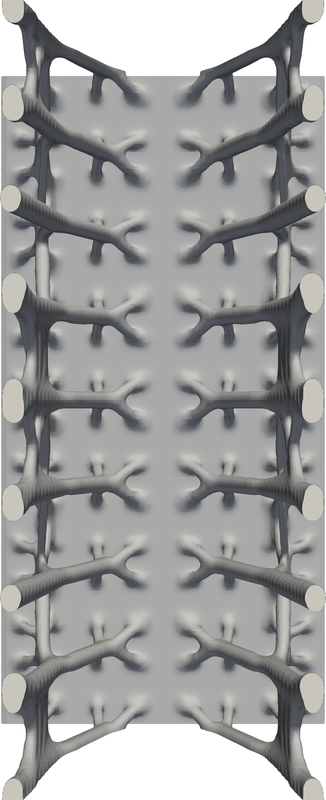}
 \end{minipage}

 \caption{Stochastic compliance optimization of a 3D bridge structure from a prescribed initial density distribution.}
 \label{fig:bridge3D_full_compliance}
\end{figure}

The final solid mechanics example demonstrates topology optimization of a 3D bridge structure with designs shown in Figures~\ref{fig:bridge3D_full_compliance_deterministic},~\ref{fig:bridge3D_full_compliance_uniform},~\ref{fig:bridge3D_full_compliance}.
The bridge is loaded with a unit vertical distributed load.
The optimization is performed only on a quarter of the domain shown in the figures.
Periodic boundary conditions are enforced along vertical planes perpendicular to the deck axis and a symmetric boundary condition is placed on a vertical plane aligned with the long deck axis.
The bridge is supported at the bottom of the domain.
The computational domain $D=(0,1)\times(0,0.5)\times(0,1)$ is discretized with approximately 20M elements, resulting in roughly 60M DOFs for the linear elasticity solver.
We employ the robust formulation~\cite{Wang2011} with thresholds of $\eta_d = 0.3$ and $\eta_e = 0.7$ to enforce a strict length scale on the design.
The filter parameter $r$ is set to 0.04, and the volume constraint is set to 7\% of the design volume.

The first two designs shown in Figure~\ref{fig:bridge3D_full_compliance_deterministic} are executed with deterministic vertical load and demonstrate the impact of the initial density distribution on the final design.
A predefined initial distribution affects the number of columns supporting the bridge and their locations.
Similar to the 2D case, we add a Mat\'ern-type random field, generated via the SPDE method, to represent the force on the bridge aligned with the deck axis.
The experiments were executed with different correlation lengths for the random field and uniform and prescribed initial distributions.
For small correlation lengths, the optimization tries to balance the forces close to the deck of the bridge by using inclined support legs connected to a well-defined column, which transfers the load to the support.
Increasing the correlation length results in the development of diagonal elements and, for very long correlation lengths, shear walls.
Since second-order effects, i.e., loss of stability, are not accounted for, the most effective way to transfer the excess horizontal force to the support would be through thin shear walls.
However, the enforced length scale does not allow the formation of such thin features, and the optimization concentrates the material in column-like elements.
For the nonuniform initial design density and short correlation lengths, the optimization process develops a single shear wall along the axis of the deck.
The columns close to the two ends of the deck transfer only the vertical load to the support and do not affect the horizontal stiffness of the bridge.
In contrast to the 2D case, the unbalanced part of the nonuniform horizontal random excitation introduces an equivalent torsion force, which requires moving the vertical supports close to the two ends of the deck.
This effect is observed for the long correlation length case shown in Figure~\ref{fig:bridge3D_full_compliance} (c).
We emphasize that the presented examples aim to demonstrate the scalability and versatility of the SPDE approach, and more realistic, physically-relevant bridge loads can be obtained using pointwise, memoryless GRF transformations or non-Gaussian random fields.
The SPDE approach and the current MFEM implementation can be utilized to generate non-Gaussian random fields with Mat\'ern-type covariances by replacing the Gaussian white noise with a non-Gaussian noise \cite{Bolin2014, Wallin2015}.
Discussions covering further generalizations to spatio-temporal processes can be found in \cite{Lindgren2022}.

\section{Conclusions}
\label{sec:conclusions}

This work aimed at incorporating realistic uncertainties into computational mechanics and design optimization workflows using Matérn-type Gaussian random fields generated via the SPDE method.
We focused on aleatoric uncertainties arising from environmental influences, variating material properties, and geometric uncertainties.
We consolidated an efficient numerical scheme, previously scattered in the literature, and provided an open-source implementation, which is now part of the widely used MFEM library.
We further identified crucial requirements for effective random field generation within the modern scientific computing landscape.
These requirements encompass the ability to handle diverse correlation lengths and field smoothness, address intricate domains and manifolds, align with finite element data structures, and computational costs on the order of the physical problem subject to uncertainties.

We find that the SPDE method effectively satisfies all the identified requirements.
While alternative methods like FFT and EOLE offer advantages in specific niche areas, the SPDE method's overall versatility and scalability outweigh these isolated shortcomings.
Its ability to extend naturally to manifolds enables uncertainty quantification of complex geometries and paves the way for design optimization under uncertainty on manifolds embedded in 3D space.
Moreover, its integration into design problems bears the potential to discover novel topological features previously undetectable with established random field discretization methods.
We demonstrated the method's capabilities through an extensive set of examples including the construction microstructures mimicking fiber materials, handling geometric uncertainties of celebral aneurysms, and optimizing topologies for thermal and structural design, even on 2D surfaces embedded in 3D.

While the SPDE method demonstrates remarkable capabilities, it has limitations.
The numerical scheme can become expensive for specific parameter regimes, particularly when the smoothness parameter $\nu$ approaches infinity, leading to a long sequence of equations in~\eqref{eq:SecondOrderPDESequence}.
Solving the linear system associated with strongly anisotropic fields can also incur significant computational costs.
Furthermore, the current implementation and theory are confined to Matérn-type covariances.
Specific use cases might encounter covariance structures diverging from the Mat\'ern kernel~\eqref{eq:MaternCovariance} or might even necessitate non-Gaussian fields.
Various (local) transformations can be applied to construct fields with different statistical properties and may alleviate some of these issues.
However, whether or not the Mat\'ern covariance defines an appropriate spatial structure for a given problem must be assessed on a case-by-case basis.
Ideally, such a decision should be supported by experimental data, aiding the identification of the covariance structure.
While the general behavior of the SPDE covariance is well understood, obtaining a closed-form expression for~\eqref{eq:CovarianceOfSPDE_eigen} on embedded manifolds and other complex domains often proves challenging.
However, even in such cases, the covariance~\eqref{eq:CovarianceOfSPDE_eigen} is typically well-approximated by~\eqref{eq:MaternCovariance} given a correlation length smaller than the characteristic features of the domain.

Our observations hold significant implications for a broad audience utilizing finite elements, including researchers in computational mechanics, design optimization, biomechanics, and inverse problems.
The SPDE method simplifies incorporating spatial uncertainties into diverse scientific computing workflows. Additionally, the open-source implementation allows researchers to leverage this powerful tool with minimal effort within existing MFEM projects.
These advancements are expected to drive the widespread adoption of the SPDE method.

Looking ahead, several promising avenues for future research present themselves.
Extending the SPDE method beyond Matérn-type covariances would broaden its applicability to encompass a broader range of real-world uncertainty scenarios.
Additionally, exploring the potential of neural operators for accelerating computations holds promise in enhancing the method's efficiency.
Furthermore, developing dedicated numerical schemes tailored to simultaneously solving the equations arising from the rational approximation in~\eqref{eq:RA_approximant} can significantly improve the method's computational performance.
Collectively, these advancements aim to further refine and enhance the SPDE method for various applications across scientific computing domains.

In summary, this work establishes the SPDE method as a powerful and versatile tool for describing aleatoric uncertainties in modern scientific computing, ultimately leading to substantial advancements in uncertainty quantification, computational mechanics, and design optimization.

\section*{Acknowledgements}

This work was performed under the auspices of the U.S.~Department of Energy by Lawrence Livermore National Laboratory under Contract DE-AC52-07NA27344, the LLNL-LDRD Program under Project tracking No.~22-ERD-009, and Differentiating Large-Scale Finite Element Applications project supported by the U.S.~Department of Energy, Office of Science, Office of Advanced Scientific Computing Research.
Release number LLNL-JRNL-861284.
This material is based upon work supported by the U.S.~Department of Energy Office of Science, Early Career Research Program under Award Number DE-SC0024335.
TD was sponsored by LLNL'S Computing Scholar Program and the Wolfgang Gentner Programme of the German Federal Ministry of Education and Research (grant no. 13E18CHA).
BW gratefully acknowledges the financial support provided by the German Science Foundation (DFG) under project number 465242983 within the priority programme \textit{SPP 2311: Robust coupling of continuum-biomechanical in silico models to establish active biological system models for later use in clinical applications -- Co-design of modeling, numerics and usability} (WO 671/20-1).
TD and BW would like to thank Markus Muhr, Stephan Lunowa, and Fabian Holzberger for insightful discussions regarding the cerebral aneurysm examples.

\appendix

\section{Implementation and Scalability}
\label{apx:implementation}

We implemented the SPDE method with MFEM~\cite{Anderson2021,andrej2024mfem}, and the
resulting MPI-parallel solver is part of MFEM's 4.6 release\footnote{MFEM 4.6,
 \href{https://mfem.org/download/}{mfem.org/download/},
 available in \texttt{miniapps/spde}}. While the numerics are non-trivial, using
the solver in MFEM-based projects only requires adding three lines of C++ code,
see lines 11 to 13 in Algorithm~\ref{alg:SPDESolver}. The statement in line 11
initializes the solver and assembles the mass and stiffness matrices in
\eqref{eq:DiscreteWeakFormSPDE}. Line 12 sets the seed for all MPI ranks. Lastly,
line 13 assembles the white noise, constructs the linear systems arising from
the rational approximation~\eqref{eq:RA_approximant}, and solves them with a
preconditioned conjugate gradient algorithm~\cite{Wathen2015,Xu2017}.

\begin{algorithm}[ht]
 \caption{Application example of the MFEM based SPDE-solver
  (assumes pre-existing finite element space object "fespace"; see MFEM
  documentation for details).
 }
 \label{alg:SPDESolver}
 \begin{algorithmic}[1]
  \LComment{Define the fractional PDE solution}
  \State \textbf{ParGridFunction} u(\&fespace);
  \State u = 0.0;
  \LComment{Define GRF parameter}
  \State const double nu = 2.0;
  \State const double lengthscale = 0.05;
  \State const int seed = 1;
  \LComment{Define the boundary conditions (defaults to Neumann).}
  \State spde::\textbf{Boundary} bc;
  \LComment{Solve the SPDE problem}
  \State spde::\textbf{SPDESolver} solver(nu, bc, \&fespace, lengthscale, lengthscale, lengthscale);
  \State solver.\textit{SetupRandomFieldGenerator}(seed);
  \State solver.\textit{GenerateRandomField}(u);
 \end{algorithmic}
\end{algorithm}

We argue that the implementation inherits its scalability from MFEM~\cite{Anderson2021,andrej2024mfem} and
\textit{hypre}~\cite{hypre} because these libraries govern the computationally expensive aspects of the algorithm, i.e., the finite element assembly and solving the linear systems.
Detailed scalability studies have been performed and presented by Fischer and co-workers~\cite{Fischer2020}.
They considered solving the equation $\left(- \nabla \cdot \mu \nabla + \beta \right) u = f$ with MFEM and other numerical libraries via a diagonally preconditioned conjugate gradient method and measured the performance using (up to) $16,384$ MPI ranks distributed among 512 nodes.
Their results demonstrate MFEM's ability to scale on top-tier supercomputing systems for equations similar to~\eqref{eq:SPDE_fullspace}.
Additional performance analyses on GPU architectures may be found in~\cite{andrej2024mfem}.
Unlike~\cite{Fischer2020}, we use \textit{hypre}'s BoomerAMG as a preconditioner.
Henson and Yang~\cite{Henson2002} investigated and demonstrated the preconditioner's effectiveness and parallel scalability for a set of toy problems with different characteristics.
These problems included the Laplace equation and related problems with anisotropy and the absence of symmetry.
They found that the preconditioner works well across the set of example problems and scales to tens of millions of unknowns and more than a thousand processors.

During the development of our solver and the examples presented in
Section~\ref{sec:num_ex}, the solver routinely ran on hundreds of cores at Lawrence Livermore National Laboratory.

\section{Sample Size Choice}
\label{apx:sample_size}

We optimize the topology with the Method of Moving Asymptotes~\cite{Svanberg1987} using a fixed set of samples.
For the heat sink design in Section~\ref{sub:heat_sink}, we used a sample size of $N=300$.
For the bride design in Section~\ref{sub:bridge}, we chose $N=400$.
To determine $N$ in a problem-specific setup, we carefully investigate the effect of the sample size on the objective function and optimized design.
Here, we show our numerical experiments justifying this choice for the heat sink optimization.
Analogous experiments were carried out for the bridge examples.

\begin{figure}[ht]
 \centering
 \vspace*{0.3cm}
 \begin{minipage}[b]{0.05\textwidth}\centering
  \hspace{0.1cm}
 \end{minipage}
 \begin{minipage}[b]{0.185\textwidth}\centering
  \small
  $N = 50$
 \end{minipage}
 \begin{minipage}[b]{0.185\textwidth}\centering
  \small
  $N = 100$
 \end{minipage}
 \begin{minipage}[b]{0.185\textwidth}\centering
  \small
  $N = 200$
 \end{minipage}
 \begin{minipage}[b]{0.185\textwidth}\centering
  \small
  $N = 300$
 \end{minipage}
 \begin{minipage}[b]{0.185\textwidth}\centering
  \small
  $N = 400$
 \end{minipage}\\ \vspace*{0.3 cm}
 \begin{minipage}[b]{0.05\textwidth}\centering
  \small
  \rotatebox{90}{
   \hspace*{1.0cm}
   (a) $l=0.2$
  }
 \end{minipage}
 \begin{minipage}[b]{0.185\textwidth}\centering
  \includegraphics[width=1.0\textwidth]{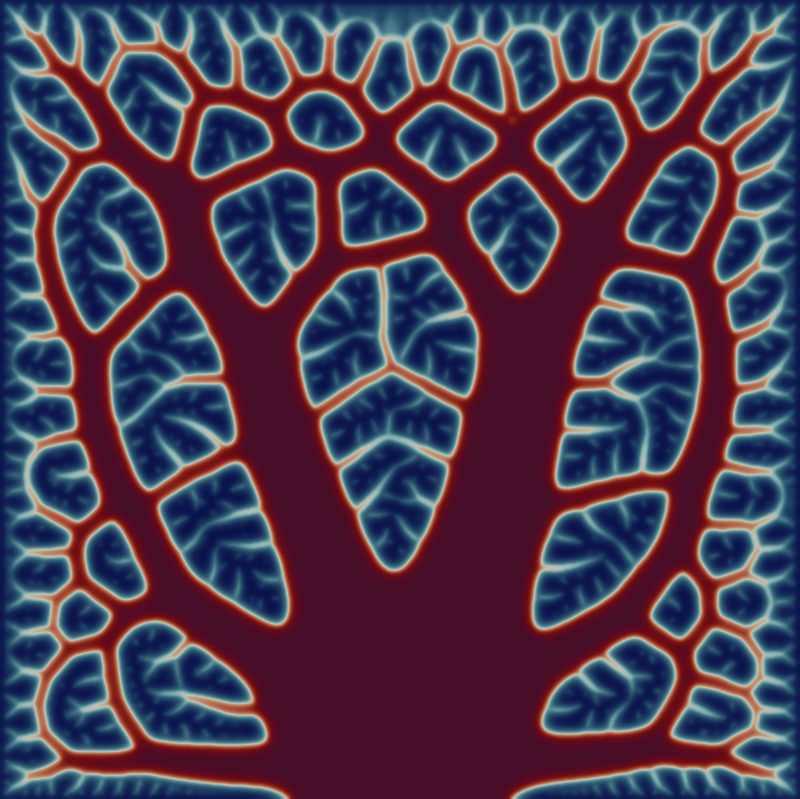}
  \footnotesize
  $j = 0.466$
 \end{minipage}
 \begin{minipage}[b]{0.185\textwidth}\centering
  \includegraphics[width=1.0\textwidth]{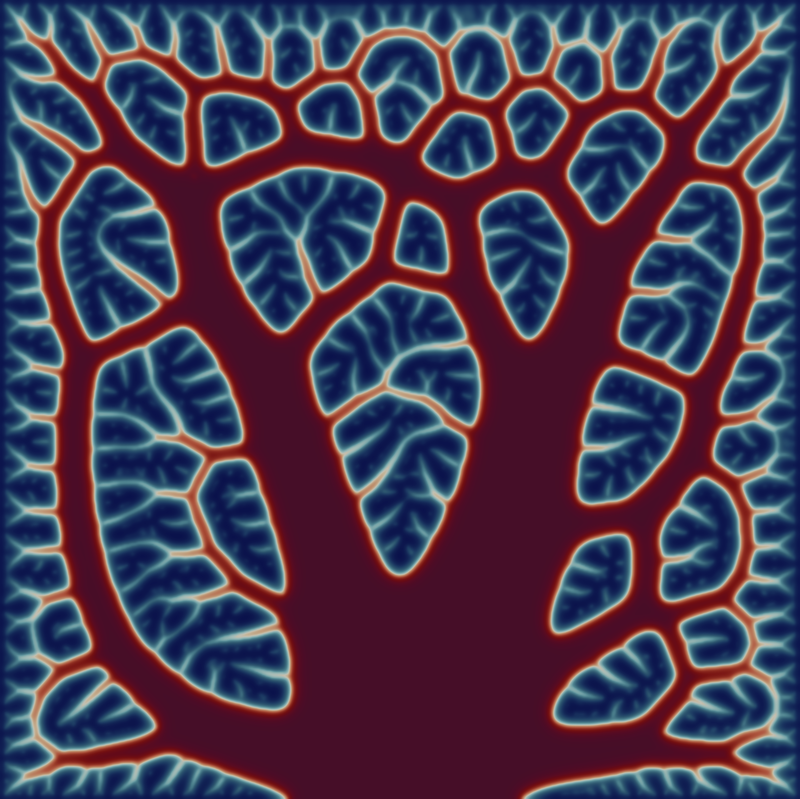}
  \footnotesize
  $j = 0.524$
 \end{minipage}
 \begin{minipage}[b]{0.185\textwidth}\centering
  \includegraphics[width=1.0\textwidth]{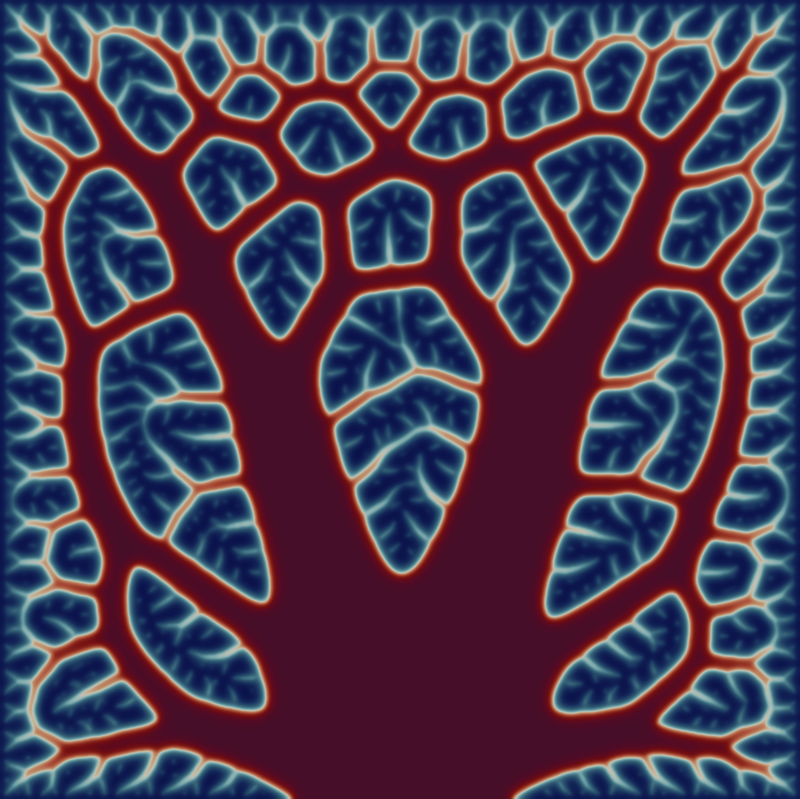}
  \footnotesize
  $j = 0.577$
 \end{minipage}
 \begin{minipage}[b]{0.185\textwidth}\centering
  \includegraphics[width=1.0\textwidth]{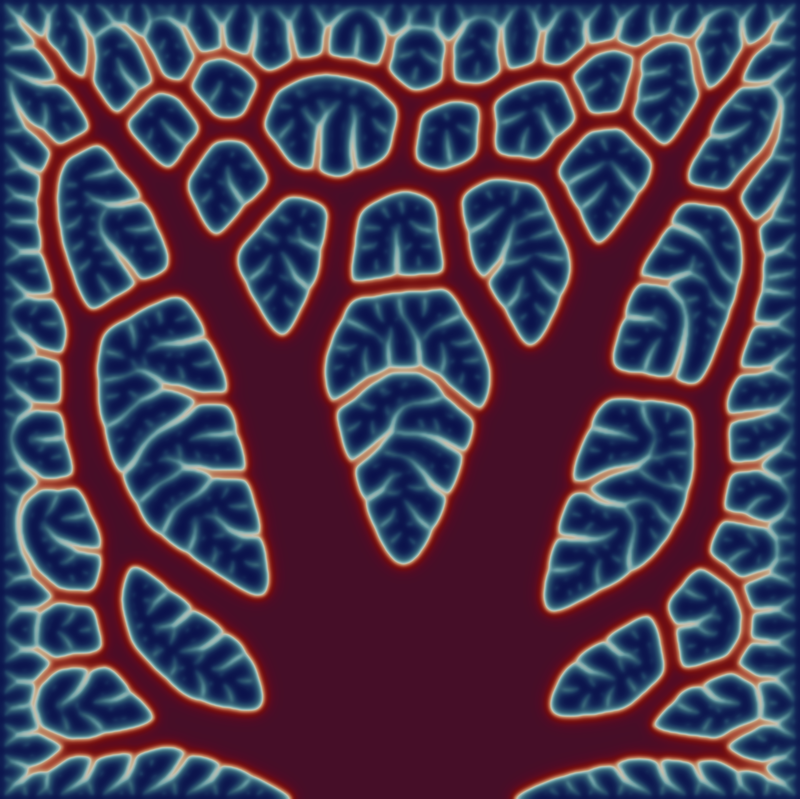}
  \footnotesize
  $j = 0.589$
 \end{minipage}
 \begin{minipage}[b]{0.185\textwidth}\centering
  \includegraphics[width=1.0\textwidth]{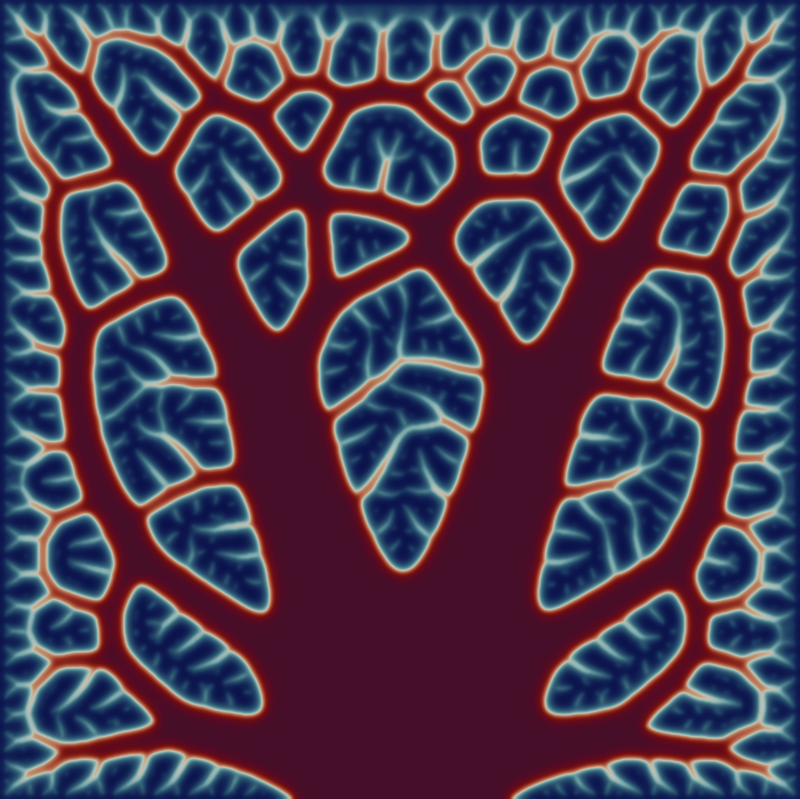}
  \footnotesize
  $j = 0.587$
 \end{minipage} \\ \vspace*{0.3cm}
 \begin{minipage}[b]{0.05\textwidth}\centering
  \small
  \rotatebox{90}{
   \hspace*{1.0cm}
   (b) $l=0.05$
  }
 \end{minipage}
 \begin{minipage}[b]{0.185\textwidth}\centering
  \includegraphics[width=1.0\textwidth]{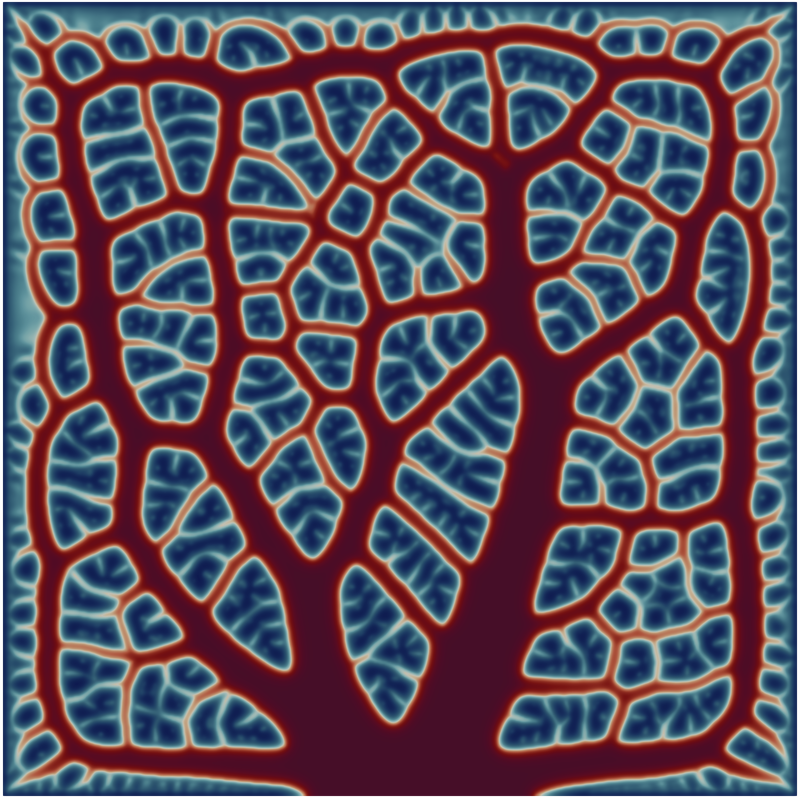}
  \footnotesize
  $j = 0.062$
 \end{minipage}
 \begin{minipage}[b]{0.185\textwidth}\centering
  \includegraphics[width=1.0\textwidth]{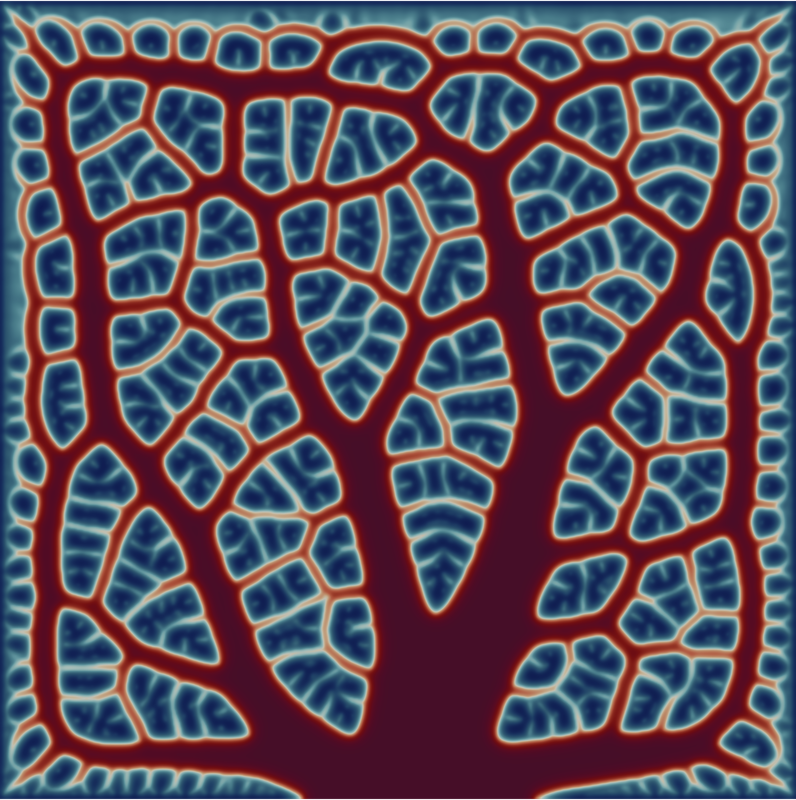}
  \footnotesize
  $j = 0.068$
 \end{minipage}
 \begin{minipage}[b]{0.185\textwidth}\centering
  \includegraphics[width=1.0\textwidth]{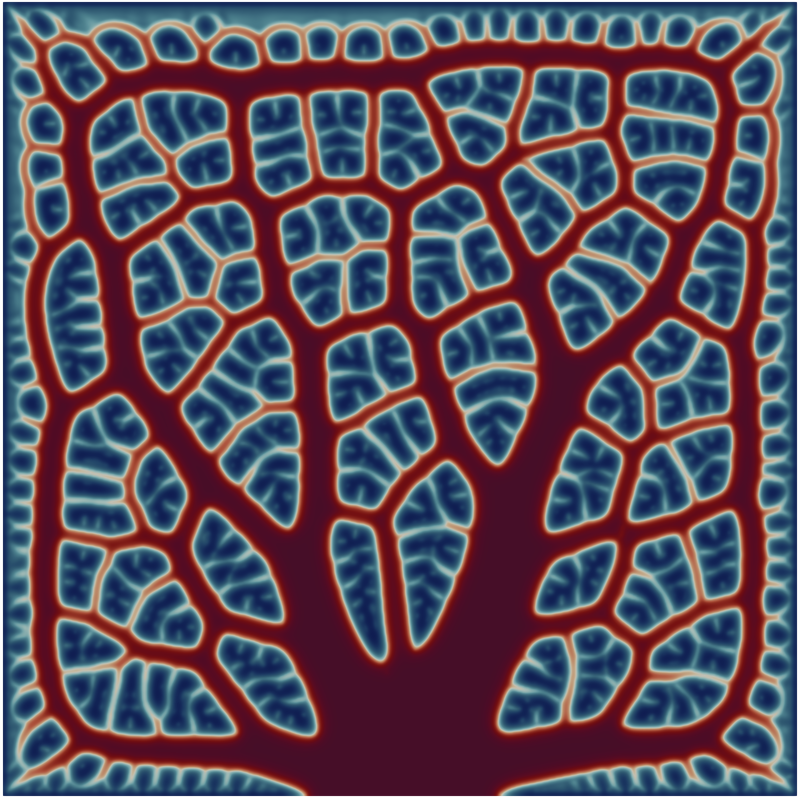}
  \footnotesize
  $j = 0.072$
 \end{minipage}
 \begin{minipage}[b]{0.185\textwidth}\centering
  \includegraphics[width=1.0\textwidth]{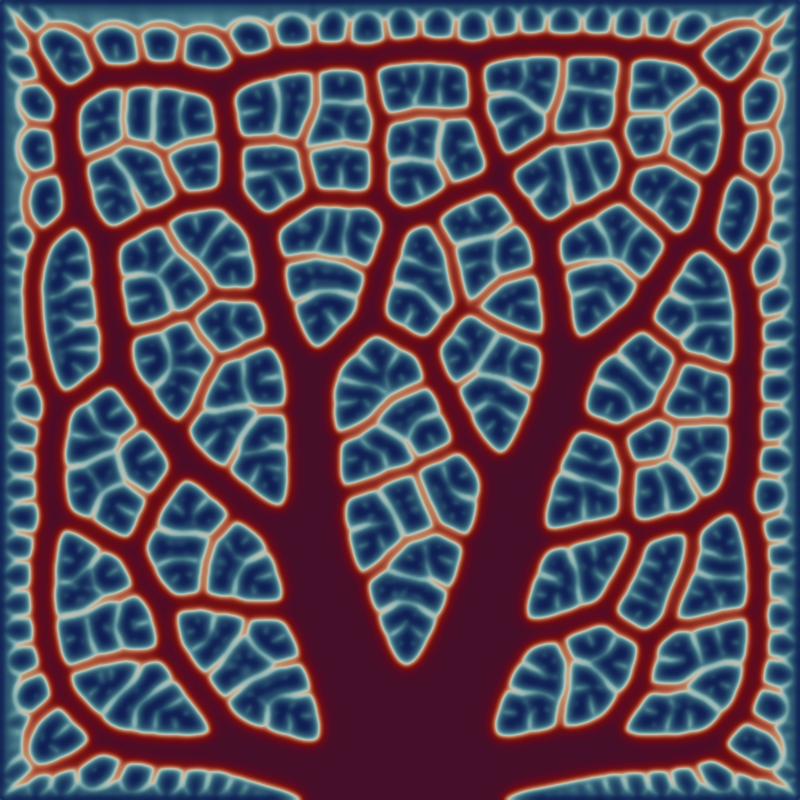}
  \footnotesize
  $j = 0.074$
 \end{minipage}
 \begin{minipage}[b]{0.185\textwidth}\centering
  \includegraphics[width=1.0\textwidth]{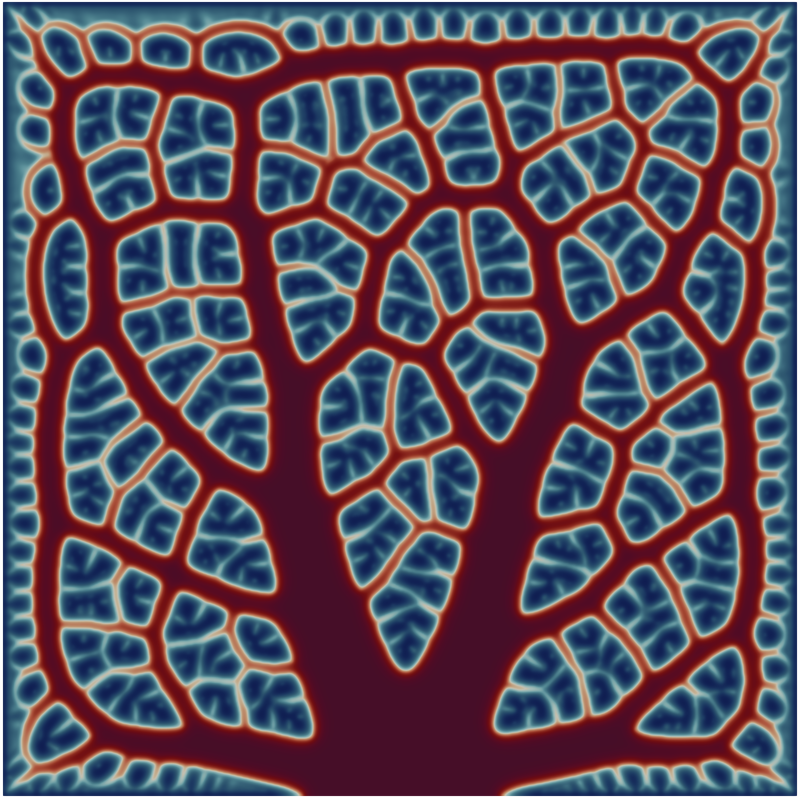}
  \footnotesize
  $j = 0.074$
 \end{minipage}
 \caption{
  Effect of the sample size on the density $\tilde{\rho}$ and objective function $j$.
  The sample size increases from left to right.
 }
 \label{fig:thermal_compliance_stochastic_sample_size}
\end{figure}

The numerical experiments proceed akin to Section~\ref{sec:stochatic_thermal_compliance} with the notable difference that we use different sample sizes $N \in \{50, 100, 200, 300, 400\}$.
We choose the correlation lengths $l=0.2$ and $l=0.05$ for our numerical experiments, representing the long and short correlation lengths considered in this work.
We expect that larger correlation lengths require fewer samples than short correlation lengths.
Figure~\ref{fig:thermal_compliance_stochastic_sample_size} illustrates the result of this experiment, confirming this hypothesis.
Comparing the cases $N=50$ and $N=400$, we find that the longer correlation length yields more similar designs than the short correlation length.
Overall, increasing the sample size increases the symmetry, i.e., the main branches resulting from the optimization become more evenly balanced.
Regarding the objective function $j$, we find that its value converges once we pass 300 samples.
This is also reflected in the design, which is very similar for the cases $N=300$ and $N=400$.
We conclude that computationally feasible increases beyond the sample size of $N=300$ barely affect the design and limit ourselves to $300$ samples for each heat sink optimization experiment.

\bibliographystyle{elsarticle-num}
\bibliography{references}

\end{document}